\documentclass[%
 amsmath,amssymb,
 aps,
]{revtex4-1}

\usepackage{graphicx}
\usepackage{dcolumn}
\usepackage{bm}
\usepackage{float}
\usepackage{rotating}

\begin{document}
\maxdeadcycles=100000
\preprint{APS/123-QED}

\begin{center} 
{\bf\Large More than six hundreds new families 
of \\ Newtonian periodic planar collisionless three-body orbits }

\vspace{0.5cm}

Xiaoming Li$^1$ and  Shijun Liao$^{1, 2, *}$ 

\vspace{0.5cm}

$^1$ School of Naval Architecture, Ocean and Civil Engineering, Shanghai Jiaotong University, China \\
$^2$ Ministry-of-Education Key Laboratory in Scientific and Engineering Computing, Shanghai 200240, China\\ 
* The corresponding author:  sjliao@sjtu.edu.cn
\end{center}

\hspace{-0.5cm} {\bf Abstract}
{\em The famous three-body problem can be traced back to Isaac Newton in  1680s.  In the 300 years since  this  ``three-body problem''  was first recognized,  only three families of periodic solutions had been found, until 2013 when \v{S}uvakov and Dmitra\v{s}inovi\'{c} [Phys. Rev. Lett. $\bm{110}$, 114301 (2013)] made a breakthrough to numerically find 13 new distinct periodic orbits, which belong  to 11 new families of Newtonian planar three-body problem with equal mass and zero angular momentum.   In this paper,  we  numerically  obtain 695  families of Newtonian periodic  planar  collisionless  orbits of three-body system with equal mass and zero angular momentum  in  case  of  initial conditions  with isosceles collinear configuration, including the well-known Figure-eight family found by Moore in 1993, the 11 families found by \v{S}uvakov and Dmitra\v{s}inovi\'{c} in 2013, and more than 600 new families that have been never reported, to the best of our knowledge.   With the definition of the average period $\bar{T} = T/L_f$, where $L_f$ is the length of the so-called ``free group element'', these 695 families  suggest that  there should  exist the quasi Kepler's third law   $ \bar{T}^* \approx 2.433 \pm 0.075$ for the considered case,  where $\bar{T}^*= \bar{T} |E|^{3/2}$ is the scale-invariant  average  period and $E$ is its total kinetic and potential energy, respectively.     The movies of these 695 periodic orbits in the real space and the corresponding close curves on the ``shape sphere''  can be found via the website:}\url{http://numericaltank.sjtu.edu.cn/three-body/three-body.htm}

\vspace{0.5cm}

\hspace{-0.5cm} {\bf PACS numbers} 45.50.Jf, 05.45.-a, 95.10.Ce

\vspace{0.5cm}

\section{Introduction}
The famous three-body problem \cite{Musielak2014} can be traced back to Isaac Newton in  1680s.   According to Poincar{\' e} \cite{Poincare1890},  a three-body system is not integrable in general.  Besides, orbits of three-body problem are often chaotic \cite{Lorenz1963}, say,  sensitive to initial conditions \cite{Poincare1890},  although there exist periodic orbits in some special  cases.  In the 300 years since  this  ``three-body problem'' \cite{Musielak2014}  was first recognized, only three families of periodic solutions had been found,  until 2013 when \v{S}uvakov and Dmitra\v{s}inovi\'{c} \cite{Suvakov2013} made a breakthrough to find 13 new distinct periodic collisionless
orbits belonging to 11 new families of Newtonian planar three-body problem with equal mass and zero angular momentum.  Before their elegant work, only three families of periodic three-body orbits were found: (1) the Lagrange-Euler family  discovered by Lagrange and Euler in 18th century; (2) the Broucke-Hadjidemetriou-H\'{e}non family \cite{Broucke1975, Hadjidemetriou1975A,Hadjidemetriou1975B, Henon1976,Henon1977}; (3) the Figure-eight family, first discovered numerically by Moore \cite{More1993} in 1993 and rediscovered by Chenciner and Montgomery \cite{Chenciner2000} in 2000, and then extended to the rotating case \cite{Nauenberg2001, Chenciner2005, Broucke2006, Nauenberg2007}. In 2014, Li and Liao \cite{Li2014} studied the stability of the periodic orbits in \cite{Suvakov2013}.  In 2015 Hudomal \cite{Hudomal2015} reported 25 families of periodic orbits, including the 11 families found in \cite{Suvakov2013}.    Some studies on topological dependence of Kepler's third law for three-body problem  were  currently reported  \cite{Dmitrasinovic2015, Jankovic2016}.

Recently, \v{S}uvakov and Dmitra\v{s}inovi\'{c} \cite{Suvakov2014a} specifically illustrated their numerical strategies  used in \cite{Suvakov2013}  for their 11 families of periodic orbits with the periods $T \leq 100$.  They suggested that more new periodic solutions are expected to be found when $T\geq 100$.   In this paper,  we used a different numerical approach to solve the same problem, i.e.  Newtonian planar three-body problem with equal mass and zero angular momentum, but gained 695 families of periodic orbits without collision, i.e. 229 families within $T \leq 100$ and 466 families within $100< T\leq 200$,  including the well-known Figure-eight family found by Moore \cite{More1993}, the 11 families found by \v{S}uvakov and Dmitra\v{s}inovi\'{c} \cite{Suvakov2013},  the 25  families mentioned in \cite{Hudomal2015},  and especially more than 600 new families that have been never reported, to the best of our knowledge.

\section{Numerical  approaches}

The motions of Newtonian planar three-body system are governed by the Newton's second law and gravitational law
\begin{equation}
  \ddot{\bm{r}}_{i}=\sum_{j=1,j\neq i}^{3} \frac{G m_{j}(\bm{r}_{j}-\bm{r}_{i})}{| \bm{r}_i-\bm{r}_j |^{3}}, \label{geq:3-body}
\end{equation}
where ${\bm r}_i$ and $m_j$ are the position vector and mass of the $i$th body $(i=1,2,3)$, $G$ is the Newtonian gravity coefficient,  and  the dot denotes the derivative with respect to the time $t$, respectively.  Like  \v{S}uvakov and Dmitra\v{s}inovi\'{c} \cite{Suvakov2013},  we consider a planar  three-body system with zero angular momentum in the case of  $G=1$, $m_1=m_2=m_3=1$, and  the initial conditions in case of the isosceles collinear configurations:
\begin{equation}
\left\{
\begin{array} {l}
  \bm{r}_1(0)=(x_1,x_2)=-\bm{r}_2(0), \;\;  \bm{r}_3(0)=(0,0), \\
 \dot{\bm{r}}_1(0)= \dot{\bm{r}}_2(0)=(v_1,v_2), \;\; \dot{\bm{r}}_3(0)=-2\dot{\bm{r}}_1(0),
 \end{array}
 \right.  \label{initial}
 \end{equation}
which are specified by the four parameters $(x_1, x_2, v_1, v_2)$.   Write $\bm{y}(t)=(\bm{r}_1(t), \dot{\bm{r}}_1(t))$.
A periodic solution with the period $T_0$ is the root of the equation $\bm{y}(T_0)-\bm{y}(0)=0$, where $T_0$ is unknown.  Note that $x_1=-1$ and $x_2=0$  correspond to the normal case considered in \cite{Suvakov2013} that regards $\bm{r}_1(0)=(-1,0)$ to be fixed.  However, unlike \v{S}uvakov and Dmitra\v{s}inovi\'{c} \cite{Suvakov2013}, we regard $x_1$ and $x_2$ as variables.    So, mathematically speaking,  we search for the periodic orbits of the same three-body problem using  a larger degree of freedom  than \v{S}uvakov and Dmitra\v{s}inovi\'{c} \cite{Suvakov2013}.

First,  like  \v{S}uvakov and Dmitra\v{s}inovi\'{c} \cite{Suvakov2013},  we use the  grid search method to  find  candidates  of  the  initial conditions $\bm{y}(0)=(x_1, x_2, v_1, v_2)$ for periodic orbits.   As is well known, the grid search method suffers from the curse of dimensionality.  In order to reduce the dimension of the search space, we set the initial positions $x_1=-1$ and $x_2=0$.  Then,  we search for the initial conditions of periodic orbits in the two dimensional plane: $v_1\in [0,1]$ and $v_2 \in [0,1]$.  We set 1000 points in each dimension and thus have one million grid points in the square search plane.  With these different $10^6$ initial conditions, the motion equations (\ref{geq:3-body}) subject to the initial conditions (\ref{initial})  are integrated up to the time $t = 100$ by means of the ODE solver dop853 developed by Hairer et al. \cite{Hairer1993},  which is based on an explicit Runge-Kutta method of order 8(5,3) in double precision with adaptive step size control.   The corresponding initial conditions and the period $T_0$  are chosen as the candidates when the return proximity function
\begin{equation}
|\bm{y}(T_0)-\bm{y}(0)| = \sqrt{\sum_{i=1}^{4}(y_i(T_0)-y_i(0))^2}   \label{def:return-proximity-function}
\end{equation}
 is less than $10^{-1}$.

Secondly, we modify  these  candidates  of the initial conditions  by  means  of  the Newton-Raphson method \cite{Farantos1995, Lara2002,  Abad2011}.    At this stage,  the motion equations are solved numerically by means of the same ODE solver dop853 \cite{Hairer1993}.   A  periodic  orbit  is  found when  the level of the return proximity function (\ref{def:return-proximity-function}) is less than $10^{-6}$.   Note that, different from the numerical approach in  \cite{Suvakov2013}, not only the initial velocity $\dot{\bm{r}}_1(0)=(v_1, v_2)$ but also the initial position $\bm{r}_1(0)=(x_1,x_2)$ are also modified.  In other words,  our numerical approach also allows $\bm{r}_1(0)=(x_1,x_2)$  to deviate from its initial guess $(-1,0)$.    With such kind of larger degree of freedom,  our approach  gives 137  families of periodic orbits,  including the well-known Figure-eight family \cite{More1993},  the 10 families found by \v{S}uvakov and Dmitra\v{s}inovi\'{c} \cite{Suvakov2013}, and lots of completely new families that have been never reported.

However,  one family reported in \cite{Suvakov2013} was {\em not}  among these 137 periodic orbits.  So, at least one periodic orbit was lost at this stage.   This is not surprising, since three-body problem is not integrable in general  \cite{Poincare1890}  and   might be  rather  sensitive  to  initial conditions, i.e. the butterfly-effect  \cite{Lorenz1963}.   For example,  Hoover  et al. \cite{Hoover2015} compared numerical simulations of a chaotic Hamiltonian system given by five symplectic and two Runge-Kutta integrators in {\em double precision}, and found that ``all  numerical methods are susceptible'', ``which severely limits the maximum time for which chaotic solutions can be accurate'', although ``all of these  integrators conserve energy almost {\em perfectly}''.  In fact,  there exist many examples which suggest that numerical noises have great influence on chaotic systems.   Currently,   some  numerical  approaches were  developed  to  gain reliable  results  of  chaotic  systems in a long (but finite) interval of time.  One of them is the so-called ``Clean Numerical Simulation'' (CNS) \cite{Liao2009, Liao2013-3b, Liao2013-Chaos,Liao2014-SciChina, Liao2015-IJBC, Lin2017},  which is based on the {\em arbitrary} order of Taylor  series method \cite{Barton1971, Corliss1982, Chang1994, Barrio2005}  in  {\em arbitrary}  precision \cite{Oyanarte1990, Viswanath2004}, and more importantly,  a check of solution verification (in a given interval of time) by comparing two simulations gained with different levels of numerical noise.

First,   we checked  the 137 periodic orbits by means of the high-order Taylor series method in the 100-digit precision with truncation errors less than $10^{-70}$, and guaranteed that they are indeed periodic orbits.    Especially,   we  further found the  additional   27  families of periodic orbits (with the periods less than 100)  by  means  of  the Newton-Raphson method \cite{Farantos1995, Lara2002,  Abad2011}  for the modifications of initial conditions  and using the high-order Taylor series method (in 100-digit precision with truncation errors less than $10^{-70}$) for the evolution of motion equations (\ref{geq:3-body}),  instead  of  the ODE solver dop853  \cite{Hairer1993}  based on the Runge-Kutta method in double precision.    In addition, we use the CNS with even smaller round-off error (in 120-digit precision) and truncation error (less than $10^{-90}$)  to guarantee the reliability of these 27  families.    It is found that one of them belongs to the 11 families found by  \v{S}uvakov and Dmitra\v{s}inovi\'{c} \cite{Suvakov2013}.

Similarly, we found 165 families within the period $100<T_0<200$, including 119 families gained  by the ODE solver dop853  \cite{Hairer1993}  in double precision and the  additional 46 families by the CNS  in the multiple precision.    Obviously,  more periodic orbits can be found within a larger period.

It is interesting that more periodic orbits can be found by means of finer search grids.  Using $2000\times 2000$ grids for the  candidates of initial conditions, we   gained  totally 498  periodic orbits  within $0<T_0<200$ in the similar way, including the 163 families within $0\leq T_0\leq 100$ by the ODE solver dop853  \cite{Hairer1993}   in double precision, the additional 33 families within $0\leq T_0\leq 100$  by the CNS  in the multiple precision,  the 182 families within $100<T_0\leq 200$ by the ODE solver dop853  in double precision, the  additional 120 families within $100<T_0\leq200$  by the CNS  in the multiple precision, respectively.

Similarly,  using $4000\times 4000$ grids, we  totally  gained 695  periodic orbits  within $0<T_0<200$, including the 192 families within $0\leq T_0\leq 100$ by the ODE solver dop853  \cite{Hairer1993} in double precision, the  additional 37 families within $0\leq T_0\leq 100$  by the CNS  in the multiple precision,  the 260 families within $100<T_0\leq 200$ by the ODE solver dop853  in double precision, and the  additional 206 families within $100<T_0\leq200$  by the CNS  in the multiple precision, respectively.   Thus, the finer the search grid,  the more periodic orbits can be found.  It is indeed a  surprise that there exist much more families of periodic orbits of three-body problem than we had thought a few years ago!

It should be emphasized that,  in case of  the search grid $4000\times 4000$,   we  found 243 more periodic orbits by means of  the CNS \cite{Liao2009, Liao2013-3b, Liao2013-Chaos,Liao2014-SciChina, Liao2015-IJBC, Lin2017} in multiple precision than  the ODE solver dop853  \cite{Hairer1993} in double precision.    It indicates that the numerical noises might  lead to great loss of  periodic orbits of three-body system.

\section{Periodic orbits of the three-body system}

\begin{table*}
\tabcolsep 0pt \caption{The initial velocities and periods $T$ of some newly-found periodic orbits of the three-body system with equal mass and zero angular momentum  in the case of the isosceles collinear configurations:  $\bm{r}_1(0)=(-1,0)=-\bm{r}_2(0)$,  $\dot{\bm{r}}_1(0)=(v_1,v_2)=\dot{\bm{r}}_2(0)$ and $\bm{r}_3(0)=(0,0)$, $\dot{\bm{r}}_3(0)=(-2v_1, -2v_2)$  when $G=1$ and $m_1=m_2=m_3=1$, where $T^*=T |E|^{3/2}$ is its scale-invariant period, $L_f$ is the length of the free group element.  Here, the superscript {\em i.c.} indicates the case of the initial conditions with {\em isosceles collinear}  configuration, due to the fact that there exist periodic orbits in many other cases.  } \label{table1} \vspace*{-12pt}
\begin{center}
\def\temptablewidth{1\textwidth}
{\rule{\temptablewidth}{1pt}}
\begin{tabular*}{\temptablewidth}{@{\extracolsep{\fill}}lccccc}
\hline
Class and number & $v_1$ & $v_2$  & $T$ & $T^*$ & $L_f$\\
\hline
I.A$^{i.c.}_{77}$ &      0.4159559963 &     0.2988672319 &   114.5882843578 &    256.902 &  104\\
I.A$^{i.c.}_{100}$ &     0.0670760777 &     0.5889627892 &   163.9101889958 &    284.971 &  124\\
I.A$^{i.c.}_{115}$ &     0.3369172422 &     0.2901238678 &   139.2379425543 &    366.661 &  148\\
I.B$^{i.c.}_{102}$ &     0.4784306757 &     0.3771895698 &   199.5148149879 &    325.727 &  134\\
I.B$^{i.c.}_{155}$ &     0.3310536467 &     0.1351316515 &   192.5759642513 &    592.936 &  238\\
II.C$^{i.c.}_{156}$ &    0.3231926176 &     0.3279135713 &   145.3798952880 &    369.993 &  150\\
\hline
\end{tabular*}
{\rule{\temptablewidth}{1pt}}
\end{center}
\end{table*}

The Montgomery's topological identification and classification method  \cite{Montgomery1998} is used here to identify these periodic orbits.  The positions $\bm{r_1}, \bm{r_2}$ and $\bm{r_3}$ of the three-body correspond to a unit vector $\bm{n}$ in the so-called ``shape sphere'' with the Cartesian components
\[ n_x=\frac{2\bm{\rho}\cdot\bm{\lambda}}{R^2}, \;\; n_y=\frac{\lambda^2-\rho^2}{R^2}, \;\;  n_z=\frac{2(\bm{\rho}\times\bm{\lambda}) \cdot \bm{e}_z }{R^2},\]
where  $   \bm{\rho}=\frac{1}{\sqrt{2}}(\bm{r_1}-\bm{r_2})$, $ \bm{\lambda}=\frac{1}{\sqrt{6}}(\bm{r_1}+\bm{r_2}-2\bm{r_3})$ and the hyper-radius $R=\sqrt{\rho^2+\lambda^2}$.     A periodic orbit of three-body system gives a closed curve on the shape sphere, which can be characterized by its topology with three punctures (two-body collision points).  With one of the punctures as the ``north pole", the sphere can be mapped onto a plane by a stereographic projection.  And a closed curve can be mapped onto a plane with two punctures and its topology can be described by the so-called ``free group element'' (word) with letters $a$ (a clockwise around right-hand side puncture),  $b$ (a counter-clockwise around left-hand side puncture) and their inverses $a^{-1}=A$ and $b^{-1}=B$.  For details, please refer  to  \cite{Suvakov2014a, Montgomery1998}.

 The periodic orbits can be divided into different classes according to their geometric and algebraic symmetries \cite{Suvakov2013}.   There are two types of geometric symmetries in the shape space:
 \begin{enumerate}
\item[(I)] the reflection symmetries of two orthogonal axes --- the equator and the zeroth meridian passing through the ``far'' collision point;
 \item[(II)] a central reflection symmetry about one point --- the intersection of the equator and the aforementioned zeroth meridian.
 \end{enumerate}
  Besides, \v{S}uvakov and Dmitra\v{s}inovi\'{c} \cite{Suvakov2013} mentioned three types of algebraic exchange symmetries for the free group elements:
   \begin{enumerate}
  \item[(A)] the free group elements are symmetric with $a \leftrightarrow A$ and $b \leftrightarrow B$;
  \item[(B)] free group elements are symmetric with $a \leftrightarrow b$ and $A \leftrightarrow B$;
  \item[(C)] free group elements are not symmetric under either (A) or (B).
  \end{enumerate}

The 695 families of the periodic collisionless orbits can be divided into five classes:  I.A, II.A,  I.B, II.B and II.C,  as listed in Table~S.~III-XXX in Supplementary material \cite{Supplement}.   Note that the class II.A  was  {\em not}  included   in \cite{Suvakov2013}.   Here,  we regard all periodic orbits (and its satellites) with the same free group element   as  one family,  so  the  ``moth I''  orbit and its satellite  ``yarn''  orbit in \cite{Suvakov2013} belong to one family in this paper.  These 695 families include the Figure-eight family \cite{More1993,Chenciner2000}, the 11 families found by  \v{S}uvakov and Dmitra\v{s}inovi\'{c} \cite{Suvakov2013} (see Table~S.I) and the 25 families reported in \cite{Hudomal2015} (11 among them was given in \cite{Suvakov2013},  see Table~S.II).   In Table~S.I and S.II, the superscript {\em i.c.} indicates the case of the initial conditions with {\em isosceles collinear}  configuration, due to the fact that there exist periodic orbits in many other cases.   Note that Rose \cite{Rose2016} currently reported 90 periodic planar collisionless orbits in the same case of  the isosceles collinear configurations (\ref{initial}), which include the Figure-eight family \cite{More1993,Chenciner2000} and many families reported in \cite{Suvakov2013, Hudomal2015}.  Even considering these,   more than 600 families among our 695 ones are new.

\begin{figure}[t]
  \centering
  \includegraphics[scale=0.3]{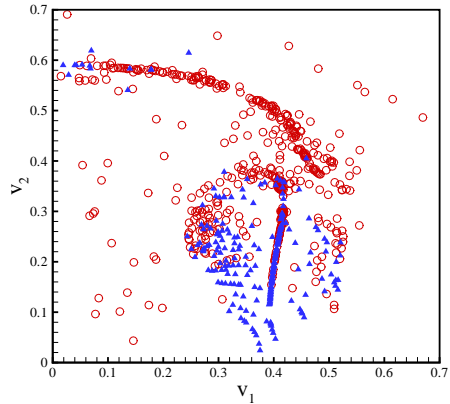}
  \caption{(color online).  The scatterplot of the initial velocities of the 695 periodic orbits. Circle: the orbits gained by ODE solver dop853 in double precision; triangle: the orbits gained by the CNS  in the multiple precision. }
  \label{v1v2}
\end{figure}

Note that the initial positions ${\bm r}_1 =(x_1,x_2)$ in Table S.III - XVI in Supplementary material \cite{Supplement} depart from $(-1,0)$ a little.  However, it is well-known that, if  ${\bm r}_i(t)$ $(i=1,2,3)$  denotes a periodic orbit with the period $T$ of a three-body system, then
\begin{equation}
{\bm r}'_i(t') = \alpha \; {\bm r}_i(t),\;\; {\bm v}'_i(t') = {\bm v}_i(t)/\sqrt{\alpha}, \;\;  t' = \alpha^{3/2} \; t, \label{scaling}
\end{equation}
 is also a periodic orbit with the period $T' = \alpha^{3/2}\; T$ for arbitrary $\alpha>0$.   Thus, through coordinate transformation and then the scaling of the spatial and temporal coordinates, we can always enforce (-1,0), (1,0) and (0,0) as the initial positions of the body-1, 2 and 3, respectively, with the initial velocities $\dot{{\bm r}}_1(0)=\dot{\bm r}_2(0)$ and $\dot{{\bm r}}_3(0)=-2 \dot{{\bm r}}_1(0)$,  corresponding to zero angular momentum.  This is the reason why we choose $-1$ and 0 as the initial guesses of $x_1$ and $x_2$ for our search approaches.    The corresponding initial velocities of the 695 periodic orbits are listed in Tables~S.XVII - XXX in Supplementary material \cite{Supplement}.  The scatterplot of the initial velocities of the 695 periodic orbits are shown in FIG.~\ref{v1v2}.  Note that two very close initial conditions can give completely different periodic orbits.   This well explains  why more periodic orbits can be found by means of finer search grids.

\begin{figure*}[!ht]
  \centering
  \includegraphics[scale=0.25]{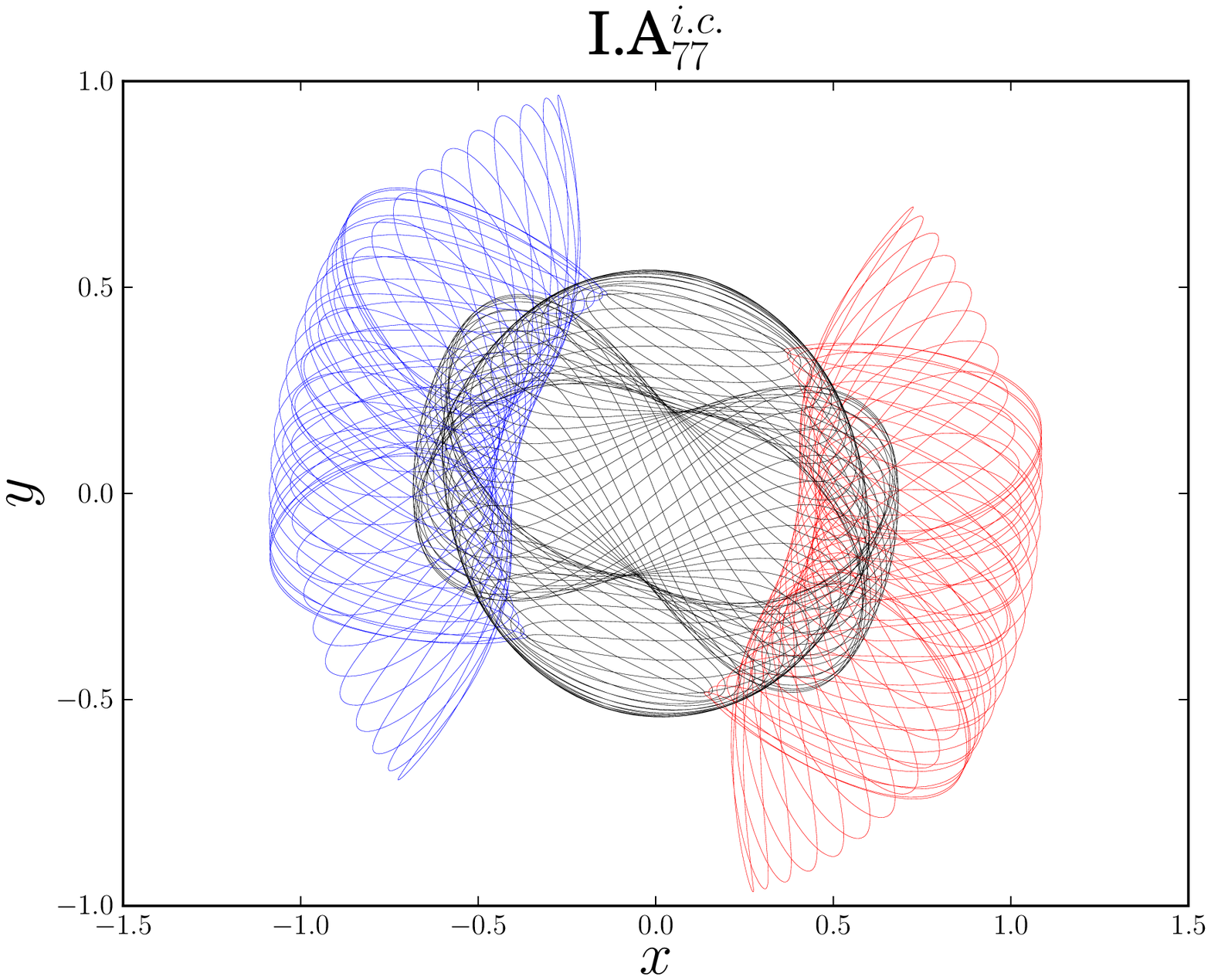}
  \includegraphics[scale=0.25]{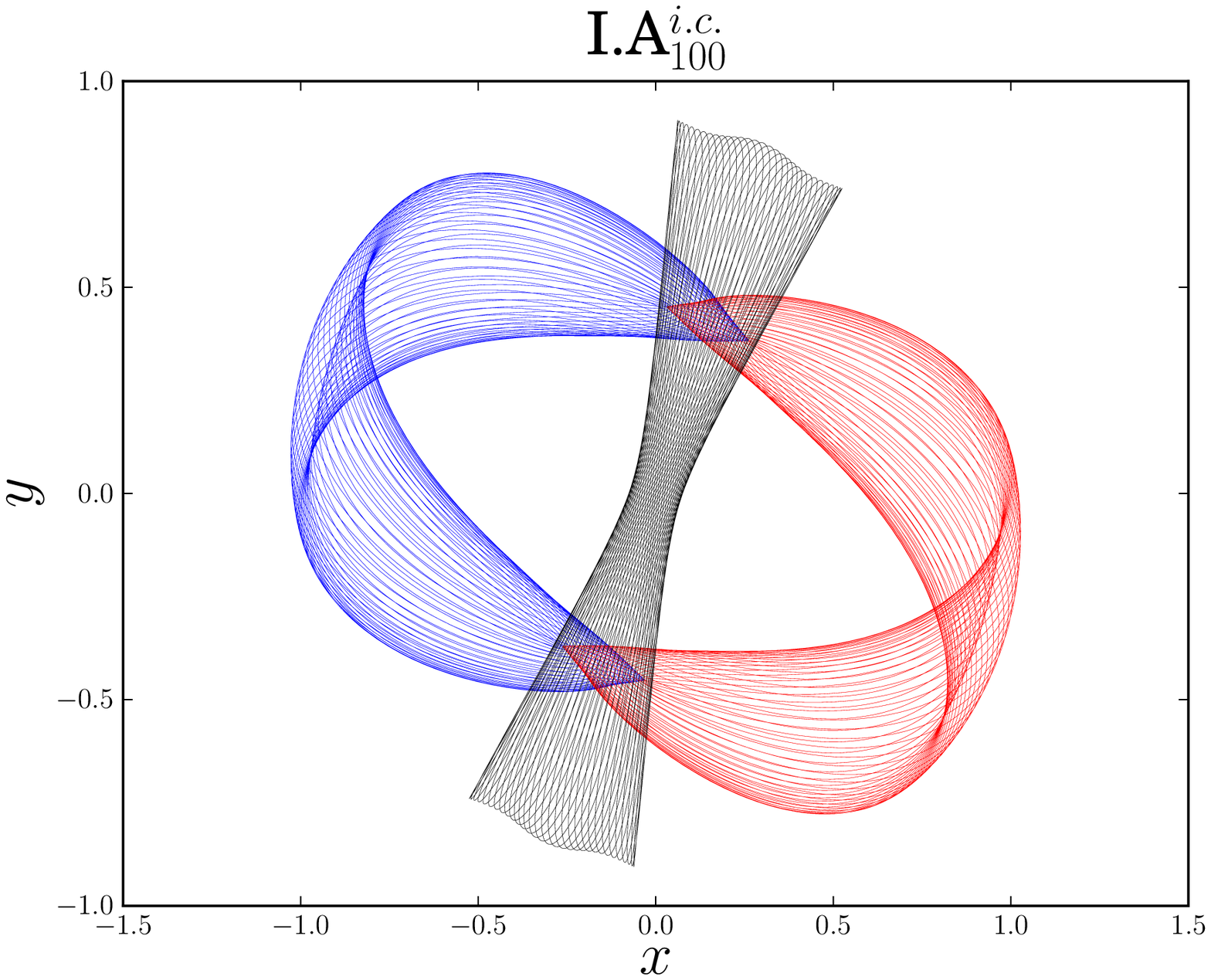}
  \includegraphics[scale=0.25]{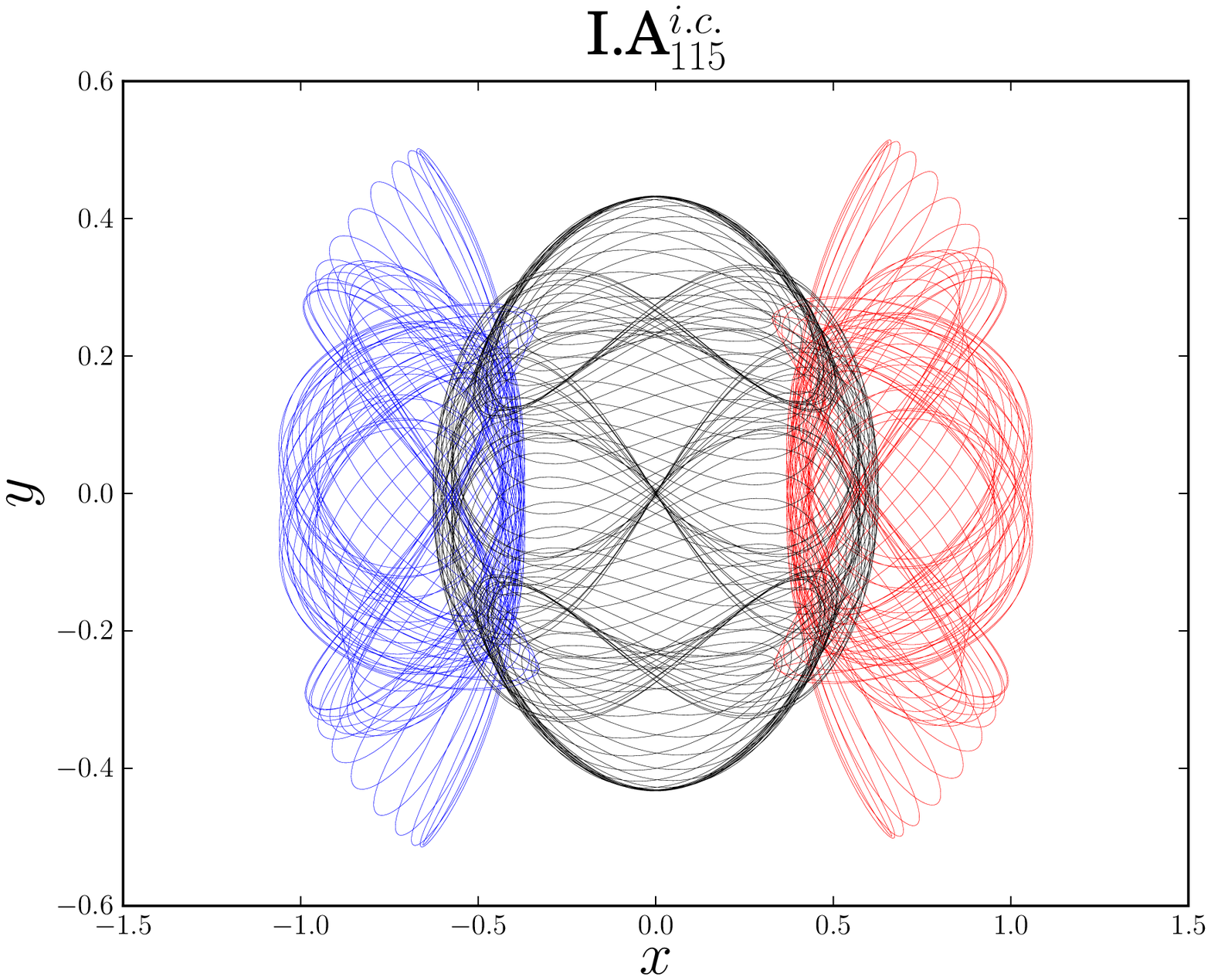}
  \includegraphics[scale=0.25]{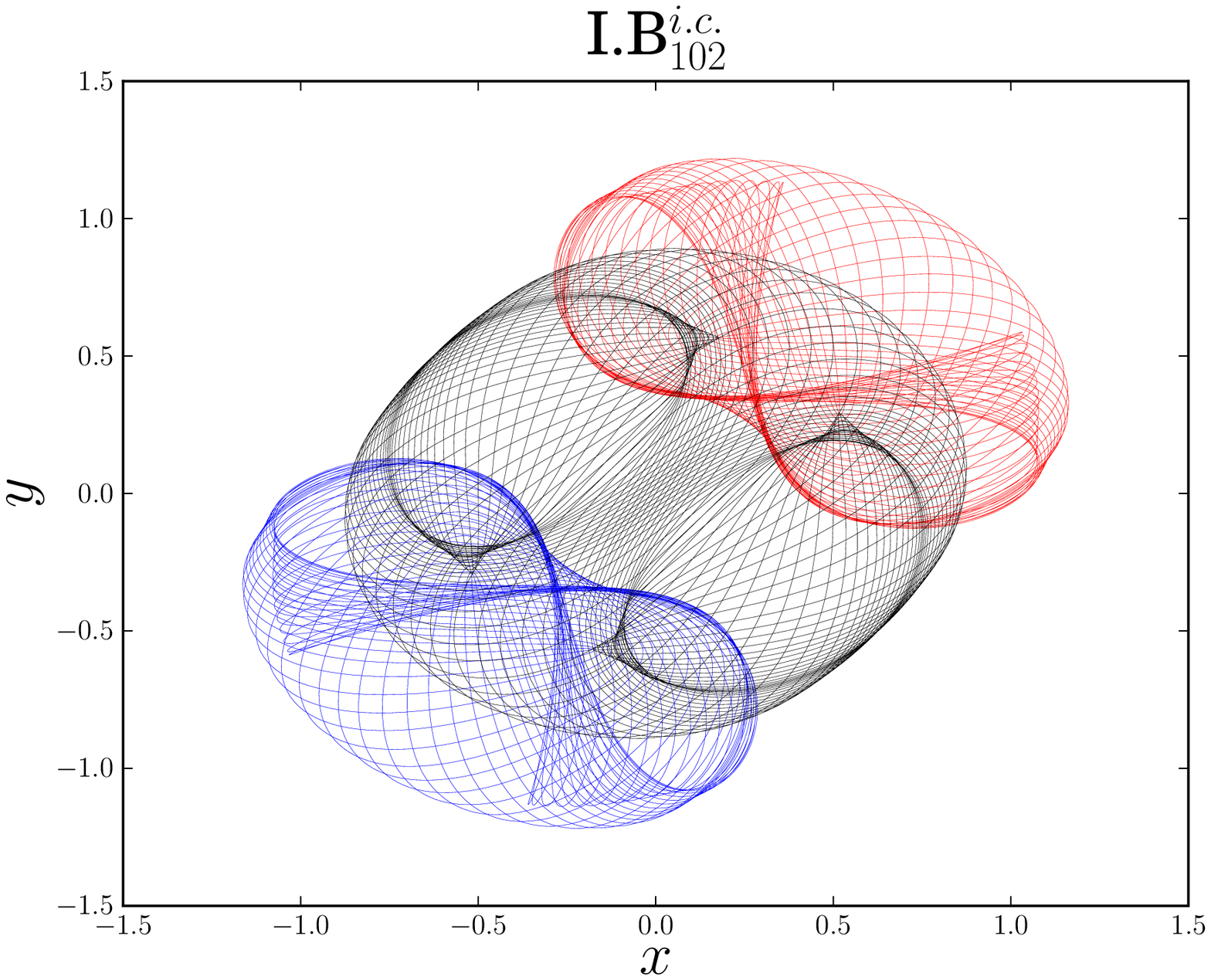}
  \includegraphics[scale=0.25]{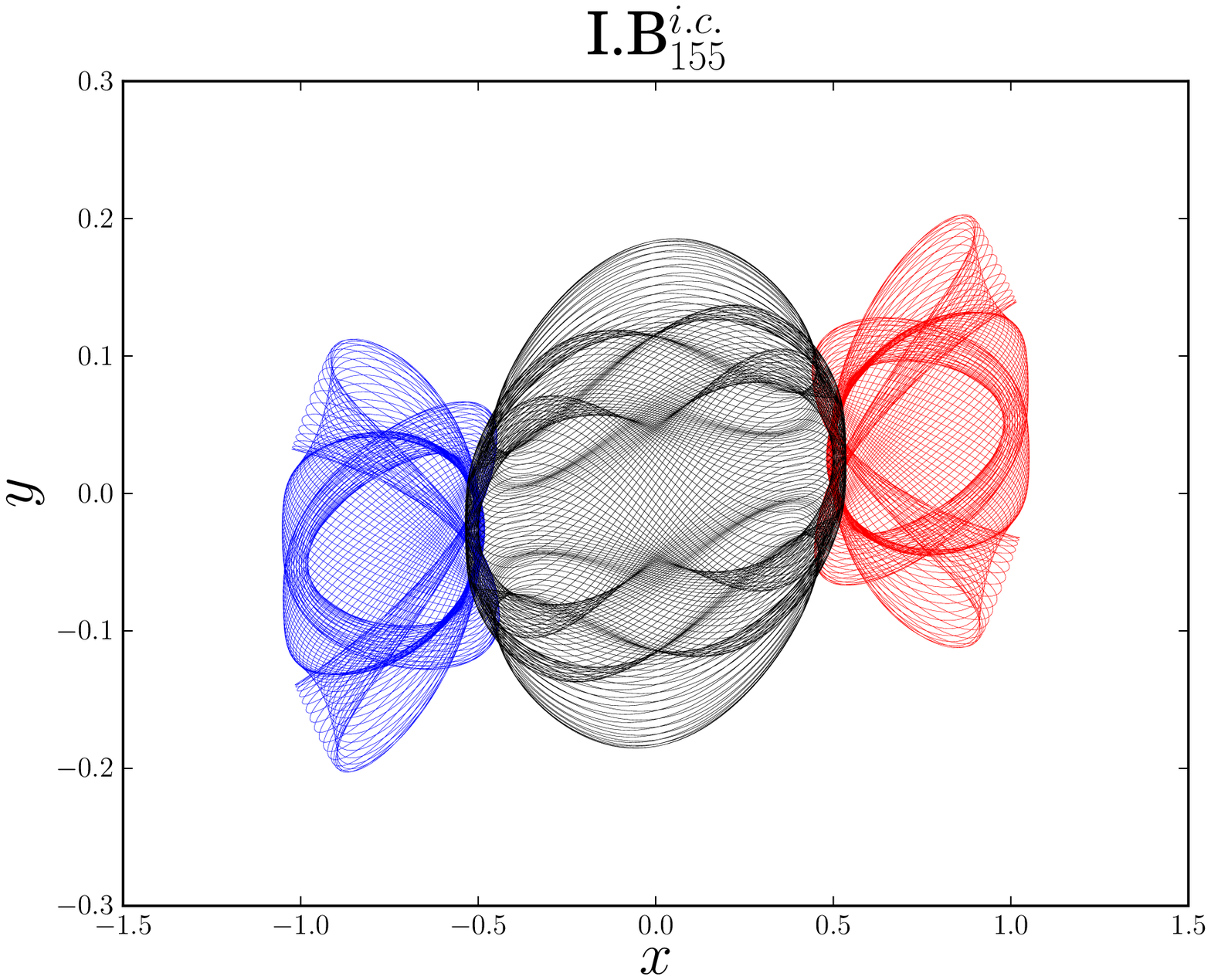}
  \includegraphics[scale=0.25]{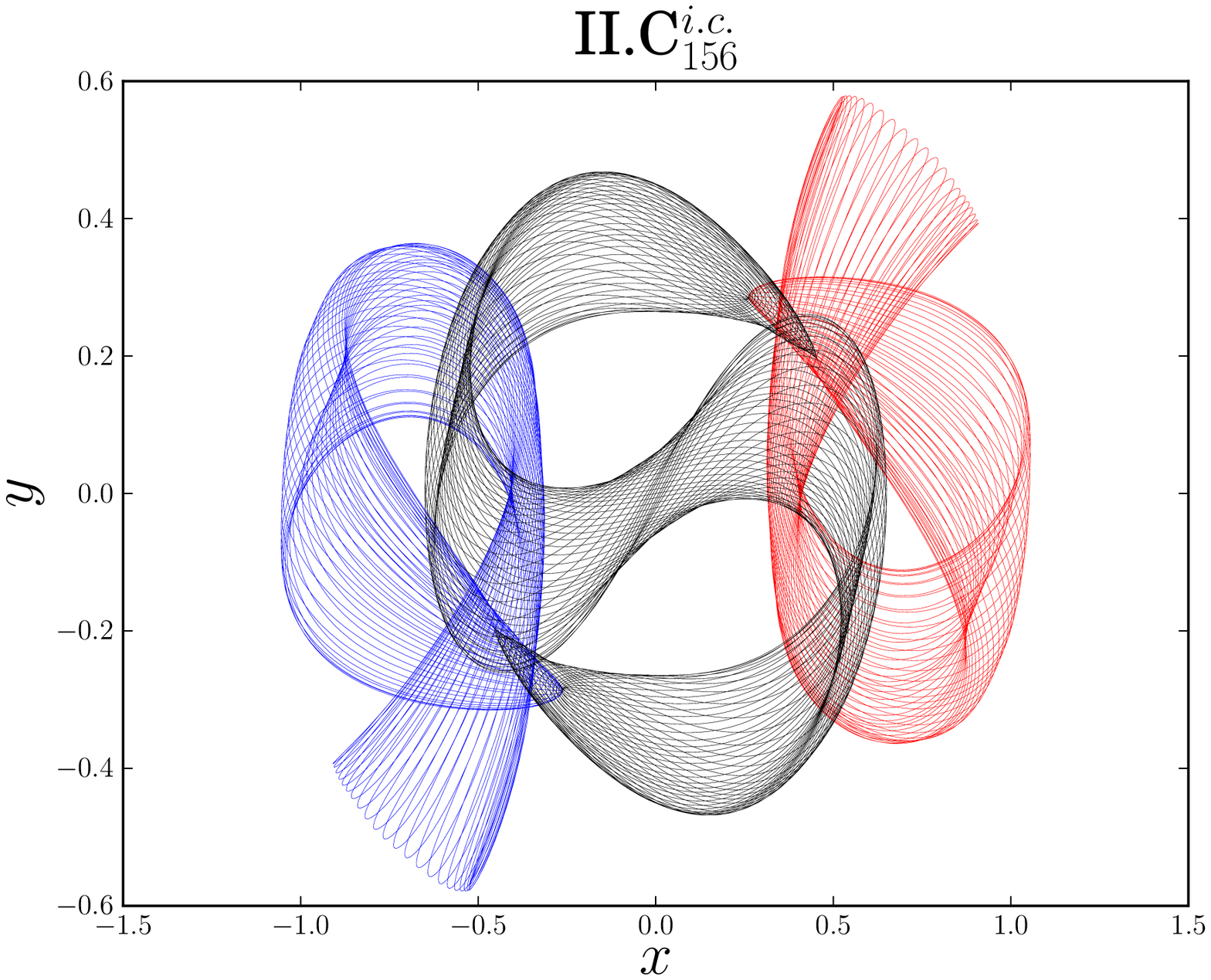}
  \caption{(color online.) Brief overview of the six newly-found families of periodic three-body orbits  in case of equal mass, zero angular momentum  and  initial conditions with isosceles collinear configuration.}
  \label{fig1}
\end{figure*}

\begin{figure}
  \centering
  \includegraphics[scale=0.3]{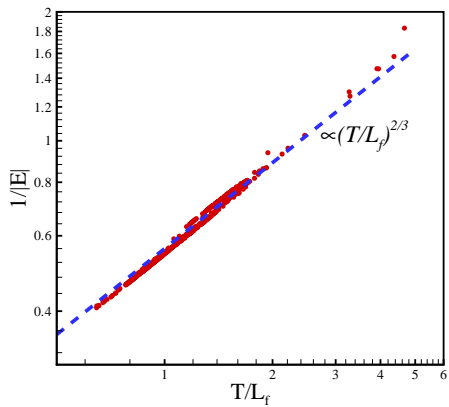}
  \caption{(color online). The inverse of total energy $1/|E|$ and average period $T/L_f$ approximately fall in with  a power law with exponent 2/3, where $L_f$ is the length of free group element.  Symbols: the 695 families of periodic orbits; dashed line: $1/|E|=0.56\cdot(T/L_f)^{2/3}$. }
  \label{T-k}
\end{figure}

The so-called ``free group elements''  of these  695 families are listed in Table S.XXXI-LIV in Supplementary material \cite{Supplement}.    Due to the limited length,  only six newly-found ones are listed in Table~\ref{table1},  and their real space orbits are shown in FIG.~\ref{fig1}.    In addition,  the real space orbits of a few families are shown in FIG.~S.1.    The movies of these 695 periodic orbits in the real space and the corresponding close curves on the shape sphere  can be found via the website: \url{http://numericaltank.sjtu.edu.cn/three-body/three-body.htm}

For a two-body system, there exists the so-called Kepler's third law $ r_a  \propto  T^{2/3}$, where $T$ is the period and $r_a$ is the semi-major axis of periodic orbit.     For a three-body system,  \v{S}uvakov and Dmitra\v{s}inovi\'{c} \cite{Suvakov2013}  mentioned  that  there should exist  the relation  $T^{2/3} |E| = $ constant,  where $E$  denotes the total kinetic and potential energy of the three-body system.   But, they pointed out that ``the constant on the right-hand side of this equation is {\em not} universal'', which may depend  on ``both of the family of the three-body orbit  and  its  angular  momentum'' \cite{Suvakov2013}.   However, with the definition of the average period $\bar{T} = T/L_f$, where $L_f$ is the length of free group element of periodic orbit of a three-body system,  the 695 families of periodic planar collisionless orbits  approximately satisfy such a generalised  Kepler's third law  $\bar{R} \propto |E|^{-1} = 0.56 \; \bar{T}^{2/3}$, as shown in FIG.~\ref{T-k}, where  $\bar{R}$ is the mean of hyper-radius of the three-body system.  In other words, the scale-invariant  average  period $\bar{T}^* = \bar{T} |E|^{3/2}$ should be approximately equal to a {\em universal} constant, i.e.  $ \bar{T}^* \approx 2.433 \pm 0.075$,  for the three-body system with equal mass and zero angular momentum in the case of initial conditions  with isosceles collinear configuration (\ref{initial}).  Note that  the scale-invariant period  $T^* = T |E|^{3/2}$  and $L_f$ (the length of free group element) are {\em invariable} under the scaling (\ref{scaling}) of the spatial and temporal coordinates for {\em arbitrary} $\alpha>0$.  So, they  are two  characteristics with important physical meanings for each family that contains an infinite number of periodic orbits corresponding to different scaling parameters $\alpha>0$ in (\ref{scaling}).

\section{Concluding remarks}

In this paper, we gain 695 families of periodic orbits of the three-body system with equal mass,  zero angular momentum and initial conditions in
the isosceles collinear configuration $\bm{r}_1=(-1,0), \bm{r}_2=(+1,0), \bm{r}_3=(0,0)$.   These 695 families include the Figure-eight family \cite{More1993,Chenciner2000}, the 11 families found by  \v{S}uvakov and Dmitra\v{s}inovi\'{c} \cite{Suvakov2013} (see Table~S.I) and the 25 families reported in \cite{Hudomal2015} (11 among them was given in \cite{Suvakov2013},  see Table~S.II).  Especially,  more than six hundreds among them are completely new and have been never reported, to the best of our knowledge.   It should be emphasized that 243 more periodic orbits are found by means of  the CNS \cite{Liao2009, Liao2013-3b, Liao2013-Chaos,Liao2014-SciChina, Liao2015-IJBC, Lin2017} in multiple precision than the ODE solver dop853  \cite{Hairer1993} in double precision.    This indicates the great potential of the CNS for complicated nonlinear dynamic systems.

It should be emphasized that,  in  the considered  initial conditions with isosceles collinear configuration,  more and more  periodic planar three-body orbits could be found by means of  finer search grids within a larger period.   Similarly,  a large number of periodic orbits can be gained in  other cases of three-body systems.  Thereafter,  a  data  base  for periodic orbits of three-body problem could be built, which is of benefit to better understandings of three-body systems, a  very  famous  problem that  can be traced back to Isaac Newton in 1680s.

\section*{Acknowledgment}
This work was carried out on TH-2 at National Supercomputer Centre in Guangzhou, China.  It is partly supported by National Natural Science Foundation of China (Approval No. 11432009).

\nocite{*}

\bibliography{ref}

\begin{thebibliography}{40}%
\makeatletter
\providecommand \@ifxundefined [1]{%
 \@ifx{#1\undefined}
}%
\providecommand \@ifnum [1]{%
 \ifnum #1\expandafter \@firstoftwo
 \else \expandafter \@secondoftwo
 \fi
}%
\providecommand \@ifx [1]{%
 \ifx #1\expandafter \@firstoftwo
 \else \expandafter \@secondoftwo
 \fi
}%
\providecommand \natexlab [1]{#1}%
\providecommand \enquote  [1]{``#1''}%
\providecommand \bibnamefont  [1]{#1}%
\providecommand \bibfnamefont [1]{#1}%
\providecommand \citenamefont [1]{#1}%
\providecommand \href@noop [0]{\@secondoftwo}%
\providecommand \href [0]{\begingroup \@sanitize@url \@href}%
\providecommand \@href[1]{\@@startlink{#1}\@@href}%
\providecommand \@@href[1]{\endgroup#1\@@endlink}%
\providecommand \@sanitize@url [0]{\catcode `\\12\catcode `\$12\catcode
  `\&12\catcode `\#12\catcode `\^12\catcode `\_12\catcode `\%12\relax}%
\providecommand \@@startlink[1]{}%
\providecommand \@@endlink[0]{}%
\providecommand \url  [0]{\begingroup\@sanitize@url \@url }%
\providecommand \@url [1]{\endgroup\@href {#1}{\urlprefix }}%
\providecommand \urlprefix  [0]{URL }%
\providecommand \Eprint [0]{\href }%
\providecommand \doibase [0]{http://dx.doi.org/}%
\providecommand \selectlanguage [0]{\@gobble}%
\providecommand \bibinfo  [0]{\@secondoftwo}%
\providecommand \bibfield  [0]{\@secondoftwo}%
\providecommand \translation [1]{[#1]}%
\providecommand \BibitemOpen [0]{}%
\providecommand \bibitemStop [0]{}%
\providecommand \bibitemNoStop [0]{.\EOS\space}%
\providecommand \EOS [0]{\spacefactor3000\relax}%
\providecommand \BibitemShut  [1]{\csname bibitem#1\endcsname}%
\let\auto@bib@innerbib\@empty
\bibitem [{\citenamefont {Musielak}\ and\ \citenamefont
  {Quarles}(2014)}]{Musielak2014}%
  \BibitemOpen
  \bibfield  {author} {\bibinfo {author} {\bibfnamefont {Z.~E.}\ \bibnamefont
  {Musielak}}\ and\ \bibinfo {author} {\bibfnamefont {B.}~\bibnamefont
  {Quarles}},\ }\href@noop {} {\bibfield  {journal} {\bibinfo  {journal} {Rep.
  Prog. Phys.}\ }\textbf {\bibinfo {volume} {77}},\ \bibinfo {pages} {065901
  (30pp)} (\bibinfo {year} {2014})}\BibitemShut {NoStop}%
\bibitem [{\citenamefont {Poincar{\' e}}(1890)}]{Poincare1890}%
  \BibitemOpen
  \bibfield  {author} {\bibinfo {author} {\bibfnamefont {J.~H.}\ \bibnamefont
  {Poincar{\' e}}},\ }\href@noop {} {\bibfield  {journal} {\bibinfo  {journal}
  {Acta Math.}\ }\textbf {\bibinfo {volume} {13}},\ \bibinfo {pages} {1}
  (\bibinfo {year} {1890})}\BibitemShut {NoStop}%
\bibitem [{\citenamefont {Lorenz}(1963)}]{Lorenz1963}%
  \BibitemOpen
  \bibfield  {author} {\bibinfo {author} {\bibfnamefont {E.~N.}\ \bibnamefont
  {Lorenz}},\ }\href@noop {} {\bibfield  {journal} {\bibinfo  {journal}
  {Journal of the Atmospheric Sciences}\ }\textbf {\bibinfo {volume} {20}},\
  \bibinfo {pages} {130} (\bibinfo {year} {1963})}\BibitemShut {NoStop}%
\bibitem [{\citenamefont {\ifmmode~\check{S}\else \v{S}\fi{}uvakov}\ and\
  \citenamefont {Dmitra\ifmmode \check{s}\else
  \v{s}\fi{}inovi\ifmmode~\acute{c}\else \'{c}\fi{}}(2013)}]{Suvakov2013}%
  \BibitemOpen
  \bibfield  {author} {\bibinfo {author} {\bibfnamefont {M.}~\bibnamefont
  {\ifmmode~\check{S}\else \v{S}\fi{}uvakov}}\ and\ \bibinfo {author}
  {\bibfnamefont {V.}~\bibnamefont {Dmitra\ifmmode \check{s}\else
  \v{s}\fi{}inovi\ifmmode~\acute{c}\else \'{c}\fi{}}},\ }\href {\doibase
  10.1103/PhysRevLett.110.114301} {\bibfield  {journal} {\bibinfo  {journal}
  {Phys. Rev. Lett.}\ }\textbf {\bibinfo {volume} {110}},\ \bibinfo {pages}
  {114301} (\bibinfo {year} {2013})}\BibitemShut {NoStop}%
\bibitem [{\citenamefont {Broucke}(1975)}]{Broucke1975}%
  \BibitemOpen
  \bibfield  {author} {\bibinfo {author} {\bibfnamefont {R.}~\bibnamefont
  {Broucke}},\ }\href {\doibase 10.1007/BF01595390} {\bibfield  {journal}
  {\bibinfo  {journal} {Celestial Mechanics}\ }\textbf {\bibinfo {volume}
  {12}},\ \bibinfo {pages} {439 } (\bibinfo {year} {1975})}\BibitemShut
  {NoStop}%
\bibitem [{\citenamefont {Hadjidemetriou}(1975)}]{Hadjidemetriou1975A}%
  \BibitemOpen
  \bibfield  {author} {\bibinfo {author} {\bibfnamefont {J.~D.}\ \bibnamefont
  {Hadjidemetriou}},\ }\href {\doibase 10.1007/BF01228563} {\bibfield
  {journal} {\bibinfo  {journal} {Celestial Mechanics}\ }\textbf {\bibinfo
  {volume} {12}},\ \bibinfo {pages} {255} (\bibinfo {year} {1975})}\BibitemShut
  {NoStop}%
\bibitem [{\citenamefont {Hadjidemetriou}\ and\ \citenamefont
  {Christides}(1975)}]{Hadjidemetriou1975B}%
  \BibitemOpen
  \bibfield  {author} {\bibinfo {author} {\bibfnamefont {J.~D.}\ \bibnamefont
  {Hadjidemetriou}}\ and\ \bibinfo {author} {\bibfnamefont {T.}~\bibnamefont
  {Christides}},\ }\href {\doibase 10.1007/BF01230210} {\bibfield  {journal}
  {\bibinfo  {journal} {Celestial mechanics}\ }\textbf {\bibinfo {volume}
  {12}},\ \bibinfo {pages} {175} (\bibinfo {year} {1975})}\BibitemShut
  {NoStop}%
\bibitem [{\citenamefont {H{\'e}non}(1976)}]{Henon1976}%
  \BibitemOpen
  \bibfield  {author} {\bibinfo {author} {\bibfnamefont {M.}~\bibnamefont
  {H{\'e}non}},\ }\href {\doibase 10.1007/BF01228647} {\bibfield  {journal}
  {\bibinfo  {journal} {Celestial mechanics}\ }\textbf {\bibinfo {volume}
  {13}},\ \bibinfo {pages} {267} (\bibinfo {year} {1976})}\BibitemShut
  {NoStop}%
\bibitem [{\citenamefont {H{\'e}non}(1977)}]{Henon1977}%
  \BibitemOpen
  \bibfield  {author} {\bibinfo {author} {\bibfnamefont {M.}~\bibnamefont
  {H{\'e}non}},\ }\href {\doibase 10.1007/BF01228465} {\bibfield  {journal}
  {\bibinfo  {journal} {Celestial mechanics}\ }\textbf {\bibinfo {volume}
  {15}},\ \bibinfo {pages} {243} (\bibinfo {year} {1977})}\BibitemShut
  {NoStop}%
\bibitem [{\citenamefont {Moore}(1993)}]{More1993}%
  \BibitemOpen
  \bibfield  {author} {\bibinfo {author} {\bibfnamefont {C.}~\bibnamefont
  {Moore}},\ }\href {\doibase 10.1103/PhysRevLett.70.3675} {\bibfield
  {journal} {\bibinfo  {journal} {Phys. Rev. Lett.}\ }\textbf {\bibinfo
  {volume} {70}},\ \bibinfo {pages} {3675} (\bibinfo {year}
  {1993})}\BibitemShut {NoStop}%
\bibitem [{\citenamefont {Chenciner}\ and\ \citenamefont
  {Montgomery}(2000)}]{Chenciner2000}%
  \BibitemOpen
  \bibfield  {author} {\bibinfo {author} {\bibfnamefont {A.}~\bibnamefont
  {Chenciner}}\ and\ \bibinfo {author} {\bibfnamefont {R.}~\bibnamefont
  {Montgomery}},\ }\href {http://www.jstor.org/stable/2661357} {\bibfield
  {journal} {\bibinfo  {journal} {Annals of Mathematics}\ }\textbf {\bibinfo
  {volume} {152}},\ \bibinfo {pages} {881} (\bibinfo {year}
  {2000})}\BibitemShut {NoStop}%
\bibitem [{\citenamefont {Nauenberg}(2001)}]{Nauenberg2001}%
  \BibitemOpen
  \bibfield  {author} {\bibinfo {author} {\bibfnamefont {M.}~\bibnamefont
  {Nauenberg}},\ }\href {\doibase
  http://dx.doi.org/10.1016/S0375-9601(01)00768-X} {\bibfield  {journal}
  {\bibinfo  {journal} {Physics Letters A}\ }\textbf {\bibinfo {volume}
  {292}},\ \bibinfo {pages} {93 } (\bibinfo {year} {2001})}\BibitemShut
  {NoStop}%
\bibitem [{\citenamefont {Chenciner}\ \emph {et~al.}(2005)\citenamefont
  {Chenciner}, \citenamefont {Fejoz},\ and\ \citenamefont
  {Montgomery}}]{Chenciner2005}%
  \BibitemOpen
  \bibfield  {author} {\bibinfo {author} {\bibfnamefont {A.}~\bibnamefont
  {Chenciner}}, \bibinfo {author} {\bibfnamefont {J.}~\bibnamefont {Fejoz}}, \
  and\ \bibinfo {author} {\bibfnamefont {R.}~\bibnamefont {Montgomery}},\
  }\href@noop {} {\bibfield  {journal} {\bibinfo  {journal} {Nonlinearity}\
  }\textbf {\bibinfo {volume} {18}},\ \bibinfo {pages} {1407} (\bibinfo {year}
  {2005})}\BibitemShut {NoStop}%
\bibitem [{\citenamefont {Broucke}\ \emph {et~al.}(2006)\citenamefont
  {Broucke}, \citenamefont {Elipe},\ and\ \citenamefont
  {Riaguas}}]{Broucke2006}%
  \BibitemOpen
  \bibfield  {author} {\bibinfo {author} {\bibfnamefont {R.}~\bibnamefont
  {Broucke}}, \bibinfo {author} {\bibfnamefont {A.}~\bibnamefont {Elipe}}, \
  and\ \bibinfo {author} {\bibfnamefont {A.}~\bibnamefont {Riaguas}},\ }\href
  {\doibase http://dx.doi.org/10.1016/j.chaos.2005.11.082} {\bibfield
  {journal} {\bibinfo  {journal} {Chaos, Solitons \& Fractals}\ }\textbf
  {\bibinfo {volume} {30}},\ \bibinfo {pages} {513 } (\bibinfo {year}
  {2006})}\BibitemShut {NoStop}%
\bibitem [{\citenamefont {Nauenberg}(2007)}]{Nauenberg2007}%
  \BibitemOpen
  \bibfield  {author} {\bibinfo {author} {\bibfnamefont {M.}~\bibnamefont
  {Nauenberg}},\ }\href {\doibase 10.1007/s10569-006-9044-7} {\bibfield
  {journal} {\bibinfo  {journal} {Celestial Mechanics and Dynamical Astronomy}\
  }\textbf {\bibinfo {volume} {97}},\ \bibinfo {pages} {1} (\bibinfo {year}
  {2007})}\BibitemShut {NoStop}%
\bibitem [{\citenamefont {Li}\ and\ \citenamefont {Liao}(2014)}]{Li2014}%
  \BibitemOpen
  \bibfield  {author} {\bibinfo {author} {\bibfnamefont {X.~M.}\ \bibnamefont
  {Li}}\ and\ \bibinfo {author} {\bibfnamefont {S.~J.}\ \bibnamefont {Liao}},\
  }\href@noop {} {\bibfield  {journal} {\bibinfo  {journal} {Sci. China - Phys.
  Mech. Astron.}\ }\textbf {\bibinfo {volume} {57}},\ \bibinfo {pages} {2121}
  (\bibinfo {year} {2014})}\BibitemShut {NoStop}%
\bibitem [{\citenamefont {Hudomal}(2015)}]{Hudomal2015}%
  \BibitemOpen
  \bibfield  {author} {\bibinfo {author} {\bibfnamefont {A.}~\bibnamefont
  {Hudomal}},\ }\emph {\bibinfo {title} {New periodic solutions to the
  three-body problem and gravitational waves}},\ \href@noop {} {\bibinfo {type}
  {{M.S.} {T}hesis}},\ \bibinfo  {school} {University of Belgrade}, \bibinfo
  {address} {Serbia} (\bibinfo {year} {2015})\BibitemShut {NoStop}%
\bibitem [{\citenamefont {Dmitra\ifmmode \check{s}\else
  \v{s}\fi{}inovi\ifmmode~\acute{c}\else \'{c}\fi{}}\ and\ \citenamefont
  {\ifmmode~\check{S}\else \v{S}\fi{}uvakov}(2015)}]{Dmitrasinovic2015}%
  \BibitemOpen
  \bibfield  {author} {\bibinfo {author} {\bibfnamefont {V.}~\bibnamefont
  {Dmitra\ifmmode \check{s}\else \v{s}\fi{}inovi\ifmmode~\acute{c}\else
  \'{c}\fi{}}}\ and\ \bibinfo {author} {\bibfnamefont {M.}~\bibnamefont
  {\ifmmode~\check{S}\else \v{S}\fi{}uvakov}},\ }\href {\doibase
  http://dx.doi.org/10.1016/j.physleta.2015.06.026} {\bibfield  {journal}
  {\bibinfo  {journal} {Physics Letters A}\ }\textbf {\bibinfo {volume}
  {379}},\ \bibinfo {pages} {1939 } (\bibinfo {year} {2015})}\BibitemShut
  {NoStop}%
\bibitem [{\citenamefont {Jankovi\ifmmode~\acute{c}\else \'{c}\fi{}}\ and\
  \citenamefont {Dmitra\ifmmode \check{s}\else
  \v{s}\fi{}inovi\ifmmode~\acute{c}\else \'{c}\fi{}}(2016)}]{Jankovic2016}%
  \BibitemOpen
  \bibfield  {author} {\bibinfo {author} {\bibfnamefont {M.~R.}\ \bibnamefont
  {Jankovi\ifmmode~\acute{c}\else \'{c}\fi{}}}\ and\ \bibinfo {author}
  {\bibfnamefont {V.}~\bibnamefont {Dmitra\ifmmode \check{s}\else
  \v{s}\fi{}inovi\ifmmode~\acute{c}\else \'{c}\fi{}}},\ }\href {\doibase
  10.1103/PhysRevLett.116.064301} {\bibfield  {journal} {\bibinfo  {journal}
  {Phys. Rev. Lett.}\ }\textbf {\bibinfo {volume} {116}},\ \bibinfo {pages}
  {064301} (\bibinfo {year} {2016})}\BibitemShut {NoStop}%
\bibitem [{\citenamefont {Dmitra\ifmmode \check{s}\else
  \v{s}\fi{}inovi\ifmmode~\acute{c}\else \'{c}\fi{}}\ and\ \citenamefont
  {\ifmmode~\check{S}\else \v{S}\fi{}uvakov}(2014)}]{Suvakov2014a}%
  \BibitemOpen
  \bibfield  {author} {\bibinfo {author} {\bibfnamefont {V.}~\bibnamefont
  {Dmitra\ifmmode \check{s}\else \v{s}\fi{}inovi\ifmmode~\acute{c}\else
  \'{c}\fi{}}}\ and\ \bibinfo {author} {\bibfnamefont {M.}~\bibnamefont
  {\ifmmode~\check{S}\else \v{S}\fi{}uvakov}},\ }\href@noop {} {\bibfield
  {journal} {\bibinfo  {journal} {American Journal of Physics}\ }\textbf
  {\bibinfo {volume} {82}},\ \bibinfo {pages} {609} (\bibinfo {year}
  {2014})}\BibitemShut {NoStop}%
\bibitem [{\citenamefont {Hairer}\ \emph {et~al.}(1993)\citenamefont {Hairer},
  \citenamefont {N{\o}rsett},\ and\ \citenamefont {Wanner}}]{Hairer1993}%
  \BibitemOpen
  \bibfield  {author} {\bibinfo {author} {\bibfnamefont {E.}~\bibnamefont
  {Hairer}}, \bibinfo {author} {\bibfnamefont {S.~P.}\ \bibnamefont
  {N{\o}rsett}}, \ and\ \bibinfo {author} {\bibfnamefont {G.}~\bibnamefont
  {Wanner}},\ }\href@noop {} {\ \textbf {\bibinfo {volume} {8}} (\bibinfo
  {year} {1993})}\BibitemShut {NoStop}%
\bibitem [{\citenamefont {Farantos}(1995)}]{Farantos1995}%
  \BibitemOpen
  \bibfield  {author} {\bibinfo {author} {\bibfnamefont {S.~C.}\ \bibnamefont
  {Farantos}},\ }\href {\doibase
  http://dx.doi.org/10.1016/0166-1280(95)04206-L} {\bibfield  {journal}
  {\bibinfo  {journal} {Journal of Molecular Structure: THEOCHEM}\ }\textbf
  {\bibinfo {volume} {341}},\ \bibinfo {pages} {91 } (\bibinfo {year}
  {1995})}\BibitemShut {NoStop}%
\bibitem [{\citenamefont {Lara}\ and\ \citenamefont {Pelaez}(2002)}]{Lara2002}%
  \BibitemOpen
  \bibfield  {author} {\bibinfo {author} {\bibfnamefont {M.}~\bibnamefont
  {Lara}}\ and\ \bibinfo {author} {\bibfnamefont {J.}~\bibnamefont {Pelaez}},\
  }\href@noop {} {\bibfield  {journal} {\bibinfo  {journal} {Astronomy and
  Astrophysics}\ }\textbf {\bibinfo {volume} {389}},\ \bibinfo {pages} {692}
  (\bibinfo {year} {2002})}\BibitemShut {NoStop}%
\bibitem [{\citenamefont {Abad}\ \emph {et~al.}(2011)\citenamefont {Abad},
  \citenamefont {Barrio},\ and\ \citenamefont {Dena}}]{Abad2011}%
  \BibitemOpen
  \bibfield  {author} {\bibinfo {author} {\bibfnamefont {A.}~\bibnamefont
  {Abad}}, \bibinfo {author} {\bibfnamefont {R.}~\bibnamefont {Barrio}}, \ and\
  \bibinfo {author} {\bibfnamefont {A.}~\bibnamefont {Dena}},\ }\href {\doibase
  10.1103/PhysRevE.84.016701} {\bibfield  {journal} {\bibinfo  {journal} {Phys.
  Rev. E}\ }\textbf {\bibinfo {volume} {84}},\ \bibinfo {pages} {016701}
  (\bibinfo {year} {2011})}\BibitemShut {NoStop}%
\bibitem [{\citenamefont {Hoover}\ and\ \citenamefont
  {Hoover}(2015)}]{Hoover2015}%
  \BibitemOpen
  \bibfield  {author} {\bibinfo {author} {\bibfnamefont {W.}~\bibnamefont
  {Hoover}}\ and\ \bibinfo {author} {\bibfnamefont {C.}~\bibnamefont
  {Hoover}},\ }\href@noop {} {\bibfield  {journal} {\bibinfo  {journal}
  {Computational Methods in Science and Technology}\ }\textbf {\bibinfo
  {volume} {21}},\ \bibinfo {pages} {109 } (\bibinfo {year}
  {2015})}\BibitemShut {NoStop}%
\bibitem [{\citenamefont {Liao}(2009)}]{Liao2009}%
  \BibitemOpen
  \bibfield  {author} {\bibinfo {author} {\bibfnamefont {S.}~\bibnamefont
  {Liao}},\ }\href@noop {} {\bibfield  {journal} {\bibinfo  {journal} {Tellus
  A}\ }\textbf {\bibinfo {volume} {61}},\ \bibinfo {pages} {550} (\bibinfo
  {year} {2009})}\BibitemShut {NoStop}%
\bibitem [{\citenamefont {Liao}(2014)}]{Liao2013-3b}%
  \BibitemOpen
  \bibfield  {author} {\bibinfo {author} {\bibfnamefont {S.}~\bibnamefont
  {Liao}},\ }\href@noop {} {\bibfield  {journal} {\bibinfo  {journal}
  {Communications in Nonlinear Science and Numerical Simulation}\ }\textbf
  {\bibinfo {volume} {19}},\ \bibinfo {pages} {601} (\bibinfo {year}
  {2014})}\BibitemShut {NoStop}%
\bibitem [{\citenamefont {Liao}(2013)}]{Liao2013-Chaos}%
  \BibitemOpen
  \bibfield  {author} {\bibinfo {author} {\bibfnamefont {S.}~\bibnamefont
  {Liao}},\ }\href@noop {} {\bibfield  {journal} {\bibinfo  {journal} {Chaos
  Solitons \& Fractals}\ }\textbf {\bibinfo {volume} {47}},\ \bibinfo {pages}
  {1} (\bibinfo {year} {2013})}\BibitemShut {NoStop}%
\bibitem [{\citenamefont {Liao}\ and\ \citenamefont
  {Wang}(2014)}]{Liao2014-SciChina}%
  \BibitemOpen
  \bibfield  {author} {\bibinfo {author} {\bibfnamefont {S.}~\bibnamefont
  {Liao}}\ and\ \bibinfo {author} {\bibfnamefont {P.}~\bibnamefont {Wang}},\
  }\href@noop {} {\bibfield  {journal} {\bibinfo  {journal} {Sci. China - Phys.
  Mech. Astron.}\ }\textbf {\bibinfo {volume} {57}},\ \bibinfo {pages} {330 }
  (\bibinfo {year} {2014})}\BibitemShut {NoStop}%
\bibitem [{\citenamefont {Liao}\ and\ \citenamefont
  {Li}(2015)}]{Liao2015-IJBC}%
  \BibitemOpen
  \bibfield  {author} {\bibinfo {author} {\bibfnamefont {S.}~\bibnamefont
  {Liao}}\ and\ \bibinfo {author} {\bibfnamefont {X.}~\bibnamefont {Li}},\
  }\href@noop {} {\bibfield  {journal} {\bibinfo  {journal} {Int. J.
  Bifurcation \& Chaos}\ }\textbf {\bibinfo {volume} {25}},\ \bibinfo {pages}
  {1530023 (11 page)} (\bibinfo {year} {2015})}\BibitemShut {NoStop}%
\bibitem [{\citenamefont {Lin}\ \emph {et~al.}(2017)\citenamefont {Lin},
  \citenamefont {Wang},\ and\ \citenamefont {Liao}}]{Lin2017}%
  \BibitemOpen
  \bibfield  {author} {\bibinfo {author} {\bibfnamefont {Z.~L.}\ \bibnamefont
  {Lin}}, \bibinfo {author} {\bibfnamefont {L.~P.}\ \bibnamefont {Wang}}, \
  and\ \bibinfo {author} {\bibfnamefont {S.~J.}\ \bibnamefont {Liao}},\
  }\href@noop {} {\bibfield  {journal} {\bibinfo  {journal} {Sci. China - Phys.
  Mech. Astron.}\ }\textbf {\bibinfo {volume} {60}},\ \bibinfo {pages} {014712}
  (\bibinfo {year} {2017})}\BibitemShut {NoStop}%
\bibitem [{\citenamefont {Barton}\ \emph {et~al.}(1971)\citenamefont {Barton},
  \citenamefont {Willem},\ and\ \citenamefont {Zahar}}]{Barton1971}%
  \BibitemOpen
  \bibfield  {author} {\bibinfo {author} {\bibfnamefont {D.}~\bibnamefont
  {Barton}}, \bibinfo {author} {\bibfnamefont {I.}~\bibnamefont {Willem}}, \
  and\ \bibinfo {author} {\bibfnamefont {R.}~\bibnamefont {Zahar}},\
  }\href@noop {} {\bibfield  {journal} {\bibinfo  {journal} {Comput. J.}\
  }\textbf {\bibinfo {volume} {14}},\ \bibinfo {pages} {243} (\bibinfo {year}
  {1971})}\BibitemShut {NoStop}%
\bibitem [{\citenamefont {Corliss}\ and\ \citenamefont
  {Chang}(1982)}]{Corliss1982}%
  \BibitemOpen
  \bibfield  {author} {\bibinfo {author} {\bibfnamefont {G.}~\bibnamefont
  {Corliss}}\ and\ \bibinfo {author} {\bibfnamefont {Y.}~\bibnamefont
  {Chang}},\ }\href@noop {} {\bibfield  {journal} {\bibinfo  {journal} {ACM
  Trans. Math. Software}\ }\textbf {\bibinfo {volume} {8}},\ \bibinfo {pages}
  {114} (\bibinfo {year} {1982})}\BibitemShut {NoStop}%
\bibitem [{\citenamefont {Chang}\ and\ \citenamefont
  {Corhss}(1994)}]{Chang1994}%
  \BibitemOpen
  \bibfield  {author} {\bibinfo {author} {\bibfnamefont {Y.~F.}\ \bibnamefont
  {Chang}}\ and\ \bibinfo {author} {\bibfnamefont {G.~F.}\ \bibnamefont
  {Corhss}},\ }\href@noop {} {\bibfield  {journal} {\bibinfo  {journal}
  {Computers Math. Applic.}\ }\textbf {\bibinfo {volume} {28}},\ \bibinfo
  {pages} {209 } (\bibinfo {year} {1994})}\BibitemShut {NoStop}%
\bibitem [{\citenamefont {Barrio}\ \emph {et~al.}(2005)\citenamefont {Barrio},
  \citenamefont {Blesa},\ and\ \citenamefont {Lara}}]{Barrio2005}%
  \BibitemOpen
  \bibfield  {author} {\bibinfo {author} {\bibfnamefont {R.}~\bibnamefont
  {Barrio}}, \bibinfo {author} {\bibfnamefont {F.}~\bibnamefont {Blesa}}, \
  and\ \bibinfo {author} {\bibfnamefont {M.}~\bibnamefont {Lara}},\ }\href@noop
  {} {\bibfield  {journal} {\bibinfo  {journal} {Computers \& Mathematics with
  Applications}\ }\textbf {\bibinfo {volume} {50}},\ \bibinfo {pages} {93}
  (\bibinfo {year} {2005})}\BibitemShut {NoStop}%
\bibitem [{\citenamefont {Oyanarte}(1990)}]{Oyanarte1990}%
  \BibitemOpen
  \bibfield  {author} {\bibinfo {author} {\bibfnamefont {P.}~\bibnamefont
  {Oyanarte}},\ }\href@noop {} {\bibfield  {journal} {\bibinfo  {journal}
  {Computer Physics Communications}\ }\textbf {\bibinfo {volume} {59}},\
  \bibinfo {pages} {345} (\bibinfo {year} {1990})}\BibitemShut {NoStop}%
\bibitem [{\citenamefont {Viswanath}(2004)}]{Viswanath2004}%
  \BibitemOpen
  \bibfield  {author} {\bibinfo {author} {\bibfnamefont {D.}~\bibnamefont
  {Viswanath}},\ }\href@noop {} {\bibfield  {journal} {\bibinfo  {journal}
  {Physica D}\ }\textbf {\bibinfo {volume} {190}},\ \bibinfo {pages} {115 }
  (\bibinfo {year} {2004})}\BibitemShut {NoStop}%
\bibitem [{\citenamefont {Montgomery}(1998)}]{Montgomery1998}%
  \BibitemOpen
  \bibfield  {author} {\bibinfo {author} {\bibfnamefont {R.}~\bibnamefont
  {Montgomery}},\ }\href {http://stacks.iop.org/0951-7715/11/i=2/a=011}
  {\bibfield  {journal} {\bibinfo  {journal} {Nonlinearity}\ }\textbf {\bibinfo
  {volume} {11}},\ \bibinfo {pages} {363} (\bibinfo {year} {1998})}\BibitemShut
  {NoStop}%
\bibitem [{\citenamefont {Li}\ and\ \citenamefont {Liao}()}]{Supplement}%
  \BibitemOpen
  \bibfield  {author} {\bibinfo {author} {\bibfnamefont {X.}~\bibnamefont
  {Li}}\ and\ \bibinfo {author} {\bibfnamefont {S.}~\bibnamefont {Liao}},\
  }\href@noop {} {\ }\bibinfo {note} {See pages 8-61 of this
  manuscript}\BibitemShut {NoStop}%
\bibitem [{\citenamefont {Rose}(2016)}]{Rose2016}%
  \BibitemOpen
  \bibfield  {author} {\bibinfo {author} {\bibfnamefont {D.}~\bibnamefont
  {Rose}},\ }\emph {\bibinfo {title} {Geometric phase and periodic orbits of
  the equal-mass, planar three-body problem with vanishing angular momentum}},\
  \href@noop {} {\bibinfo {type} {{Ph.D.} {T}hesis}},\ \bibinfo  {school}
  {University of Sydney}, \bibinfo {address} {Australia} (\bibinfo {year}
  {2016})\BibitemShut {NoStop}%
\end{thebibliography}%

\newpage

\setcounter{table}{0}
\setcounter{figure}{0}
\renewcommand{\figurename}{FIG. S.}
\renewcommand{\tablename}{Table S.}

\begin{center}

{\bf \Large Supplementary information for\\  ``More than six hundreds new families
of Newtonian \\  periodic planar collisionless three-body orbits"}

\author{Xiaoming Li$^1$}
\vspace{0.5cm}

Xiaoming Li$^1$ and  Shijun Liao$^{1, 2, *}$ 

\vspace{0.5cm}

$^1$ School of Naval Architecture, Ocean and Civil Engineering, Shanghai Jiaotong University, China \\
$^2$ Ministry-of-Education Key Laboratory in Scientific and Engineering Computing, Shanghai 200240, China\\ 
* The corresponding author:  sjliao@sjtu.edu.cn
\end{center}

\begin{figure*}[b]
  \centering
\includegraphics[scale=0.26]{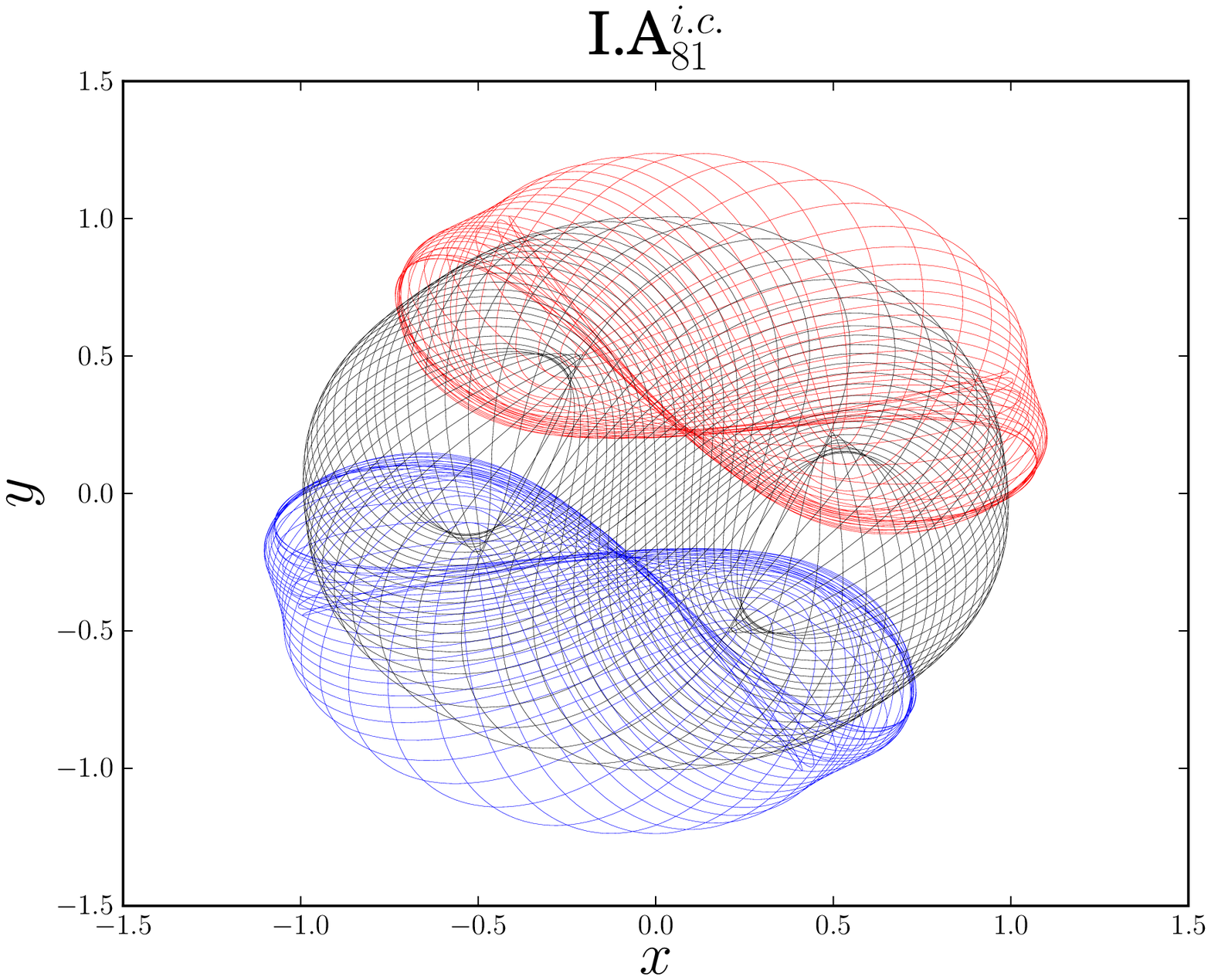}
\includegraphics[scale=0.26]{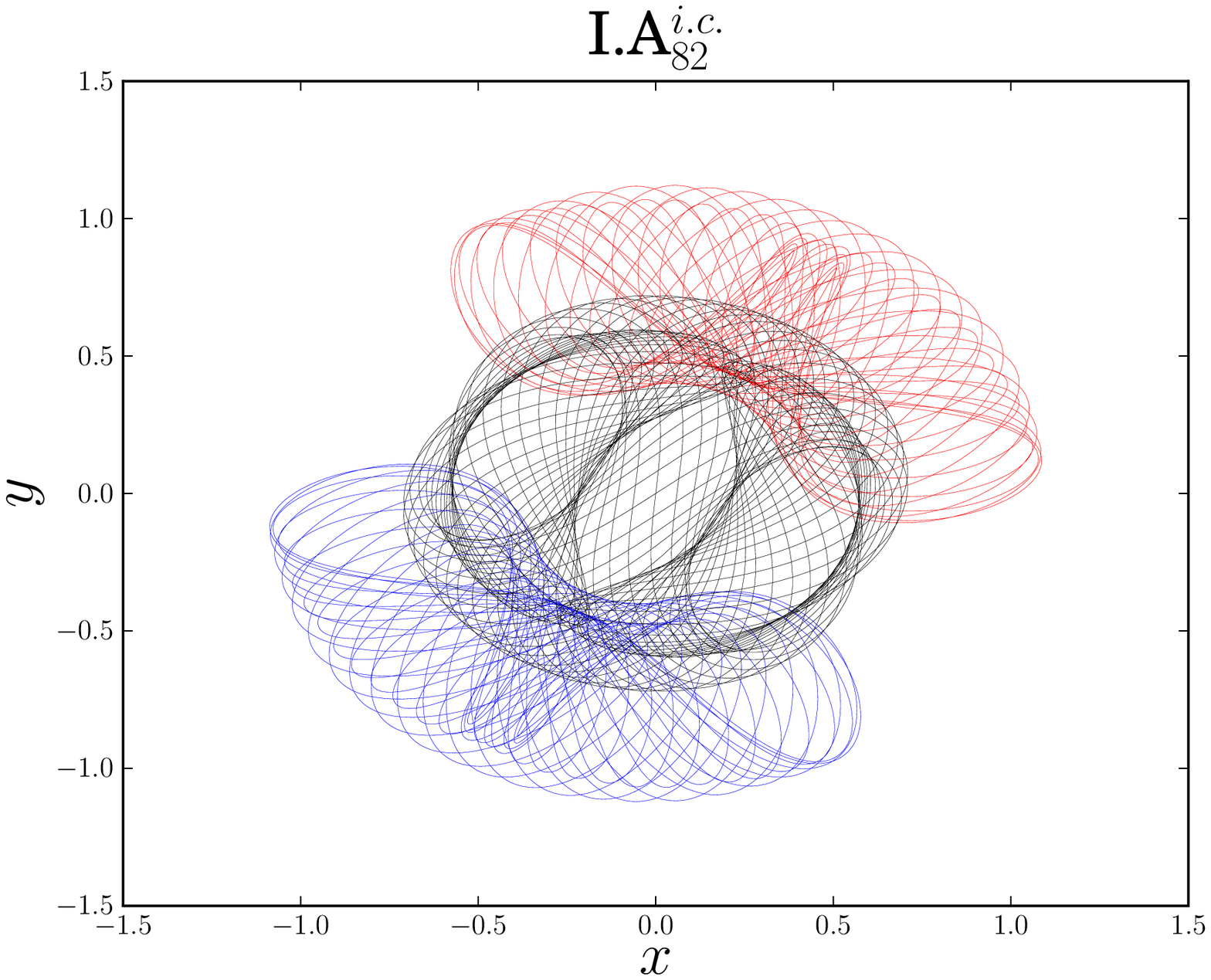}
\includegraphics[scale=0.26]{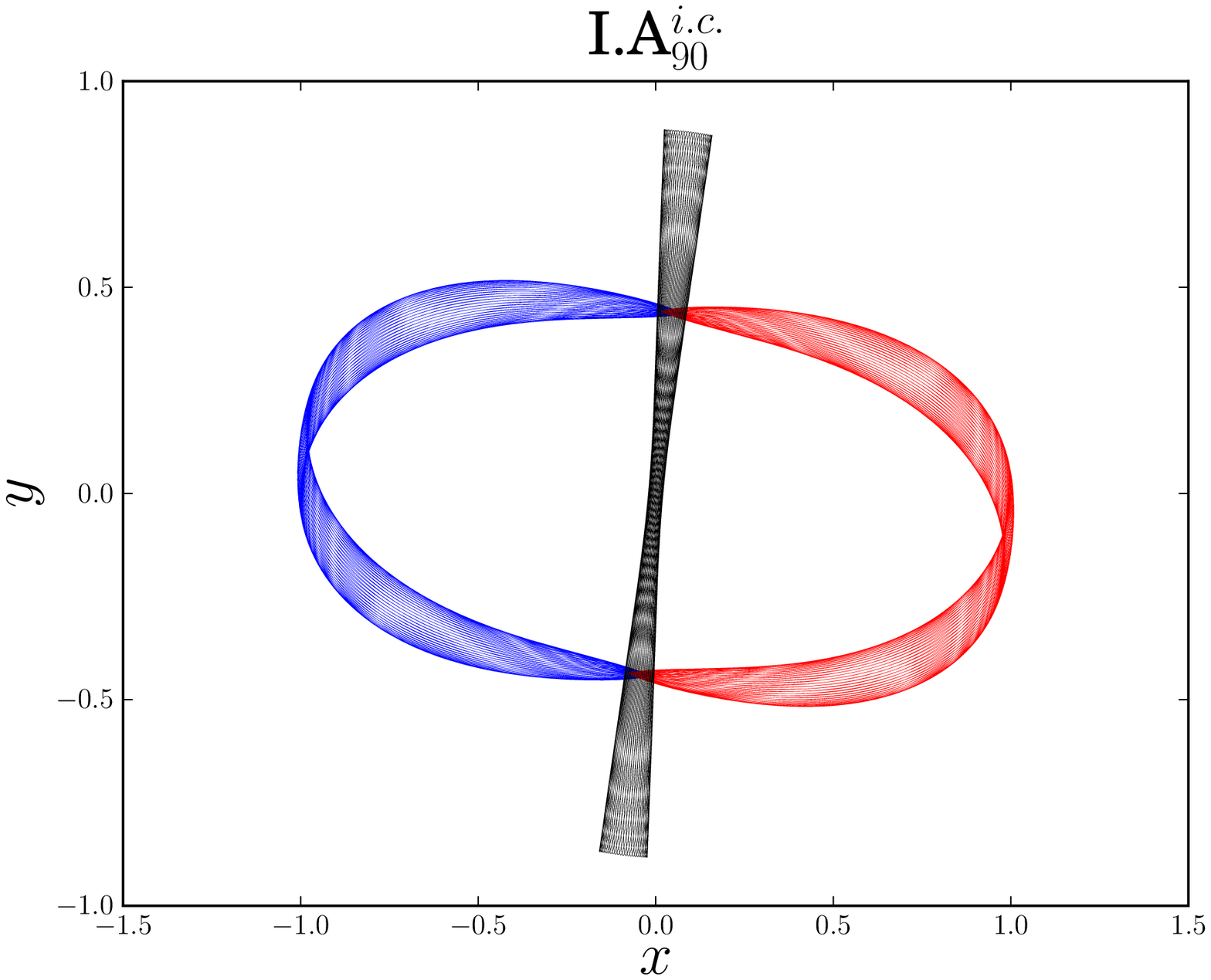}
\includegraphics[scale=0.26]{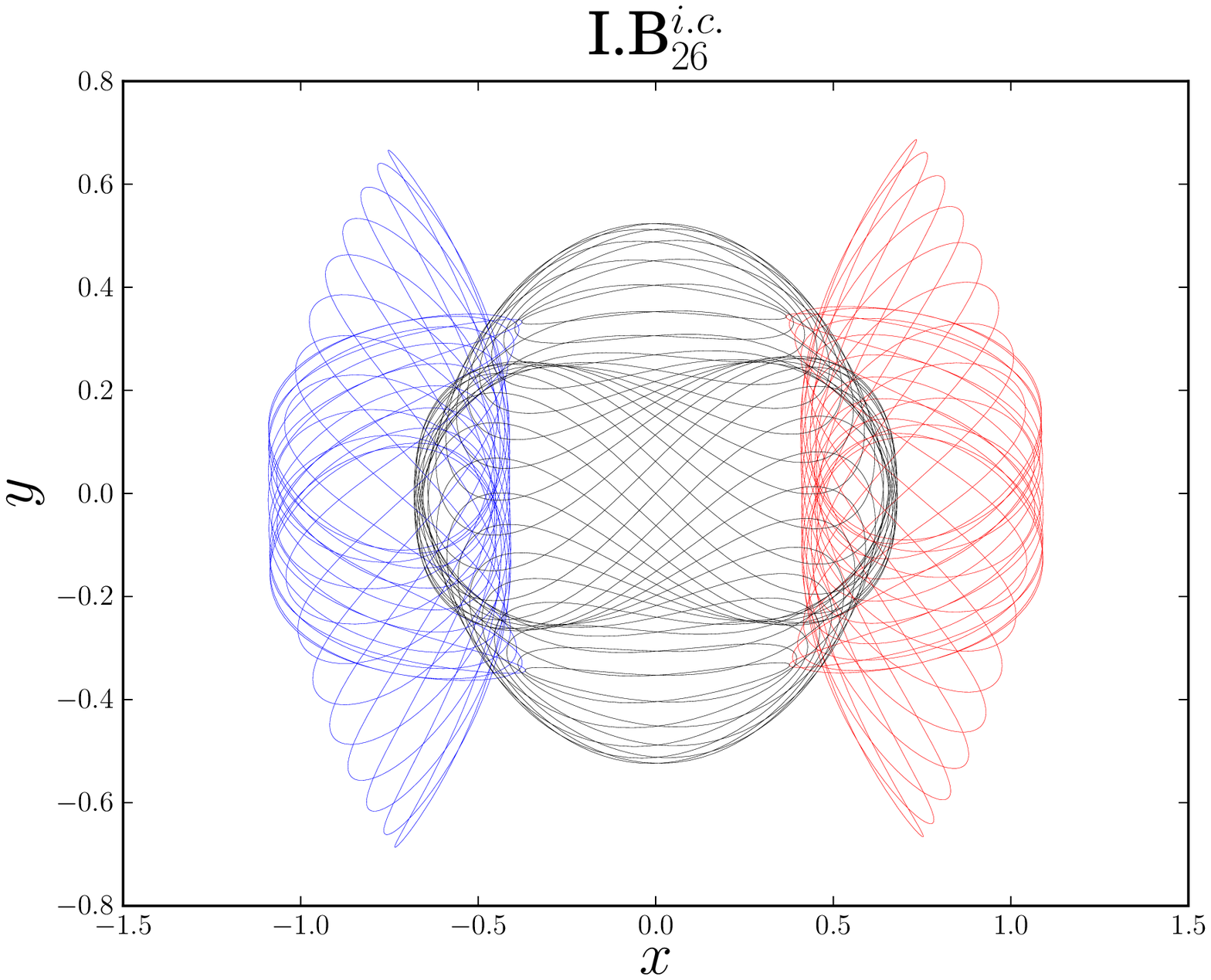}
\includegraphics[scale=0.26]{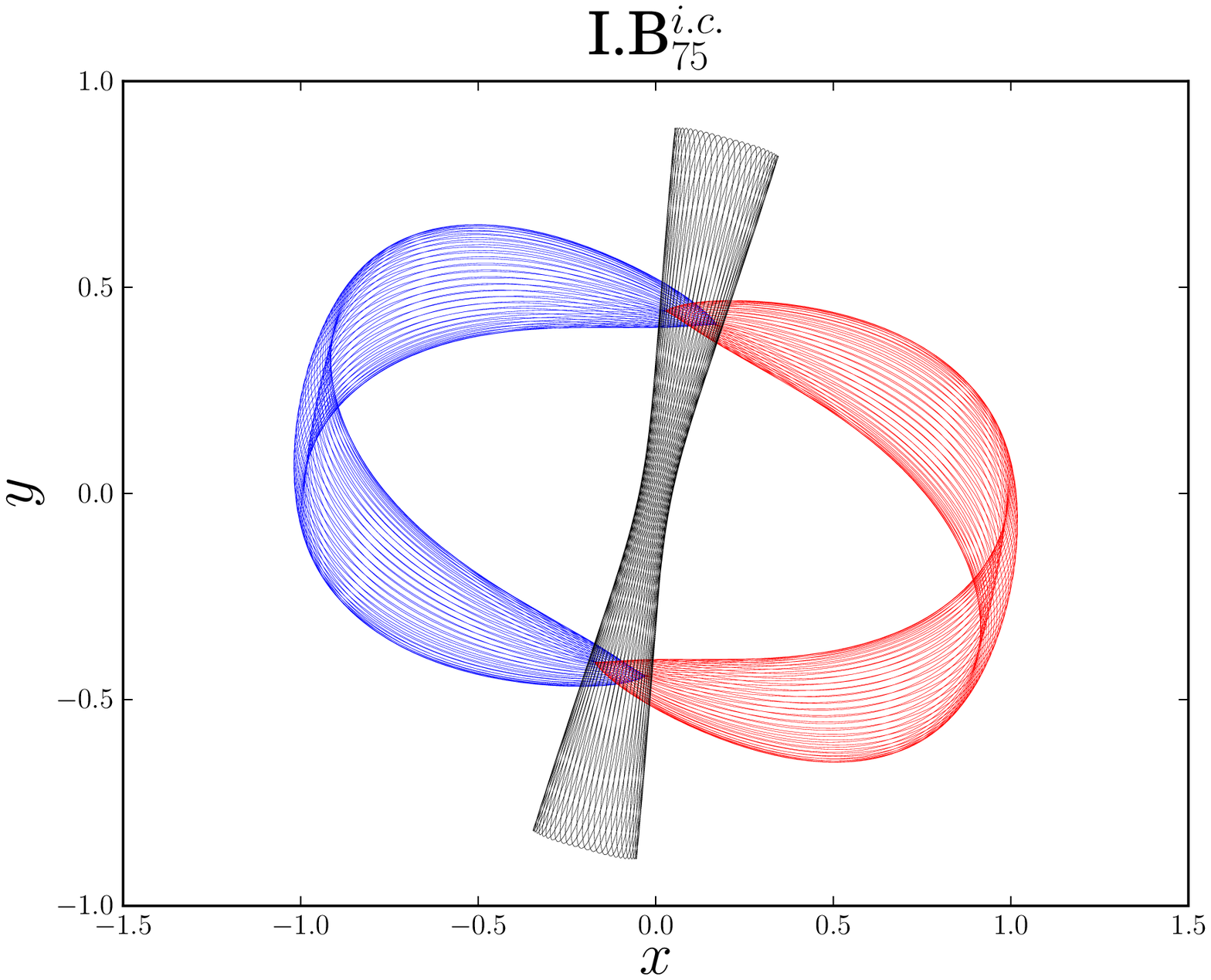}
\includegraphics[scale=0.26]{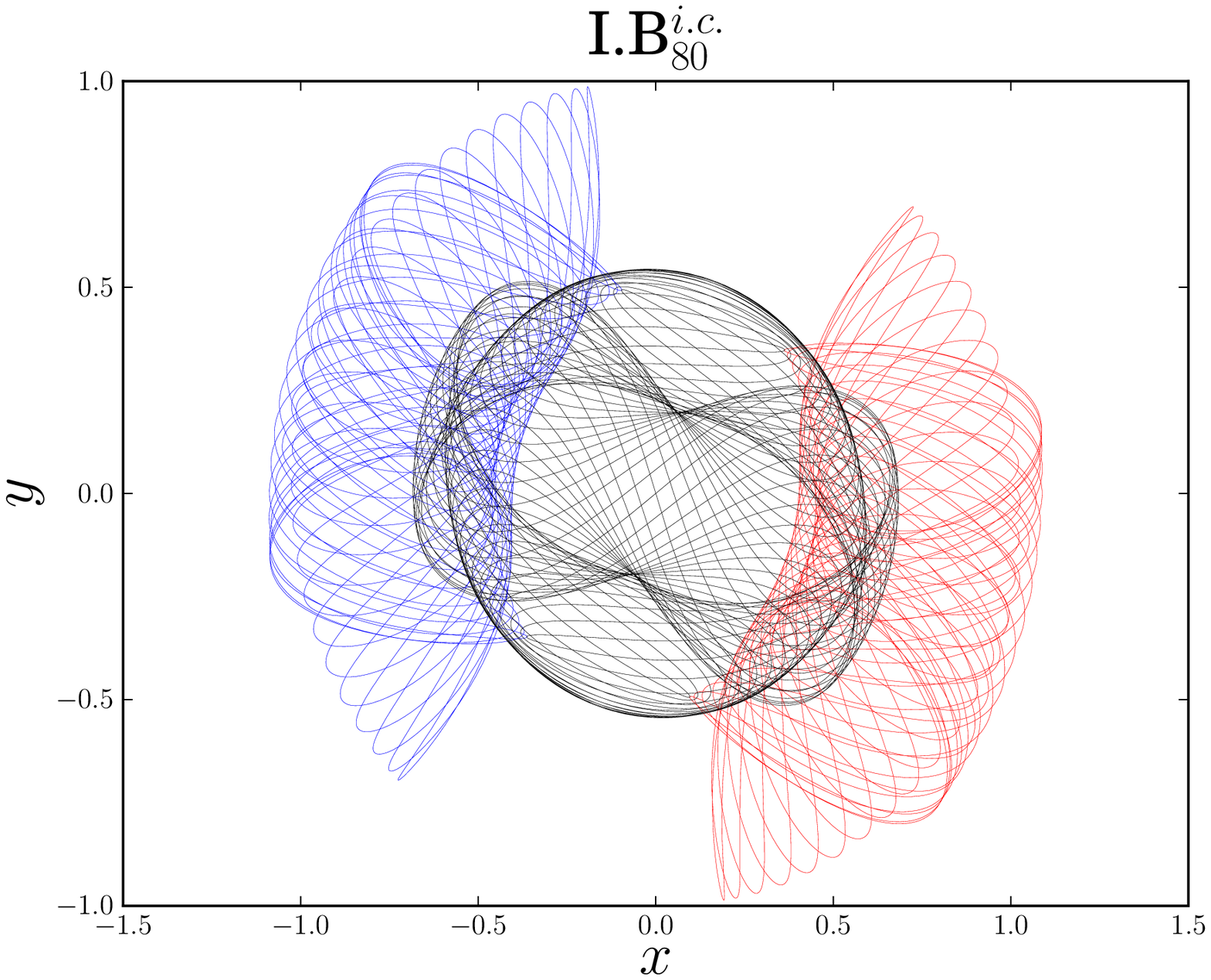}
\includegraphics[scale=0.26]{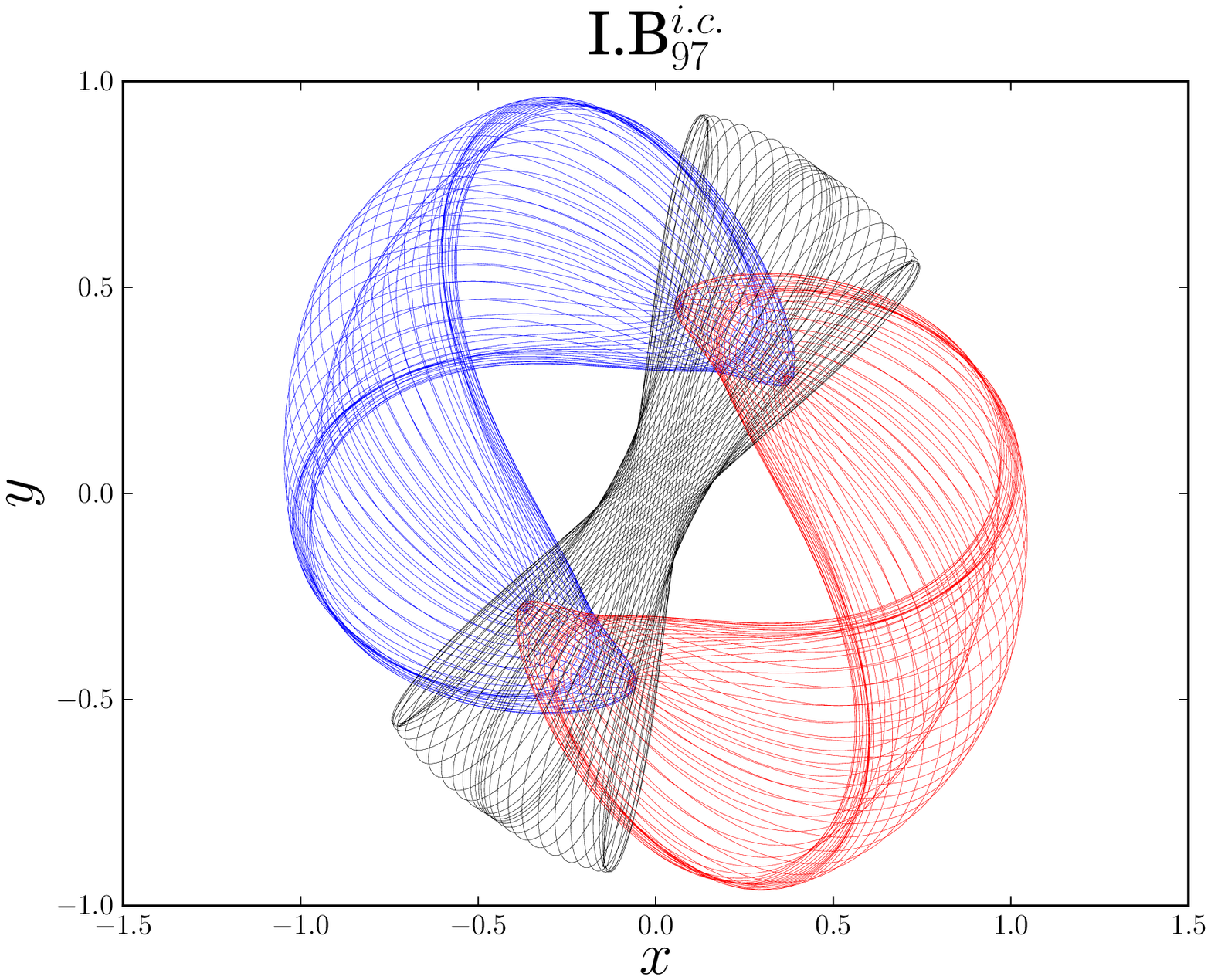}
\includegraphics[scale=0.26]{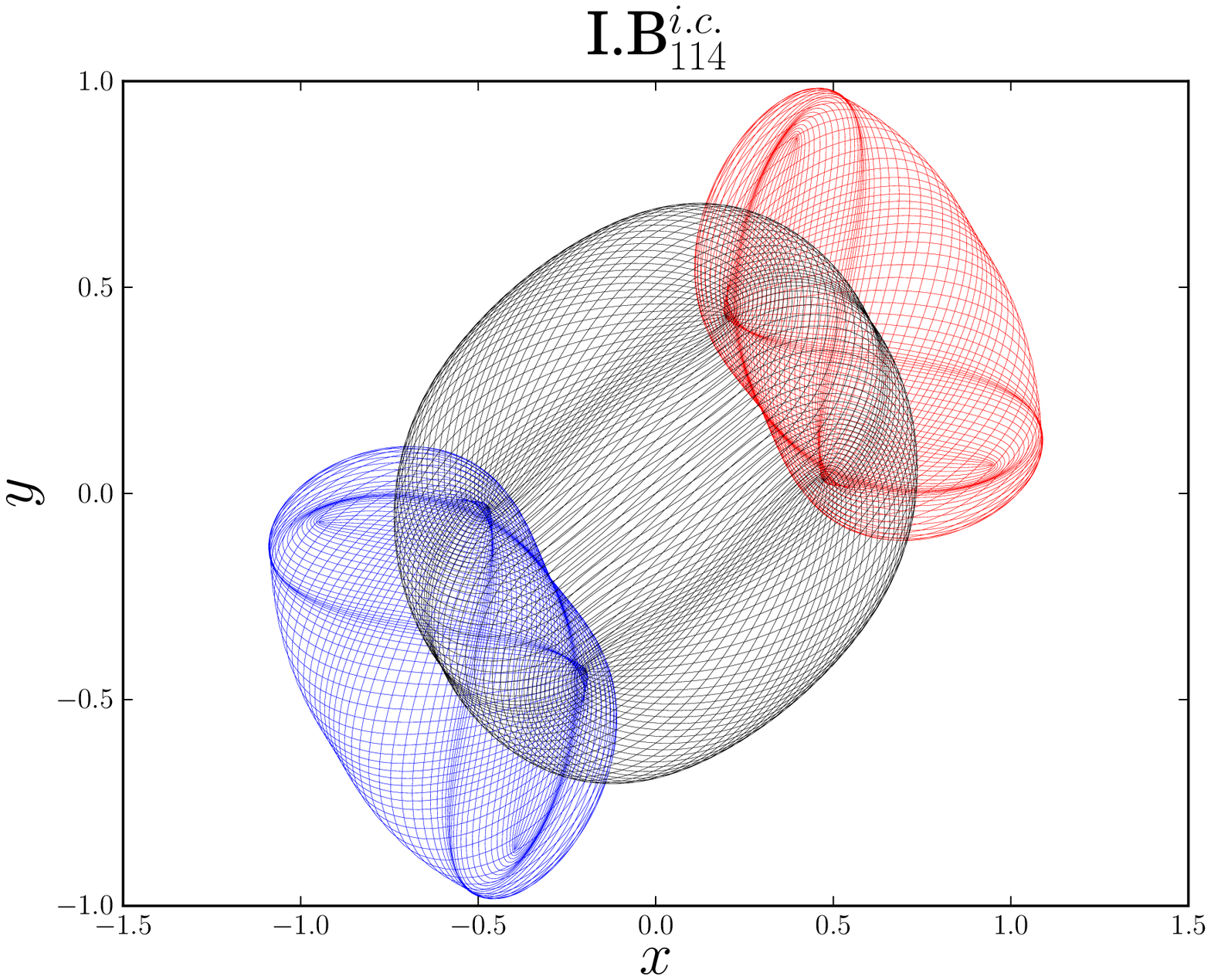}
\includegraphics[scale=0.26]{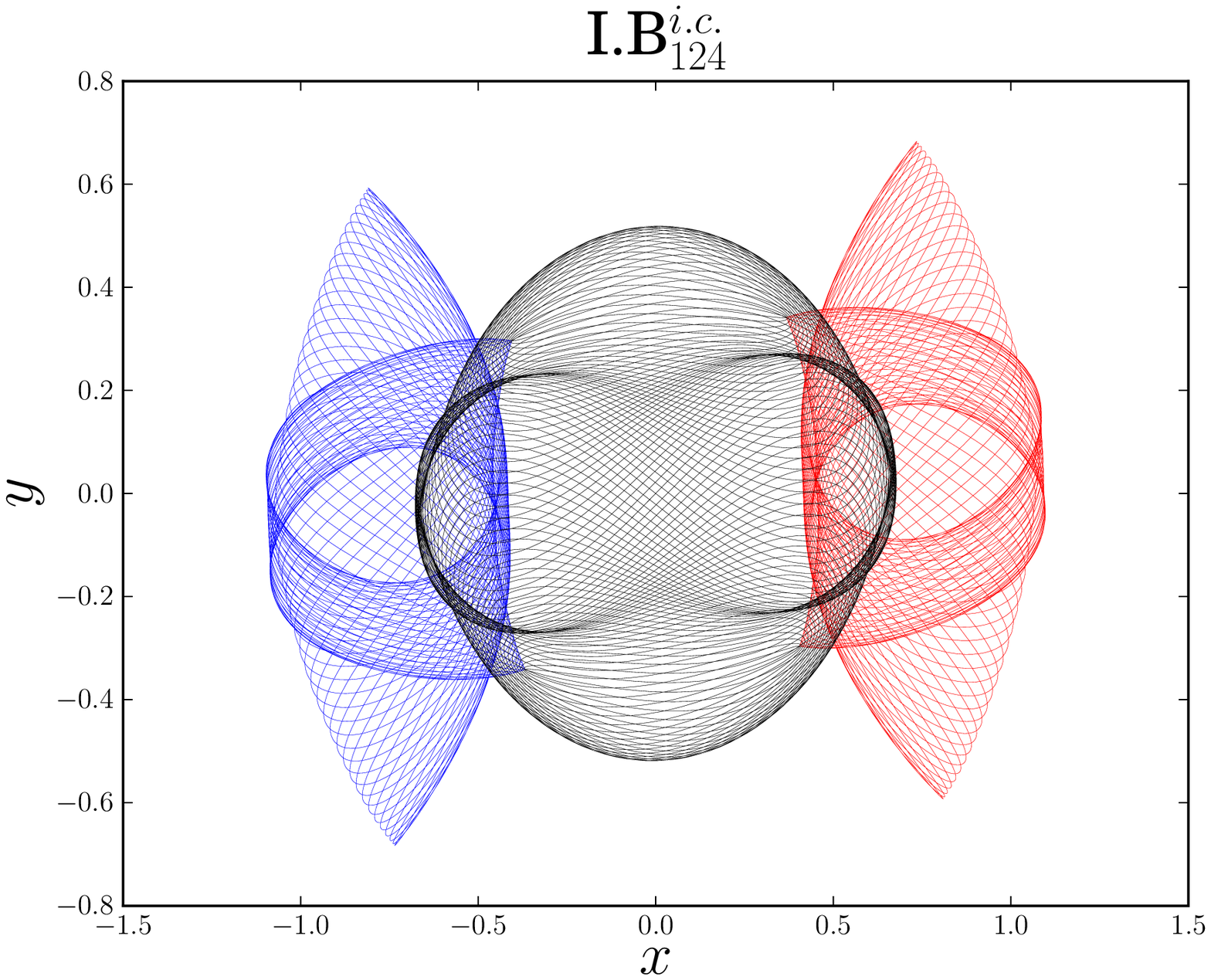}
\includegraphics[scale=0.26]{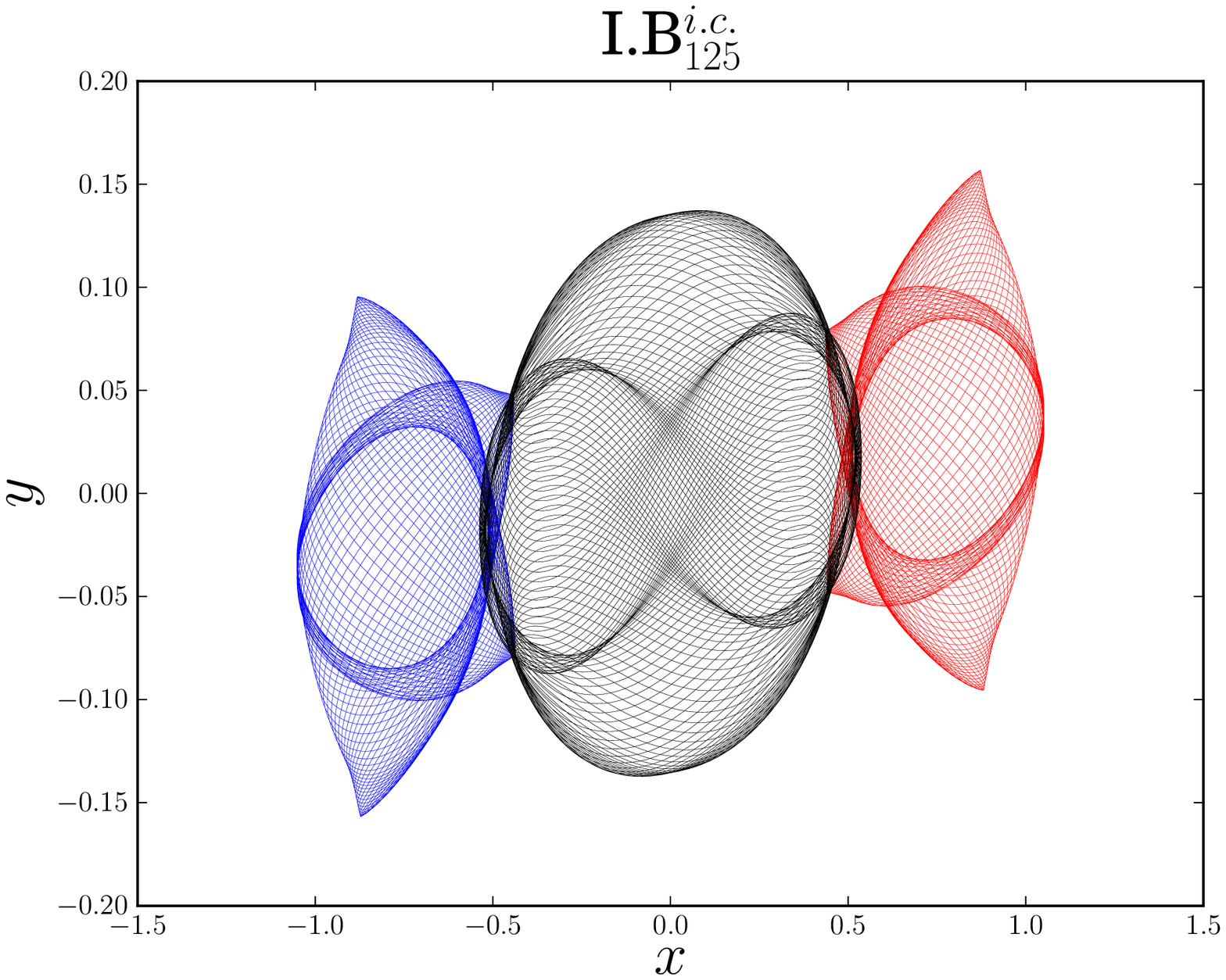}
\includegraphics[scale=0.26]{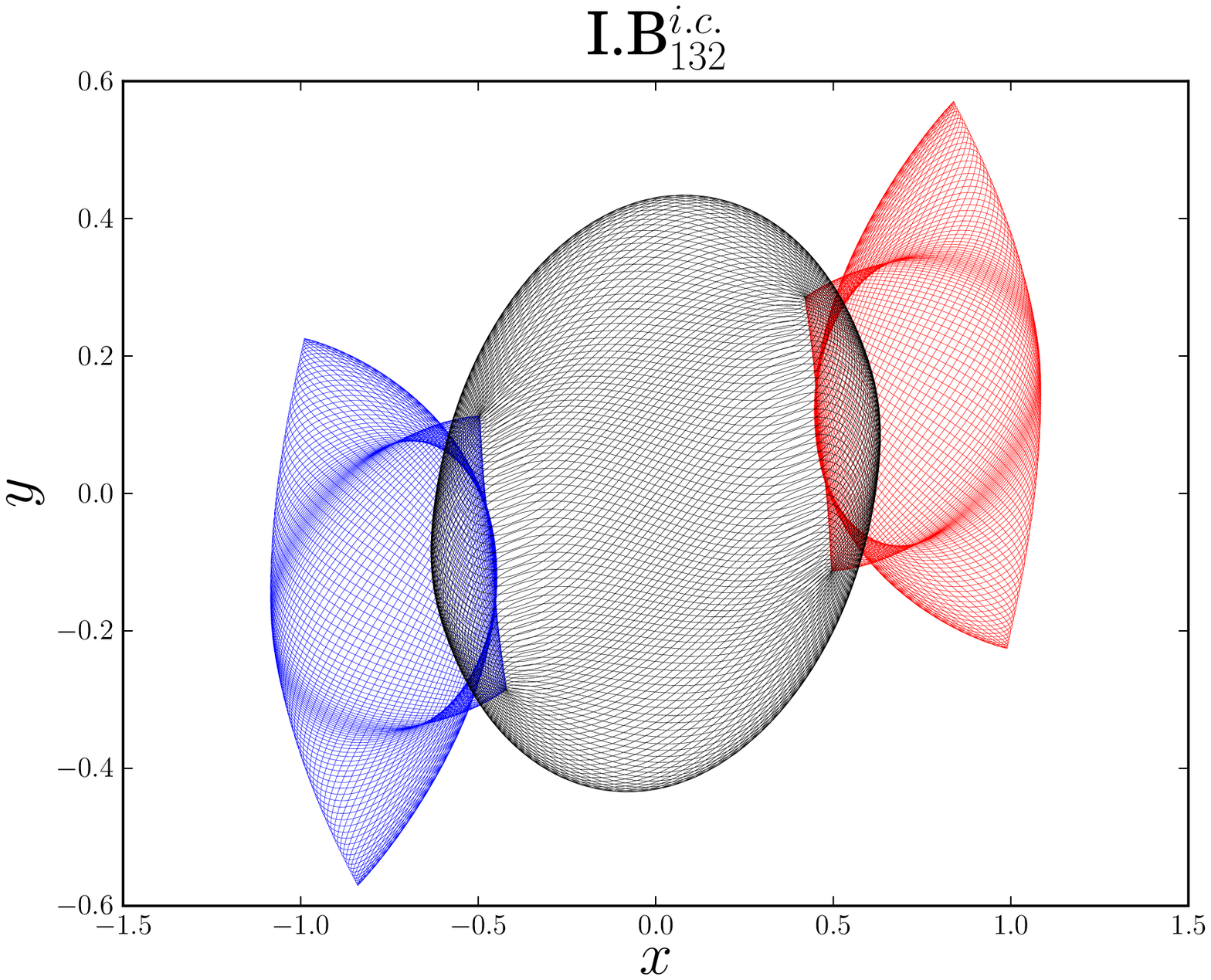}
\includegraphics[scale=0.26]{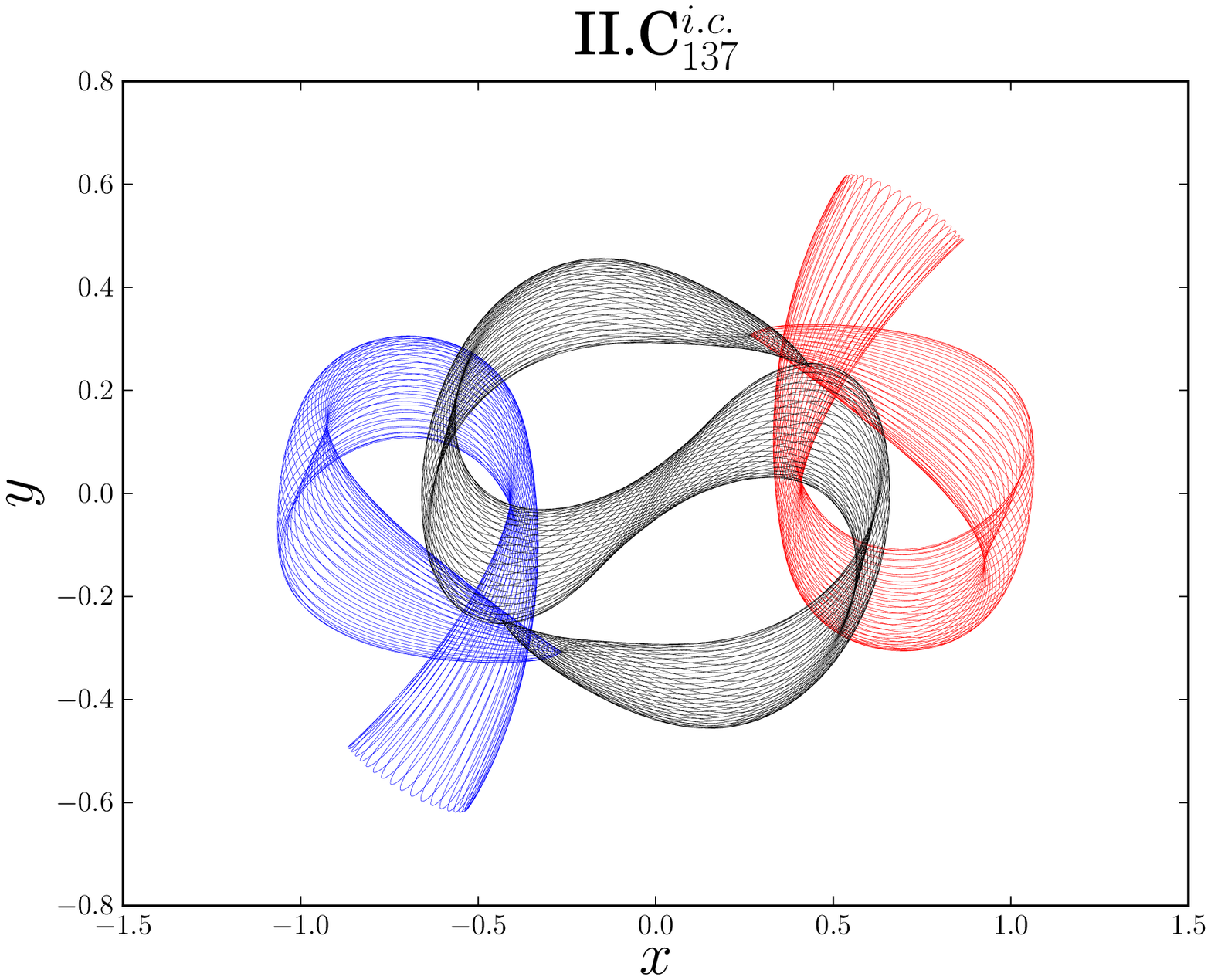}
   \caption{(color online) Periodic orbits of the three-body system  in the case of $\bm{r}_1(0)=(-1,0)=-\bm{r}_2(0)$,  $\dot{\bm{r}}_1(0)=(v_1,v_2)=\dot{\bm{r}}_2(0)$ and $\bm{r}_3(0)=(0,0)$, $\dot{\bm{r}}_3(0)=(-2v_1, -2v_2)$, where $G=1$ and $m_1=m_2=m_3=1$.}
  \label{fig-S1}
\end{figure*}

\begin{table*}
\tabcolsep 0pt \caption{The Figure-eight and the periodic orbits reported in \cite{Suvakov2013} and their names defined in this letter.}\label{table-S1} \vspace*{-6pt}
\def\temptablewidth{1\textwidth}
{\rule{\temptablewidth}{1pt}}
\begin{tabular*}{\temptablewidth}{@{\extracolsep{\fill}}lcl}
\hline
Class and name  & free group element & Class and number in this letter\\
\hline
I.A figure-eight & abAB & \hspace{2.0cm} I.A$^{i.c.}_{1}$ \\
I.A butterfly I and II & $(ab)^2(AB)^2$  &\hspace{2.0cm}  I.A$^{i.c.}_{2}$\\
I.A bumblebee & $b^2(ABab)^2A^2(baBA)^2baB^2(abAB)^2a^2(BAba)^2BA$ &\hspace{2.0cm}  I.A$^{i.c.}_{17}$\\
I.B moth I & $baBABabABA$  &\hspace{2.0cm}  I.B$^{i.c.}_{1}$\\
I.B butterfly III & $(ab)^2(ABA)(ba)^2(BAB)$ &\hspace{2.0cm}  I.B$^{i.c.}_{2}$\\
I.B moth II & $(abAB)^2A(baBA)^2B$&\hspace{2.0cm}  I.B$^{i.c.}_{5}$ \\
I.B moth III & $ABabaBABabaBAbabABAbab$ &\hspace{2.0cm}  I.B$^{i.c.}_{6}$ \\
I.B goggles & $(ab)^2ABBA(ba)^2BAAB$ &\hspace{2.0cm}  I.B$^{i.c.}_{3}$ \\
I.B butterfly IV & $((ba)^2(AB)^2)^6A((ba)^2(BA)^2)^6B$ &\hspace{2.0cm}  I.B$^{i.c.}_{49}$ \\
I.B dragonfly & $b^2(ABabAB)a^2(BAbaBA)$&\hspace{2.0cm}  I.B$^{i.c.}_{4}$\\
II.C yin-yang I (a) and (b) & $(ab)^2ABAba(BAB)$ &\hspace{2.0cm}  II.C$^{i.c.}_{1}$\\
II.C yin-yang \uppercase\expandafter{\romannumeral2} (a) and (b) & $(abaBAB)^3abaBAbab(ABAbab)^3(AB)^2$ &\hspace{2.0cm}  II.C$^{i.c.}_{27}$ \\

\hline
\end{tabular*}
{\rule{\temptablewidth}{1pt}}
\end{table*}

\begin{table*}
\tabcolsep 0pt \caption{The 14 among 25 families reported in \cite{Hudomal2015} and their names defined in this letter.  The other 11 families was the same as listed in Table S. I.   }\label{table-S1} \vspace*{-6pt}
\def\temptablewidth{1\textwidth}
{\rule{\temptablewidth}{1pt}}
\begin{tabular*}{\temptablewidth}{@{\extracolsep{\fill}}ll}
\hline
Class and name in [15] & Class and number in this letter\\
\hline
I.5.A	&		I.A$^{i.c.}_{5}$	\\
I.8.A	&		I.A$^{i.c.}_{10}$	\\
I.12.A	&		I.A$^{i.c.}_{21}$	\\
II.6.A	&		I.B$^{i.c.}_{7}$	\\
II.8.A	&		II.C$^{i.c.}_{10}$	\\
III.15.A.$\beta$	&		II.C$^{i.c.}_{35}$	\\
IVa.6.A	&		I.B$^{i.c.}_{8}$	\\
IVa.8.A	&		I.B$^{i.c.}_{13}$	\\
IVb.11.A	&		I.B$^{i.c.}_{17}$	\\
IVb.15.A	&		I.B$^{i.c.}_{59}$	\\
IVb.19.A	&		I.B$^{i.c.}_{38}$	\\
IVc.12.A.$\beta$	&		II.C$^{i.c.}_{29}$	\\
IVc.17.A	&		I.B$^{i.c.}_{33}$	\\
IVc.19.A	&		I.B$^{i.c.}_{37}$	\\

\hline
\end{tabular*}
{\rule{\temptablewidth}{1pt}}
\end{table*}

\begin{table*}
\tabcolsep 0pt \caption{Initial conditions and periods $T_0$ of the periodic three-body orbits for class I.A in case of the isosceles collinear configurations:  $\bm{r}_1(0)=(x_1,x_2)=-\bm{r}_2(0)$,  $\dot{\bm{r}}_1(0)=(v_1,v_2)=\dot{\bm{r}}_2(0)$ and $\bm{r}_3(0)=(0,0)$, $\dot{\bm{r}}_3(0)=(-2v_1, -2v_2)$ when $G=1$ and $m_1=m_2=m_3=1$ by means of the search grid $1000 \times 1000$  in the interval $T_0 \in [0,100]$.  Here, the superscript {\em i.c.} indicates the case of the initial conditions with {\em isosceles collinear}  configuration, due to the fact that there exist periodic orbits in many other cases. } \label{table-S2} \vspace*{-12pt}
\begin{center}
\def\temptablewidth{1\textwidth}
{\rule{\temptablewidth}{1pt}}
\begin{tabular*}{\temptablewidth}{@{\extracolsep{\fill}}lccccc}
\hline
Class and number & $x_1$ & $x_2$ & $v_1$ & $v_2$  & $T_0$\\
\hline
I.A$^{i.c.}_1$ &    -1.0024277970 &     0.0041695061 &     0.3489048974 &     0.5306305100 &     6.3490473929\\
I.A$^{i.c.}_2$ &    -1.0005576155 &    -0.0029240248 &     0.3064392516 &     0.1263673939 &     6.2399303613\\
I.A$^{i.c.}_3$ &    -0.9826997937 &    -0.0315203302 &     0.6030557226 &     0.5466747521 &    36.3842947805\\
I.A$^{i.c.}_4$ &    -0.9964559619 &     0.0051918896 &     0.5406827989 &     0.3392497094 &    26.7762404680\\
I.A$^{i.c.}_5$ &    -0.9720963681 &    -0.0704958088 &     0.3964121098 &     0.2935656454 &    19.9650778710\\
I.A$^{i.c.}_6$ &    -0.9999327738 &    -0.0008223485 &     0.4422571974 &     0.4238918567 &    35.8298692301\\
I.A$^{i.c.}_7$ &    -0.9977227094 &     0.0070021715 &     0.1222984917 &     0.1004607487 &    15.6909239194\\
I.A$^{i.c.}_8$ &    -0.9994554344 &    -0.0011995337 &     0.4091701188 &     0.3634132078 &    33.8401262309\\
I.A$^{i.c.}_9$ &    -0.9987682260 &    -0.0031341566 &     0.5251116596 &     0.2519279723 &    43.7891455476\\
I.A$^{i.c.}_{10}$ &    -1.0135928065 &     0.0369002019 &     0.4191609820 &     0.2663106893 &    34.9824204443\\
I.A$^{i.c.}_{11}$ &    -1.0019260810 &     0.0020911852 &     0.4369275314 &     0.4443694754 &    49.7389219446\\
I.A$^{i.c.}_{12}$&    -1.0005140398 &    -0.0001156871 &     0.1181029973 &     0.5846001931 &    48.3095979022\\
I.A$^{i.c.}_{13}$ &    -1.0020551612 &     0.0043916449 &     0.4775162156 &     0.3764636910 &    53.6371875645\\
I.A$^{i.c.}_{14}$ &    -1.3147318816 &     0.7132945256 &     0.3713149918 &    -0.0060886985 &    63.5000850224\\
I.A$^{i.c.}_{15}$ &    -0.9929272851 &    -0.0134998193 &     0.3925874406 &     0.3595501581 &    45.6518283178\\
I.A$^{i.c.}_{16}$ &    -0.9985908494 &    -0.0022240402 &     0.4333321596 &     0.4619833415 &    63.9809834569\\
I.A$^{i.c.}_{17}$ &    -1.0074958476 &     0.0081648176 &     0.1883232887 &     0.5834831526 &    64.2532204831\\
I.A$^{i.c.}_{18}$ &    -1.0035505512 &     0.0069356949 &     0.4600346659 &     0.4054759860 &    66.7080558750\\
I.A$^{i.c.}_{19}$ &    -0.9975373781 &    -0.0067760199 &     0.4137670183 &     0.2945195624 &    47.7502255665\\
I.A$^{i.c.}_{20}$ &    -1.0023416890 &     0.0034690771 &     0.0987694721 &     0.5606002909 &    53.8791265298\\
I.A$^{i.c.}_{21}$ &    -0.9944059892 &    -0.0160342344 &     0.4053380361 &     0.2509213812 &    48.0899422464\\
I.A$^{i.c.}_{22}$ &    -1.0051040678 &     0.0094190694 &     0.4322835083 &     0.4671381437 &    79.0046932990\\
I.A$^{i.c.}_{23}$ &    -1.0005990303 &     0.0006513647 &     0.4139830510 &     0.3474219895 &    61.9003772551\\
I.A$^{i.c.}_{24}$ &    -1.0022464885 &    -0.0005079367 &     0.4897501985 &     0.4042162458 &    85.6738583738\\
I.A$^{i.c.}_{25}$ &    -0.9996665302 &    -0.0003500779 &     0.1981083820 &     0.5762280189 &    73.2401681272\\
I.A$^{i.c.}_{26}$ &    -0.9946395214 &    -0.0199130330 &     0.3963758249 &     0.1931593194 &    48.2910929099\\
I.A$^{i.c.}_{27}$ &    -1.0128122834 &    -0.0049944126 &     0.4110801302 &     0.2954958184 &    62.5062638452\\
I.A$^{i.c.}_{28}$ &    -0.9983642281 &    -0.0029915248 &     0.0473206001 &     0.5908205345 &    78.9582358872\\
I.A$^{i.c.}_{29}$ &    -1.0012940616 &     0.0040175714 &     0.3460993325 &     0.3914034723 &    67.5194380903\\
I.A$^{i.c.}_{30}$ &    -1.0001268768 &    -0.0002768989 &     0.2201401036 &     0.5718474139 &    85.8099256919\\
I.A$^{i.c.}_{31}$ &    -1.0012673904 &     0.0023779503 &     0.4447554832 &     0.3810128399 &    82.5188557424\\
I.A$^{i.c.}_{32}$ &    -0.9877512802 &    -0.0430400097 &     0.3964569812 &     0.2378986017 &    61.3902686322\\
I.A$^{i.c.}_{33}$ &    -1.0172548423 &     0.0480118437 &     0.4209823631 &     0.3230893830 &    77.0997278616\\
I.A$^{i.c.}_{34}$ &    -1.0007553891 &     0.0016489085 &     0.4577398645 &     0.4065094686 &    95.9895491214\\
I.A$^{i.c.}_{35}$ &    -1.0049087422 &     0.0155206574 &     0.3987835408 &     0.1626357009 &    63.1005696104\\
I.A$^{i.c.}_{36}$ &    -1.0012474113 &     0.0030153067 &     0.4166996961 &     0.2957105307 &    74.9297037600\\
I.A$^{i.c.}_{37}$ &    -1.0002982171 &    -0.0008001808 &     0.1034021115 &     0.5907158341 &    92.3300512832\\
I.A$^{i.c.}_{38}$ &    -0.9977885909 &    -0.0055266460 &     0.4124824487 &     0.2735925337 &    75.7782102494\\
I.A$^{i.c.}_{39}$ &    -1.0004497905 &     0.0025599575 &     0.3976005814 &     0.3697808266 &    85.9927903991\\
I.A$^{i.c.}_{40}$ &    -0.9983014289 &    -0.0053078120 &     0.4064430497 &     0.2412089372 &    76.0294647048\\
I.A$^{i.c.}_{41}$ &    -0.9679361060 &    -0.1104962526 &     0.0672594569 &     0.6030209734 &    98.8657694414\\
I.A$^{i.c.}_{42}$ &    -0.9984824236 &    -0.0062824747 &     0.4005780292 &     0.2049425482 &    76.2294965022\\
I.A$^{i.c.}_{43}$ &    -1.0016614175 &     0.0039687130 &     0.4166059221 &     0.2962897857 &    88.2694596541\\
I.A$^{i.c.}_{44}$ &    -1.0001378487 &     0.0004564237 &     0.4189450671 &     0.2768485374 &    90.4399218097\\
I.A$^{i.c.}_{45}$ &    -0.9999058470 &     0.0041754333 &     0.3967334937 &     0.1569524375 &    76.5828835512\\
I.A$^{i.c.}_{46}$ &    -1.0056329618 &     0.0176753713 &     0.4127890336 &     0.2437542476 &    90.7691206861\\
I.A$^{i.c.}_{47}$ &    -0.9999108695 &     0.0015640018 &     0.4050458197 &     0.2219737238 &    90.1694565806\\

\hline
\end{tabular*}
{\rule{\temptablewidth}{1pt}}
\end{center}
\end{table*}

\begin{table*}
\tabcolsep 0pt \caption{Initial conditions and periods $T_0$ of the additional periodic three-body orbits for class I.A in case of the isosceles collinear configurations:  $\bm{r}_1(0)=(x_1,x_2)=-\bm{r}_2(0)$,  $\dot{\bm{r}}_1(0)=(v_1,v_2)=\dot{\bm{r}}_2(0)$ and $\bm{r}_3(0)=(0,0)$, $\dot{\bm{r}}_3(0)=(-2v_1, -2v_2)$ when $G=1$ and $m_1=m_2=m_3=1$ by means of the search grid $2000 \times 2000$ in the interval $T_0\in[0,200]$. Here, the superscript {\em i.c.} indicates the case of the initial conditions with {\em isosceles collinear}  configuration, due to the fact that there exist periodic orbits in many other cases. } \label{table-S2} \vspace*{-12pt}
\begin{center}
\def\temptablewidth{1\textwidth}
{\rule{\temptablewidth}{1pt}}
\begin{tabular*}{\temptablewidth}{@{\extracolsep{\fill}}lccccc}
\hline
Class and number & $x_1$ & $x_2$ & $v_1$ & $v_2$  & $T_0$\\
\hline
I.A$^{i.c.}_{48}$ &    -0.9970717514 &    -0.0066112020 &     0.2725401879 &     0.3471898950 &    36.8668830688\\
I.A$^{i.c.}_{49}$ &    -0.9993055583 &    -0.0017691191 &     0.5194976715 &     0.2069377318 &    61.0016940649\\
I.A$^{i.c.}_{50}$ &    -1.0014860102 &     0.0035079629 &     0.0500296227 &     0.5578404300 &    61.6443686231\\
I.A$^{i.c.}_{51}$ &    -1.0003917330 &     0.0010316214 &     0.3243735471 &     0.3661650449 &    61.7343870348\\
I.A$^{i.c.}_{52}$ &    -0.9999472207 &    -0.0001988820 &     0.4391549244 &     0.4491122766 &   107.1619509562\\
I.A$^{i.c.}_{53}$ &    -0.9991600773 &    -0.0006509487 &     0.4156894051 &     0.3447838776 &    90.1556159544\\
I.A$^{i.c.}_{54}$ &    -0.9995311212 &    -0.0003195682 &     0.2590432485 &     0.5657693024 &   112.8882799518\\
I.A$^{i.c.}_{55}$ &    -0.9942436902 &    -0.0077536363 &     0.4349257105 &     0.4616090902 &   120.6329484414\\
I.A$^{i.c.}_{56}$ &    -1.0005998776 &     0.0015362559 &     0.4717844516 &     0.3847751320 &   112.8824953036\\
I.A$^{i.c.}_{57}$ &    -0.9990550745 &     0.0000214205 &     0.1539712676 &     0.5808239605 &   114.2435641789\\
I.A$^{i.c.}_{58}$ &    -1.0021284758 &     0.0070204955 &     0.4190491165 &     0.4632198988 &   130.5436756627\\
I.A$^{i.c.}_{59}$ &    -1.0024466663 &     0.0064183595 &     0.4622726354 &     0.4013337272 &   126.6408258652\\
I.A$^{i.c.}_{60}$ &    -1.0004429249 &     0.0003273169 &     0.4261273054 &     0.4686843083 &   134.9469165376\\
I.A$^{i.c.}_{61}$ &    -0.9993038772 &    -0.0018298048 &     0.4110726779 &     0.3448242796 &   103.3700141300\\
I.A$^{i.c.}_{62}$ &    -1.0004962140 &     0.0006339779 &     0.4540087935 &     0.4147429148 &   139.0561512750\\
I.A$^{i.c.}_{63}$ &    -0.9995955668 &     0.0003135150 &     0.1809577601 &     0.5804879735 &   128.8529242838\\
I.A$^{i.c.}_{64}$ &    -0.9984538562 &     0.0004862707 &     0.0619642344 &     0.5891940141 &   120.9777119720\\
I.A$^{i.c.}_{65}$ &    -0.9956418854 &    -0.0105005200 &     0.4141715883 &     0.3035751262 &   100.7877228319\\
I.A$^{i.c.}_{66}$ &    -1.0005184687 &     0.0012193846 &     0.4759236762 &     0.3827497380 &   143.6038518334\\
I.A$^{i.c.}_{67}$ &    -1.0027183274 &     0.0056043614 &     0.4203795432 &     0.3547405295 &   118.0321541279\\
I.A$^{i.c.}_{68}$ &    -0.9997382689 &    -0.0013890352 &     0.3995945246 &     0.1903871326 &    90.3209435951\\
I.A$^{i.c.}_{69}$ &    -1.0030615840 &     0.0029505323 &     0.1952108166 &     0.5761734350 &   141.5760530626\\
I.A$^{i.c.}_{70}$ &    -0.9974378050 &    -0.0037666950 &     0.4299422880 &     0.4756088609 &   164.2039933205\\
I.A$^{i.c.}_{71}$ &    -0.9999530579 &    -0.0014466165 &     0.0952040907 &     0.5891189652 &   134.1575359173\\
I.A$^{i.c.}_{72}$ &    -1.0000801372 &     0.0010493574 &     0.3954312115 &     0.1504314002 &    90.5729328818\\
I.A$^{i.c.}_{73}$ &    -0.9992695612 &    -0.0005990245 &     0.4394776080 &     0.4506573818 &   164.4329403163\\
I.A$^{i.c.}_{74}$ &    -0.9997010964 &    -0.0007706723 &     0.3671737738 &     0.3841633179 &   119.6700426523\\
I.A$^{i.c.}_{75}$ &    -0.9994269613 &    -0.0017062739 &     0.3934480301 &     0.3873560874 &   127.2910319418\\
I.A$^{i.c.}_{76}$ &    -0.9990954797 &    -0.0020776197 &     0.3424315738 &     0.3727640281 &   111.4779241093\\
I.A$^{i.c.}_{77}$ &    -1.0052602165 &     0.0113635581 &     0.4181957794 &     0.2933664631 &   115.5046794627\\
I.A$^{i.c.}_{78}$ &    -1.0006370989 &    -0.0001585675 &     0.4068197040 &     0.2365440059 &   104.0559804341\\
I.A$^{i.c.}_{79}$ &    -0.9999910898 &    -0.0000007173 &     0.4589992222 &     0.4050055624 &   155.7639788636\\
I.A$^{i.c.}_{80}$ &    -1.0009759682 &     0.0030336107 &     0.4145778447 &     0.2853135672 &   116.7249893380\\
I.A$^{i.c.}_{81}$ &    -0.9987323215 &    -0.0021372611 &     0.4372689963 &     0.4569217324 &   178.7637976321\\
I.A$^{i.c.}_{82}$ &    -1.0004745792 &     0.0022002834 &     0.4132704922 &     0.3425300636 &   132.0514626019\\
I.A$^{i.c.}_{83}$ &    -1.0003921311 &     0.0004041641 &     0.4562613128 &     0.4115241891 &   169.0502867565\\
I.A$^{i.c.}_{84}$ &    -0.9999930696 &    -0.0026459611 &     0.3980700649 &     0.1812380577 &   104.3133382091\\
I.A$^{i.c.}_{85}$ &    -0.9984524060 &    -0.0055933406 &     0.4114699919 &     0.2699715248 &   117.2421369067\\
I.A$^{i.c.}_{86}$ &    -0.9993317122 &    -0.0020823613 &     0.3915485623 &     0.3742081549 &   137.5779781384\\
I.A$^{i.c.}_{87}$ &    -0.9960374283 &    -0.0103668006 &     0.4157355486 &     0.3035215110 &   127.5757265907\\
I.A$^{i.c.}_{88}$ &    -1.0020959234 &     0.0015663655 &     0.1389742240 &     0.5826068438 &   157.8418810354\\
I.A$^{i.c.}_{89}$ &    -1.0004561512 &    -0.0023140361 &     0.3967894160 &     0.2499782377 &   115.5242647425\\
I.A$^{i.c.}_{90}$ &    -1.0000894168 &     0.0002236977 &     0.0189957894 &     0.5900458849 &   152.0107762653\\
I.A$^{i.c.}_{91}$ &    -0.9992929543 &    -0.0038341920 &     0.3943290913 &     0.1464817627 &   104.4249785588\\
I.A$^{i.c.}_{92}$ &    -1.0018040115 &     0.0045975362 &     0.4156341608 &     0.2871611522 &   130.5396768630\\
I.A$^{i.c.}_{93}$ &    -1.0001658064 &    -0.0009545069 &     0.4047227950 &     0.2247643321 &   117.9451594108\\
I.A$^{i.c.}_{94}$ &    -1.0017458041 &     0.0014484003 &     0.4949739396 &     0.3806199303 &   196.3371049765\\
I.A$^{i.c.}_{95}$ &    -1.0004497215 &     0.0004384296 &     0.4099489458 &     0.4856847826 &   199.5671098720\\
I.A$^{i.c.}_{96}$ &    -0.9955024461 &    -0.0175078718 &     0.3527823980 &     0.2168382661 &   109.3867930164\\
I.A$^{i.c.}_{97}$ &    -0.9930705403 &    -0.0046556491 &     0.1436034540 &     0.5787455174 &   163.1696143768\\
I.A$^{i.c.}_{98}$ &    -0.9982439671 &    -0.0065502776 &     0.4364331018 &     0.1920468901 &   125.4417151404\\
I.A$^{i.c.}_{99}$ &    -1.0036674955 &     0.0114128366 &     0.3718377385 &     0.2740663503 &   122.1136696888\\
I.A$^{i.c.}_{100}$ &    -1.0000757550 &    -0.0000623332 &     0.0670368292 &     0.5889446609 &   163.9288153507\\

\hline
\end{tabular*}
{\rule{\temptablewidth}{1pt}}
\end{center}
\end{table*}

\begin{table*}
\tabcolsep 0pt \caption{Initial conditions and periods $T_0$ of the additional periodic three-body orbits for class I.A in case of the isosceles collinear configurations:  $\bm{r}_1(0)=(x_1,x_2)=-\bm{r}_2(0)$,  $\dot{\bm{r}}_1(0)=(v_1,v_2)=\dot{\bm{r}}_2(0)$ and $\bm{r}_3(0)=(0,0)$, $\dot{\bm{r}}_3(0)=(-2v_1, -2v_2)$ when $G=1$ and $m_1=m_2=m_3=1$ by means of the search grid $2000 \times 2000$ in the interval $T_0\in[0,200]$. Here, the superscript {\em i.c.} indicates the case of the initial conditions with {\em isosceles collinear}  configuration, due to the fact that there exist periodic orbits in many other cases. } \label{table-S2} \vspace*{-12pt}
\begin{center}
\def\temptablewidth{1\textwidth}
{\rule{\temptablewidth}{1pt}}
\begin{tabular*}{\temptablewidth}{@{\extracolsep{\fill}}lccccc}
\hline
Class and number & $x_1$ & $x_2$ & $v_1$ & $v_2$  & $T_0$\\
\hline
I.A$^{i.c.}_{101}$ &    -0.9980498750 &    -0.0035543947 &     0.4014889642 &     0.3688822853 &   153.1411771347\\
I.A$^{i.c.}_{102}$ &    -1.0025774193 &     0.0085296896 &     0.4117581696 &     0.2507644030 &   132.0010501147\\
I.A$^{i.c.}_{103}$ &    -0.9998693745 &    -0.0005516272 &     0.3975360756 &     0.1727271043 &   118.2509473803\\
I.A$^{i.c.}_{104}$ &    -0.9944129406 &    -0.0101009703 &     0.4045769173 &     0.3558077656 &   154.0649614429\\
I.A$^{i.c.}_{105}$ &    -0.9998144792 &    -0.0016980364 &     0.3651998879 &     0.1503923517 &   113.4719679745\\
I.A$^{i.c.}_{106}$ &    -0.9992926040 &    -0.0024378359 &     0.3283160692 &     0.2489191201 &   116.6417925924\\
I.A$^{i.c.}_{107}$ &    -1.0003559141 &    -0.0001742931 &     0.0920632794 &     0.5887350519 &   176.6936322233\\
I.A$^{i.c.}_{108}$ &    -1.0000759501 &     0.0009414908 &     0.3944719826 &     0.1398365126 &   118.5237477284\\
I.A$^{i.c.}_{109}$ &    -1.0006996213 &     0.0025734444 &     0.4169961719 &     0.2751525936 &   145.6545713368\\
I.A$^{i.c.}_{110}$ &    -0.9995686642 &    -0.0016179019 &     0.4031229886 &     0.2150896308 &   131.7859413641\\
I.A$^{i.c.}_{111}$ &    -0.9975007296 &    -0.0022271183 &     0.1807778616 &     0.5806937842 &   195.0937396077\\
I.A$^{i.c.}_{112}$ &    -0.9971617354 &    -0.0064908809 &     0.4092181923 &     0.3522382038 &   169.7646004853\\
I.A$^{i.c.}_{113}$ &    -0.9989792122 &    -0.0027842513 &     0.4153763867 &     0.2941712804 &   157.1836031770\\
I.A$^{i.c.}_{114}$ &    -1.0002217406 &     0.0011763499 &     0.3970289336 &     0.1657108263 &   132.2809838085\\
I.A$^{i.c.}_{115}$ &    -1.0086187694 &     0.0253661576 &     0.3425773592 &     0.2803116893 &   141.1088058748\\
I.A$^{i.c.}_{116}$ &    -0.9967019842 &    -0.0102228962 &     0.4772799155 &     0.2079941937 &   166.7612378927\\
I.A$^{i.c.}_{117}$ &    -0.9997318224 &     0.0015975999 &     0.1271720027 &     0.5842968041 &   199.3798927212\\
I.A$^{i.c.}_{118}$ &    -0.9958384074 &    -0.0079560647 &     0.3573510047 &     0.3805021232 &   164.4220924074\\
I.A$^{i.c.}_{119}$ &    -0.9999600489 &    -0.0002696990 &     0.3734739228 &     0.3815627662 &   171.6768976897\\
I.A$^{i.c.}_{120}$ &    -0.9999830568 &    -0.0004047452 &     0.3940040329 &     0.1365476687 &   132.4845615563\\
I.A$^{i.c.}_{121}$ &    -1.0000068462 &     0.0001539206 &     0.0399282382 &     0.5900134926 &   194.6234464267\\
I.A$^{i.c.}_{122}$ &    -0.9998855972 &    -0.0005895799 &     0.3365331437 &     0.1547319436 &   122.1298182577\\
I.A$^{i.c.}_{123}$ &    -0.9985954571 &    -0.0012733756 &     0.4147236820 &     0.3602760285 &   184.6195724713\\
I.A$^{i.c.}_{124}$ &    -0.9998714565 &    -0.0004642179 &     0.4020368267 &     0.2061879841 &   145.7969827626\\
I.A$^{i.c.}_{125}$ &    -1.0029626748 &     0.0097523094 &     0.3005946215 &     0.2609647259 &   135.2240397448\\
I.A$^{i.c.}_{126}$ &    -0.9973886171 &    -0.0103202322 &     0.3134767516 &     0.2061233272 &   129.0733455925\\
I.A$^{i.c.}_{127}$ &    -0.9998681069 &    -0.0024957082 &     0.3681697214 &     0.0549351475 &   129.1378047016\\
I.A$^{i.c.}_{128}$ &    -0.9994323909 &    -0.0004841463 &     0.4094012853 &     0.2507612159 &   159.0882991197\\
I.A$^{i.c.}_{129}$ &    -0.9996659019 &    -0.0016075960 &     0.3961087209 &     0.1615911956 &   146.1290208130\\
I.A$^{i.c.}_{130}$ &    -0.9969470109 &    -0.0253802656 &     0.4814933573 &     0.2023642181 &   185.9176802561\\
I.A$^{i.c.}_{131}$ &    -1.0002876188 &     0.0017766155 &     0.3940015070 &     0.1324912273 &   146.5309672408\\
I.A$^{i.c.}_{132}$ &    -1.0002923083 &     0.0017364899 &     0.2848253321 &     0.2571432568 &   144.7999029532\\
I.A$^{i.c.}_{133}$ &    -1.0000978869 &     0.0015260698 &     0.3614477306 &     0.0656590456 &   141.3376678772\\
I.A$^{i.c.}_{134}$ &    -0.9996791095 &    -0.0007360924 &     0.4051712145 &     0.2265898069 &   173.3010029488\\
I.A$^{i.c.}_{135}$ &    -1.0014325470 &     0.0087096270 &     0.2810416717 &     0.2283347205 &   146.8446468171\\
I.A$^{i.c.}_{136}$ &    -1.0000891131 &     0.0004413046 &     0.4089772966 &     0.2473799317 &   186.9853647996\\
I.A$^{i.c.}_{137}$ &    -0.9994516060 &     0.0017634143 &     0.2760222213 &     0.2508728684 &   155.3273630692\\
I.A$^{i.c.}_{138}$ &    -0.9999827462 &    -0.0001335022 &     0.3000121740 &     0.1955223337 &   152.1560300490\\
I.A$^{i.c.}_{139}$ &    -1.0001424259 &     0.0004770024 &     0.2760294383 &     0.2669517338 &   164.9309074218\\
I.A$^{i.c.}_{140}$ &    -0.9999832002 &    -0.0001069975 &     0.3060093674 &     0.1575218584 &   155.4592003087\\
I.A$^{i.c.}_{141}$ &    -1.0000172201 &     0.0007018679 &     0.2739672827 &     0.2264723307 &   157.9043880548\\
I.A$^{i.c.}_{142}$ &    -0.9977811572 &    -0.0065320744 &     0.2701193366 &     0.2824432141 &   172.7281498960\\
I.A$^{i.c.}_{143}$ &    -1.0000425619 &     0.0000461826 &     0.2690794848 &     0.2460572203 &   166.5520520337\\
I.A$^{i.c.}_{144}$ &    -1.0017356249 &     0.0010995344 &     0.2813821520 &     0.2925415750 &   185.4915045092\\
I.A$^{i.c.}_{145}$ &    -0.9994108565 &    -0.0010114065 &     0.2841969062 &     0.2868493123 &   183.9837349836\\
I.A$^{i.c.}_{146}$ &    -1.0002991684 &     0.0032894033 &     0.3408512038 &     0.0867477923 &   176.5621908497\\
I.A$^{i.c.}_{147}$ &    -1.0004406713 &     0.0143729532 &     0.2775430457 &     0.2145017126 &   182.6267578048\\
\hline
\end{tabular*}
{\rule{\temptablewidth}{1pt}}
\end{center}
\end{table*}

\begin{table*}
\tabcolsep 0pt \caption{Initial conditions and periods $T_0$ of the additional periodic three-body orbits for class I.A in case of the isosceles collinear configurations:  $\bm{r}_1(0)=(x_1,x_2)=-\bm{r}_2(0)$,  $\dot{\bm{r}}_1(0)=(v_1,v_2)=\dot{\bm{r}}_2(0)$ and $\bm{r}_3(0)=(0,0)$, $\dot{\bm{r}}_3(0)=(-2v_1, -2v_2)$ when $G=1$ and $m_1=m_2=m_3=1$ by means of the search grid $4000 \times 4000$ in the interval $T_0\in[0,200]$. Here, the superscript {\em i.c.} indicates the case of the initial conditions with {\em isosceles collinear}  configuration, due to the fact that there exist periodic orbits in many other cases. } \label{table-S2} \vspace*{-12pt}
\begin{center}
\def\temptablewidth{1\textwidth}
{\rule{\temptablewidth}{1pt}}
\begin{tabular*}{\temptablewidth}{@{\extracolsep{\fill}}lccccc}
\hline
Class and number & $x_1$ & $x_2$ & $v_1$ & $v_2$  & $T_0$\\
\hline
I.A$^{i.c.}_{148}$ &    -0.9997771257 &    -0.0002473119 &     0.5236496560 &     0.2616880709 &    77.5192495477\\
I.A$^{i.c.}_{149}$ &    -1.0003276612 &    -0.0046764359 &     0.5240291126 &     0.2419442235 &    97.2283803910\\
I.A$^{i.c.}_{150}$ &    -1.0001365993 &    -0.0001024192 &     0.2485307852 &     0.2285701131 &    62.8272093696\\
I.A$^{i.c.}_{151}$ &    -1.0011932354 &     0.0026579310 &     0.4725635439 &     0.4144381911 &   127.6385623751\\
I.A$^{i.c.}_{152}$ &    -0.9997888700 &    -0.0004156209 &     0.4243404003 &     0.4908383162 &   136.0334473922\\
I.A$^{i.c.}_{153}$ &    -0.9999970864 &    -0.0000698084 &     0.3302687175 &     0.4705097113 &   107.3477264984\\
I.A$^{i.c.}_{154}$ &    -1.0028440006 &     0.0028315528 &     0.2550610016 &     0.5584274943 &   121.7041905837\\
I.A$^{i.c.}_{155}$ &    -0.9988577716 &    -0.0044220900 &     0.4810718014 &     0.3762778922 &   131.4859876501\\
I.A$^{i.c.}_{156}$ &    -1.0000601785 &     0.0000990093 &     0.4299823231 &     0.4772123993 &   179.4479904302\\
I.A$^{i.c.}_{157}$ &    -1.0002768983 &    -0.0007520763 &     0.2237420794 &     0.5717534898 &   178.2998359050\\
I.A$^{i.c.}_{158}$ &    -1.0000217334 &     0.0002546750 &     0.4722610930 &     0.3824580968 &   189.2780117600\\
I.A$^{i.c.}_{159}$ &    -0.9999032893 &     0.0001429653 &     0.2339885337 &     0.5689497825 &   190.5737853201\\
I.A$^{i.c.}_{160}$ &    -1.0000095770 &    -0.0000508014 &     0.4067473890 &     0.2352410471 &   131.6929897452\\
I.A$^{i.c.}_{161}$ &    -1.0000495745 &     0.0006453791 &     0.4665421446 &     0.2180480782 &   147.4522493752\\
I.A$^{i.c.}_{162}$ &    -1.0000426017 &     0.0000187878 &     0.1082898505 &     0.5839486776 &   185.8855555382\\
I.A$^{i.c.}_{163}$ &    -0.9995887984 &    -0.0079664292 &     0.3731155324 &     0.0401978873 &   116.4087777408\\
I.A$^{i.c.}_{164}$ &    -1.0004267819 &     0.0013647477 &     0.3291910007 &     0.2063233102 &   119.0599265875\\
I.A$^{i.c.}_{165}$ &    -0.9999878406 &    -0.0001761561 &     0.4000265728 &     0.1917264238 &   132.0429946006\\
I.A$^{i.c.}_{166}$ &    -0.9999714963 &    -0.0001453364 &     0.4122528073 &     0.2657939808 &   158.9999654145\\
I.A$^{i.c.}_{167}$ &    -1.0003148302 &     0.0017415967 &     0.3903578537 &     0.3737245354 &   189.4893206556\\
I.A$^{i.c.}_{168}$ &    -0.9999199514 &    -0.0004721503 &     0.3217887881 &     0.1561334749 &   132.6859977155\\
I.A$^{i.c.}_{169}$ &    -1.0000125952 &     0.0000463745 &     0.2959944852 &     0.2812361270 &   143.8040011025\\
I.A$^{i.c.}_{170}$ &    -0.9996218265 &    -0.0005091139 &     0.4131581036 &     0.2700715404 &   172.6949997903\\
I.A$^{i.c.}_{171}$ &    -1.0001981919 &     0.0008246015 &     0.4011840317 &     0.1984649469 &   159.8269797325\\
I.A$^{i.c.}_{172}$ &    -1.0001394240 &     0.0006210875 &     0.3984550628 &     0.1785263235 &   159.9940944136\\
I.A$^{i.c.}_{173}$ &    -0.9997525003 &    -0.0007101984 &     0.4132836168 &     0.2857891233 &   185.0489803002\\
I.A$^{i.c.}_{174}$ &    -0.9999874839 &    -0.0000755450 &     0.4920036102 &     0.1412749052 &   190.8516932081\\
I.A$^{i.c.}_{175}$ &    -0.9999114242 &    -0.0004703121 &     0.3957809554 &     0.1566414374 &   160.1480611499\\
I.A$^{i.c.}_{176}$ &    -0.9998591137 &    -0.0004200685 &     0.2815649267 &     0.2753881482 &   153.8738823880\\
I.A$^{i.c.}_{177}$ &    -0.9997769345 &    -0.0003218791 &     0.4155027855 &     0.2962515130 &   197.4047907632\\
I.A$^{i.c.}_{178}$ &    -1.0000078280 &     0.0000400743 &     0.3124969319 &     0.1567416978 &   143.9081477719\\
I.A$^{i.c.}_{179}$ &    -1.0001381860 &     0.0007881308 &     0.3937112266 &     0.1302084767 &   160.4815990874\\
I.A$^{i.c.}_{180}$ &    -0.9997373818 &    -0.0011997947 &     0.3975886957 &     0.1741695765 &   173.8524950814\\
I.A$^{i.c.}_{181}$ &    -1.0000004214 &    -0.0000432343 &     0.4065005549 &     0.2337501024 &   187.1600010053\\
I.A$^{i.c.}_{182}$ &    -1.0000276359 &     0.0001358270 &     0.3934947638 &     0.1281763613 &   174.4372376545\\
I.A$^{i.c.}_{183}$ &    -1.0000112548 &     0.0001830277 &     0.4514898921 &     0.0972823987 &   192.4740379970\\
I.A$^{i.c.}_{184}$ &    -0.9997876888 &    -0.0009321740 &     0.3993218264 &     0.1875709775 &   187.6312302736\\
I.A$^{i.c.}_{185}$ &    -1.0000675533 &     0.0006135606 &     0.3467279859 &     0.0823886189 &   164.8024707359\\
I.A$^{i.c.}_{186}$ &    -0.9998271876 &    -0.0008393849 &     0.3950456786 &     0.1495711374 &   188.0610439390\\
I.A$^{i.c.}_{187}$ &    -1.0000225782 &     0.0000462496 &     0.2704952147 &     0.2617245850 &   176.2009667320\\
I.A$^{i.c.}_{188}$ &    -1.0002443833 &     0.0007374491 &     0.2703855732 &     0.2975565398 &   183.6929612378\\
I.A$^{i.c.}_{189}$ &    -0.9998887181 &    -0.0017030496 &     0.2715532933 &     0.2231920595 &   169.7774994588\\
I.A$^{i.c.}_{190}$ &    -0.9998690037 &    -0.0007537235 &     0.3932849400 &     0.1265479325 &   188.3752620421\\
\hline
\end{tabular*}
{\rule{\temptablewidth}{1pt}}
\end{center}
\end{table*}

\begin{table*}
\tabcolsep 0pt \caption{Initial conditions and periods $T_0$ of the periodic three-body orbits for class I.B and II.B in case of the isosceles collinear configurations:  $\bm{r}_1(0)=(x_1,x_2)=-\bm{r}_2(0)$,  $\dot{\bm{r}}_1(0)=(v_1,v_2)=\dot{\bm{r}}_2(0)$ and $\bm{r}_3(0)=(0,0)$, $\dot{\bm{r}}_3(0)=(-2v_1, -2v_2)$ when $G=1$ and $m_1=m_2=m_3=1$ by means of the search grid $1000 \times 1000$  in the interval $T_0 \in [0,100]$.  Here, the superscript {\em i.c.} indicates the case of the initial conditions with {\em isosceles collinear}  configuration, due to the fact that there exist periodic orbits in many other cases. } \label{table-S3} \vspace*{-12pt}
\begin{center}
\def\temptablewidth{1\textwidth}
{\rule{\temptablewidth}{1pt}}
\begin{tabular*}{\temptablewidth}{@{\extracolsep{\fill}}lccccc}
\hline
Class and number & $x_1$ & $x_2$ & $v_1$ & $v_2$  & $T_0$\\
\hline
I.B$^{i.c.}_{1}$ &    -0.9989071137 &    -0.0001484864 &     0.4646402601 &     0.3963456869 &    14.8698954200\\
I.B$^{i.c.}_{2}$ &    -1.1770534081 &    -0.5225957568 &     0.2446132140 &     0.3305126876 &    20.2667904949\\
I.B$^{i.c.}_{3}$ &    -0.9996174046 &     0.0028671814 &     0.0836823874 &     0.1276739661 &    10.4589089289\\
I.B$^{i.c.}_{4}$ &    -0.9859387540 &    -0.0288038725 &     0.0637911125 &     0.5950102821 &    20.8385742723\\
I.B$^{i.c.}_{5}$ &    -0.9997857309 &    -0.0003533584 &     0.4390528231 &     0.4531713698 &    28.6600597106\\
I.B$^{i.c.}_{6}$ &    -1.0043366457 &     0.0085104316 &     0.3857847594 &     0.3732858410 &    26.0089024096\\
I.B$^{i.c.}_{7}$ &    -0.9995693352 &    -0.0012868539 &     0.1855324998 &     0.5790776181 &    33.6197306295\\
I.B$^{i.c.}_{8}$ &    -1.0012380956 &     0.0018757155 &     0.4297132024 &     0.4742147154 &    42.9096450556\\
I.B$^{i.c.}_{9}$ &    -0.9869994424 &    -0.0329384519 &     0.4080709664 &     0.2901166521 &    27.1495760745\\
I.B$^{i.c.}_{10}$ &    -0.9999193334 &    -0.0002717176 &     0.3980126012 &     0.1762535971 &    27.8200493126\\
I.B$^{i.c.}_{11}$ &    -1.0010436520 &     0.0029859847 &     0.2318334372 &     0.5686672120 &    46.1393431256\\
I.B$^{i.c.}_{12}$ &    -0.9912922078 &    -0.0169276058 &     0.4238823678 &     0.3638554791 &    41.8211429240\\
I.B$^{i.c.}_{13}$ &    -0.9955318366 &    -0.0057317839 &     0.3560089302 &     0.5008649399 &    49.3004616047\\
I.B$^{i.c.}_{14}$ &    -1.0000328444 &     0.0003978205 &     0.4098850694 &     0.4390517046 &    54.0599247160\\
I.B$^{i.c.}_{15}$ &    -0.9820178956 &    -0.0439094593 &     0.4045186605 &     0.3089615770 &    40.0817543149\\
I.B$^{i.c.}_{16}$ &    -0.9793316018 &    -0.0343328490 &     0.2415686630 &     0.5788831000 &    57.4029776575\\
I.B$^{i.c.}_{17}$ &    -0.9852740589 &    -0.0640855916 &     0.3872339266 &     0.1809512432 &    40.9983837590\\
I.B$^{i.c.}_{18}$ &    -0.9954789776 &    -0.0090056972 &     0.4069832833 &     0.3543167976 &    53.9614235401\\
I.B$^{i.c.}_{19}$ &    -1.0003135452 &     0.0005407130 &     0.2778712767 &     0.5598575722 &    72.3315394965\\
I.B$^{i.c.}_{20}$ &    -1.0080911586 &     0.0212978481 &     0.4217594107 &     0.2832267055 &    55.5464441478\\
I.B$^{i.c.}_{21}$ &    -0.9999536614 &    -0.0001750829 &     0.4109939999 &     0.4030802063 &    71.3800291841\\
I.B$^{i.c.}_{22}$ &    -0.9686776407 &    -0.0608196023 &     0.3991611243 &     0.2827980458 &    52.9430984041\\
I.B$^{i.c.}_{23}$ &    -1.0025047383 &     0.0016376144 &     0.1294943334 &     0.5833026191 &    75.8581653606\\
I.B$^{i.c.}_{24}$ &    -1.0092396488 &     0.0396030643 &     0.4075995186 &     0.1878601935 &    56.3933245589\\
I.B$^{i.c.}_{25}$ &    -0.9991779003 &    -0.0023293910 &     0.4524794248 &     0.4173526023 &    87.5501725839\\
I.B$^{i.c.}_{26}$ &    -1.0120493442 &     0.0327654798 &     0.4224042986 &     0.2810350321 &    69.3540290982\\
I.B$^{i.c.}_{27}$ &    -0.9958261057 &    -0.0180235617 &     0.4746698475 &     0.3848074753 &    92.0041366553\\
I.B$^{i.c.}_{28}$ &    -1.0012573546 &     0.0022897677 &     0.4290649202 &     0.4627444811 &    98.8359617418\\
I.B$^{i.c.}_{29}$ &    -1.0016467940 &     0.0007588511 &     0.1554933398 &     0.5786962880 &    87.0908236211\\
I.B$^{i.c.}_{30}$ &    -0.9910057079 &    -0.0251737916 &     0.3331962796 &     0.2861128695 &    60.4345154079\\
I.B$^{i.c.}_{31}$ &    -1.0091765176 &     0.0201754445 &     0.4157410078 &     0.2569669976 &    70.1042225630\\
I.B$^{i.c.}_{32}$ &    -0.9998130400 &    -0.0053468765 &     0.3750465942 &     0.0261694007 &    54.9599616485\\
I.B$^{i.c.}_{33}$ &    -0.9961710033 &    -0.0074269978 &     0.4107912477 &     0.3486701313 &    82.1507596077\\
I.B$^{i.c.}_{34}$ &    -0.9962761072 &    -0.0172343437 &     0.4737986398 &     0.1688124971 &    79.5685727648\\
I.B$^{i.c.}_{35}$ &    -0.9919037360 &    -0.0188701087 &     0.4100106195 &     0.3071172647 &    80.1888154296\\
I.B$^{i.c.}_{36}$ &    -0.9990532732 &    -0.0043820662 &     0.3204762525 &     0.2061939336 &    62.0002762587\\
I.B$^{i.c.}_{37}$ &    -1.0025451082 &     0.0013669961 &     0.4039796728 &     0.3652787606 &    94.1439172549\\
I.B$^{i.c.}_{38}$ &    -1.0002032247 &     0.0006453603 &     0.3939791513 &     0.1344605101 &    69.7600199827\\
I.B$^{i.c.}_{39}$ &    -0.9997462078 &    -0.0011028163 &     0.2901713884 &     0.2612727959 &    69.7400917634\\
I.B$^{i.c.}_{40}$ &    -0.9902765570 &    -0.0307512535 &     0.4894639489 &     0.2365260776 &    97.2705134964\\
I.B$^{i.c.}_{41}$ &    -1.0000822875 &     0.0003330901 &     0.3650078095 &     0.0603887946 &    67.6590068332\\
I.B$^{i.c.}_{42}$ &    -1.0020843723 &     0.0171698170 &     0.4206012554 &     0.2909314421 &    95.0451372238\\
I.B$^{i.c.}_{43}$ &    -0.9973221994 &    -0.0091941904 &     0.4017883631 &     0.2174805839 &    82.9624661836\\
I.B$^{i.c.}_{44}$ &    -1.0018919789 &     0.0061735438 &     0.3010143797 &     0.2902774095 &    80.5992058621\\
I.B$^{i.c.}_{45}$ &    -0.9894653911 &    -0.0608004298 &     0.4883359248 &     0.1669356426 &    99.0978825987\\
I.B$^{i.c.}_{46}$ &    -1.0000325339 &     0.0003221382 &     0.3029750443 &     0.1969660658 &    73.2099935849\\
I.B$^{i.c.}_{47}$ &    -0.9982822070 &    -0.0052079203 &     0.4095156641 &     0.2589467566 &    96.6187276779\\
I.B$^{i.c.}_{48}$ &    -1.0003976445 &     0.0030603406 &     0.3938702763 &     0.1281065458 &    83.7700535678\\
I.B$^{i.c.}_{49}$ &    -1.0017029160 &     0.0041736578 &     0.3501405955 &     0.0778141747 &    79.6790864881\\
I.B$^{i.c.}_{50}$ &    -0.9996145603 &    -0.0017302155 &     0.4010821936 &     0.2015884635 &    97.1896385782\\
I.B$^{i.c.}_{51}$ &    -1.0002894289 &     0.0009291890 &     0.2668487750 &     0.2777035009 &    89.3197319785\\
I.B$^{i.c.}_{52}$ &    -0.9992941098 &    -0.0024372579 &     0.2931395758 &     0.1922320514 &    84.7295120544\\
I.B$^{i.c.}_{53}$ &    -1.0012336598 &     0.0060424877 &     0.3977399716 &     0.1646648075 &    97.6151914585\\
I.B$^{i.c.}_{54}$ &    -0.9986567557 &    -0.0089510385 &     0.3924139983 &     0.1289496286 &    97.5107476946\\
I.B$^{i.c.}_{55}$ &    -0.9999453732 &    -0.0008999222 &     0.4030162257 &     0.0473267119 &    99.9599973732\\
II.B$^{i.c.}_{1}$ &    -0.9795529757 &    -0.0288228080 &     0.3849595037 &     0.5254027227 &    93.5540138135\\
II.B$^{i.c.}_{2}$ &    -1.0007518096 &     0.0076238027 &     0.2537011397 &     0.2647081365 &    99.9335612373\\
\hline
\end{tabular*}
{\rule{\temptablewidth}{1pt}}
\end{center}
\end{table*}

\begin{table*}
\tabcolsep 0pt \caption{Initial conditions and periods $T_0$  of the additional periodic three-body orbits for class I.B in case of the isosceles collinear configurations:  $\bm{r}_1(0)=(x_1,x_2)=-\bm{r}_2(0)$,  $\dot{\bm{r}}_1(0)=(v_1,v_2)=\dot{\bm{r}}_2(0)$ and $\bm{r}_3(0)=(0,0)$, $\dot{\bm{r}}_3(0)=(-2v_1, -2v_2)$ when $G=1$ and $m_1=m_2=m_3=1$ by means of the search grid $2000 \times 2000$ in the interval $T_0\in[0,200]$. Here, the superscript {\em i.c.} indicates the case of the initial conditions with {\em isosceles collinear}  configuration, due to the fact that there exist periodic orbits in many other cases.} \label{table-S3} \vspace*{-12pt}
\begin{center}
\def\temptablewidth{1\textwidth}
{\rule{\temptablewidth}{1pt}}
\begin{tabular*}{\temptablewidth}{@{\extracolsep{\fill}}lccccc}
\hline
Class and number & $x_1$ & $x_2$ & $v_1$ & $v_2$  & $T_0$\\
\hline
I.B$^{i.c.}_{56}$ &    -1.0002539139 &     0.0003964491 &     0.4014357492 &     0.4983198944 &    68.5797898945\\
I.B$^{i.c.}_{57}$ &    -1.0002918598 &     0.0005162892 &     0.2843267154 &     0.5553811591 &    84.0188438872\\
I.B$^{i.c.}_{58}$ &    -1.0037288781 &     0.0072403740 &     0.4408153221 &     0.3413502480 &    73.3719699527\\
I.B$^{i.c.}_{59}$ &    -0.9999874974 &    -0.0001760590 &     0.3945156168 &     0.1425073802 &    55.7600087000\\
I.B$^{i.c.}_{60}$ &    -1.0032127030 &     0.0084987577 &     0.4828415456 &     0.2757669517 &    93.4730106172\\
I.B$^{i.c.}_{61}$ &    -0.9965295426 &    -0.0075988006 &     0.4543611179 &     0.4152838622 &   117.0474628000\\
I.B$^{i.c.}_{62}$ &    -0.9992398026 &    -0.0009095812 &     0.1994100058 &     0.5755815749 &   112.4722550040\\
I.B$^{i.c.}_{63}$ &    -1.0006360462 &     0.0020005131 &     0.4737305335 &     0.3806842909 &   121.4880405179\\
I.B$^{i.c.}_{64}$ &    -1.0037383087 &     0.0043289996 &     0.4335673948 &     0.4441418635 &   126.6737879541\\
I.B$^{i.c.}_{65}$ &    -1.0009006707 &     0.0012634524 &     0.4332923347 &     0.4279910325 &   126.8142978100\\
I.B$^{i.c.}_{66}$ &    -1.0026993767 &     0.0035227297 &     0.1125527722 &     0.5831916044 &   117.6765954500\\
I.B$^{i.c.}_{67}$ &    -0.9999425775 &     0.0000768146 &     0.2160601821 &     0.5721576171 &   125.1988535687\\
I.B$^{i.c.}_{68}$ &    -0.9987881484 &    -0.0023279547 &     0.4133711321 &     0.3611869593 &   109.2132506445\\
I.B$^{i.c.}_{69}$ &    -0.9892675968 &    -0.0098535078 &     0.1295677145 &     0.5870031876 &   127.3291233343\\
I.B$^{i.c.}_{70}$ &    -0.9981021442 &    -0.0045601502 &     0.4145887363 &     0.3010225376 &   107.5845185672\\
I.B$^{i.c.}_{71}$ &    -0.9969180310 &    -0.0064532805 &     0.4582538461 &     0.4110300518 &   147.3494679161\\
I.B$^{i.c.}_{72}$ &    -0.9984717487 &    -0.0018770837 &     0.4352804810 &     0.4603165735 &   156.8591408369\\
I.B$^{i.c.}_{73}$ &    -0.9996886255 &    -0.0007076083 &     0.4131464970 &     0.2854659107 &   109.6432001517\\
I.B$^{i.c.}_{74}$ &    -0.9981956906 &     0.0003457765 &     0.1525755985 &     0.5814443076 &   141.2508479790\\
I.B$^{i.c.}_{75}$ &    -0.9998451703 &    -0.0002385752 &     0.0425440524 &     0.5900891856 &   136.8348605033\\
I.B$^{i.c.}_{76}$ &    -0.9984446498 &    -0.0031579058 &     0.4114797284 &     0.2658506221 &   110.3688900131\\
I.B$^{i.c.}_{77}$ &    -0.9969511140 &    -0.0087182083 &     0.5104917926 &     0.2415013359 &   136.1812489117\\
I.B$^{i.c.}_{78}$ &    -0.9998843707 &    -0.0002206419 &     0.4571327947 &     0.4151729660 &   161.5210185089\\
I.B$^{i.c.}_{79}$ &    -1.0002381265 &     0.0022777498 &     0.4084247701 &     0.2410714966 &   110.8814543365\\
I.B$^{i.c.}_{80}$ &    -0.9988064593 &    -0.0038394841 &     0.4152528993 &     0.3006714463 &   121.0343128196\\
I.B$^{i.c.}_{81}$ &    -0.9979999075 &    -0.0026638779 &     0.1760284506 &     0.5832452411 &   156.9142601013\\
I.B$^{i.c.}_{82}$ &    -0.9996882260 &    -0.0009580399 &     0.5140634278 &     0.2574240103 &   152.2149896883\\
I.B$^{i.c.}_{83}$ &    -0.9996432709 &    -0.0014415475 &     0.4036670877 &     0.2183774966 &   110.9670504099\\
I.B$^{i.c.}_{84}$ &    -1.0101674750 &     0.0129295943 &     0.4479931056 &     0.4407237171 &   191.7887663629\\
I.B$^{i.c.}_{85}$ &    -0.9990037112 &    -0.0042397434 &     0.3993390700 &     0.1926968678 &   111.0361910180\\
I.B$^{i.c.}_{86}$ &    -1.0000735237 &     0.0007222188 &     0.4134917820 &     0.2698682101 &   124.4239062885\\
I.B$^{i.c.}_{87}$ &    -1.0006610026 &    -0.0050723417 &     0.1036264092 &     0.5951708348 &   165.8944687717\\
I.B$^{i.c.}_{88}$ &    -1.0002537675 &     0.0005826233 &     0.4159412331 &     0.2987512669 &   134.4420097789\\
I.B$^{i.c.}_{89}$ &    -0.9996395834 &    -0.0120707186 &     0.3941262105 &     0.3746706579 &   145.8399204128\\
I.B$^{i.c.}_{90}$ &    -0.9996283785 &    -0.0014023641 &     0.4092077154 &     0.2515480983 &   124.5360012697\\
I.B$^{i.c.}_{91}$ &    -1.0003175663 &     0.0005523869 &     0.5128479903 &     0.2177071172 &   154.3113210889\\
I.B$^{i.c.}_{92}$ &    -1.0002779857 &     0.0015887018 &     0.3964162716 &     0.1594794112 &   111.4439694026\\
I.B$^{i.c.}_{93}$ &    -1.0012504928 &     0.0043191425 &     0.4159649266 &     0.2884285730 &   137.2413742278\\
I.B$^{i.c.}_{94}$ &    -0.9986892732 &    -0.0016039817 &     0.4536988129 &     0.4157176741 &   189.8592960198\\
I.B$^{i.c.}_{95}$ &    -1.0056439166 &     0.0107509046 &     0.4061605156 &     0.3464068422 &   148.6793726851\\
I.B$^{i.c.}_{96}$ &    -0.9999658166 &    -0.0007955780 &     0.3930104491 &     0.1225855601 &   111.6774645026\\
I.B$^{i.c.}_{97}$ &    -1.0000883608 &    -0.0003492696 &     0.1227234223 &     0.5846410415 &   172.2085677576\\
I.B$^{i.c.}_{98}$ &    -0.9999040745 &    -0.0008781531 &     0.2582208699 &     0.2321796298 &   103.4367480061\\
I.B$^{i.c.}_{99}$ &    -1.0000556499 &     0.0006225664 &     0.4024766549 &     0.2074180795 &   124.9910283095\\
I.B$^{i.c.}_{100}$ &    -1.0000494838 &     0.0011719082 &     0.3989609262 &     0.0582842312 &   113.4776603332\\
I.B$^{i.c.}_{101}$ &    -1.0010764811 &     0.0018462823 &     0.4371716048 &     0.3596386575 &   172.6034826383\\
I.B$^{i.c.}_{102}$ &    -0.9986909282 &    -0.0029160875 &     0.4776389982 &     0.3788321639 &   199.1244476752\\
I.B$^{i.c.}_{103}$ &    -0.9984608233 &    -0.0025762661 &     0.4057869552 &     0.3661449525 &   161.1802577307\\
I.B$^{i.c.}_{104}$ &    -0.9962750934 &    -0.0022501208 &     0.1384550273 &     0.5846750226 &   183.5493071713\\
I.B$^{i.c.}_{105}$ &    -1.0001068187 &    -0.0001568929 &     0.0531044986 &     0.5910224471 &   180.2407776594\\
I.B$^{i.c.}_{106}$ &    -0.9992310702 &    -0.0013696524 &     0.4153061286 &     0.2928500474 &   150.4051707788\\
I.B$^{i.c.}_{107}$ &    -1.0002357061 &    -0.0004243590 &     0.4073724820 &     0.2398411068 &   138.6255364703\\
I.B$^{i.c.}_{108}$ &    -1.0000203169 &    -0.0003399367 &     0.3450414864 &     0.3809660005 &   155.6892635177\\
I.B$^{i.c.}_{109}$ &    -0.9954951275 &    -0.0135387866 &     0.4021656036 &     0.2262624478 &   137.8434569870\\
I.B$^{i.c.}_{110}$ &    -1.0001758522 &     0.0004980640 &     0.4219647889 &     0.2767867125 &   153.8849104674\\
\hline
\end{tabular*}
{\rule{\temptablewidth}{1pt}}
\end{center}
\end{table*}

\begin{table*}
\tabcolsep 0pt \caption{Initial conditions and periods $T_0$  of the additional periodic three-body orbits for class I.B , II.A and II.B in case of the isosceles collinear configurations:  $\bm{r}_1(0)=(x_1,x_2)=-\bm{r}_2(0)$,  $\dot{\bm{r}}_1(0)=(v_1,v_2)=\dot{\bm{r}}_2(0)$ and $\bm{r}_3(0)=(0,0)$, $\dot{\bm{r}}_3(0)=(-2v_1, -2v_2)$ when $G=1$ and $m_1=m_2=m_3=1$ by means of the search grid $2000\times 2000$ in the interval $T_0 \in [0,200]$.  Here, the superscript {\em i.c.} indicates the case of the initial conditions with {\em isosceles collinear}  configuration, due to the fact that there exist periodic orbits in many other cases.} \label{table-S3} \vspace*{-12pt}
\begin{center}
\def\temptablewidth{1\textwidth}
{\rule{\temptablewidth}{1pt}}
\begin{tabular*}{\temptablewidth}{@{\extracolsep{\fill}}lccccc}
\hline
Class and number & $x_1$ & $x_2$ & $v_1$ & $v_2$  & $T_0$\\
\hline
I.B$^{i.c.}_{111}$ &    -0.9995598709 &    -0.0028033650 &     0.3926777610 &     0.1209618602 &   125.5785779369\\
I.B$^{i.c.}_{112}$ &    -0.9983422660 &    -0.0010365811 &     0.1497698858 &     0.5826159253 &   195.9692738094\\
I.B$^{i.c.}_{113}$ &    -0.9997884893 &    -0.0007804195 &     0.2561986794 &     0.2263614210 &   115.2728148741\\
I.B$^{i.c.}_{114}$ &    -0.9982430890 &    -0.0029148571 &     0.4115636525 &     0.3625777811 &   176.6720112959\\
I.B$^{i.c.}_{115}$ &    -0.9987559943 &    -0.0068492477 &     0.3999965410 &     0.2021625106 &   138.6805240527\\
I.B$^{i.c.}_{116}$ &    -0.9993456123 &    -0.0070361658 &     0.3966658413 &     0.0681214863 &   127.1169091459\\
I.B$^{i.c.}_{117}$ &    -1.0067196917 &     0.0134165934 &     0.4155572680 &     0.3417145025 &   179.7861955530\\
I.B$^{i.c.}_{118}$ &    -0.9999710006 &     0.0039277510 &     0.0171159353 &     0.5682085458 &   184.4730981050\\
I.B$^{i.c.}_{119}$ &    -1.0010234067 &     0.0030037797 &     0.2614305076 &     0.2902326525 &   130.9660200253\\
I.B$^{i.c.}_{120}$ &    -1.0005347699 &     0.0009174238 &     0.4180680912 &     0.3575603864 &   193.2835707650\\
I.B$^{i.c.}_{121}$ &    -1.0024964299 &     0.0082415100 &     0.4144162905 &     0.2641322087 &   166.5252104245\\
I.B$^{i.c.}_{122}$ &    -0.9978032583 &    -0.0083368986 &     0.4016517199 &     0.2155070920 &   152.2205826637\\
I.B$^{i.c.}_{123}$ &    -0.9997548266 &    -0.0017332726 &     0.3925866373 &     0.1185643487 &   139.5767121863\\
I.B$^{i.c.}_{124}$ &    -0.9997295496 &    -0.0007394978 &     0.4145740205 &     0.2952835324 &   177.0641287459\\
I.B$^{i.c.}_{125}$ &    -0.9997795828 &    -0.0021846744 &     0.3191205294 &     0.1025588432 &   126.8408161369\\
I.B$^{i.c.}_{126}$ &    -1.0009687762 &     0.0038899785 &     0.4006806870 &     0.1905482510 &   153.1139415466\\
I.B$^{i.c.}_{127}$ &    -1.0001893896 &     0.0002438133 &     0.4129456859 &     0.2843192169 &   178.2683162111\\
I.B$^{i.c.}_{128}$ &    -1.0001115924 &     0.0005390181 &     0.3974376304 &     0.1702560678 &   153.1039871874\\
I.B$^{i.c.}_{129}$ &    -1.0031325766 &     0.0104011922 &     0.4090503903 &     0.2335230095 &   167.1068766699\\
I.B$^{i.c.}_{130}$ &    -0.9999340750 &    -0.0003060210 &     0.4045228909 &     0.2220880800 &   166.4790085882\\
I.B$^{i.c.}_{131}$ &    -1.0007644007 &     0.0004448313 &     0.3947420686 &     0.1455774004 &   153.4843582464\\
I.B$^{i.c.}_{132}$ &    -1.0001651737 &     0.0004613373 &     0.4109547717 &     0.2583088533 &   179.9070021508\\
I.B$^{i.c.}_{133}$ &    -0.9995280721 &    -0.0033062283 &     0.3921601515 &     0.1176190696 &   153.4547997052\\
I.B$^{i.c.}_{134}$ &    -1.0031415863 &     0.0100685632 &     0.4101688706 &     0.2402047198 &   180.9425230631\\
I.B$^{i.c.}_{135}$ &    -1.0002373201 &     0.0025255333 &     0.3954276198 &     0.0719975702 &   155.0769827467\\
I.B$^{i.c.}_{136}$ &    -0.9999249628 &    -0.0006747417 &     0.4115303906 &     0.2630998503 &   193.5999063349\\
I.B$^{i.c.}_{137}$ &    -0.9969227534 &    -0.0108057586 &     0.3422305650 &     0.2645114186 &   173.7636973825\\
I.B$^{i.c.}_{138}$ &    -1.0011330103 &     0.0045965355 &     0.4016847912 &     0.1964917110 &   180.9339094365\\
I.B$^{i.c.}_{139}$ &    -0.9988551405 &    -0.0045334389 &     0.3485759191 &     0.2115653847 &   166.6679442717\\
I.B$^{i.c.}_{140}$ &    -0.9994446288 &    -0.0039890503 &     0.3911948593 &     0.1168384072 &   167.1873920200\\
I.B$^{i.c.}_{141}$ &    -0.9993344638 &    -0.0029685873 &     0.3982182390 &     0.1820423914 &   180.6251574896\\
I.B$^{i.c.}_{142}$ &    -1.0037533818 &     0.0215897780 &     0.4358996148 &     0.1380431384 &   193.9327624187\\
I.B$^{i.c.}_{143}$ &    -0.9996210593 &    -0.0018371887 &     0.3961315475 &     0.1621739986 &   180.9043129263\\
I.B$^{i.c.}_{144}$ &    -1.0000664657 &     0.0002009450 &     0.4025133509 &     0.2084253094 &   194.4239784344\\
I.B$^{i.c.}_{145}$ &    -0.9986248493 &    -0.0078908547 &     0.3934494324 &     0.1427290316 &   180.8941838391\\
I.B$^{i.c.}_{146}$ &    -1.0001158556 &     0.0002832367 &     0.3190722889 &     0.2559959937 &   181.2542458960\\
I.B$^{i.c.}_{147}$ &    -0.9998655062 &    -0.0005934375 &     0.3320022735 &     0.2077235131 &   175.9445389716\\
I.B$^{i.c.}_{148}$ &    -0.9997456898 &    -0.0010536696 &     0.4000806760 &     0.1933785481 &   194.5050977993\\
I.B$^{i.c.}_{149}$ &    -0.9998856403 &    -0.0006450623 &     0.3940346555 &     0.1372366370 &   195.2038005603\\
I.B$^{i.c.}_{150}$ &    -0.9976490894 &    -0.0069161011 &     0.3119633845 &     0.2848961067 &   198.9997040841\\
I.B$^{i.c.}_{151}$ &    -1.0000232451 &     0.0009402990 &     0.3030922643 &     0.2510799924 &   190.0209371227\\
I.B$^{i.c.}_{152}$ &    -1.0001279531 &     0.0008944424 &     0.3449415632 &     0.1307302825 &   182.8556649490\\
I.B$^{i.c.}_{153}$ &    -1.0011363466 &     0.0023444109 &     0.3021469646 &     0.2644600497 &   199.8917448561\\
I.B$^{i.c.}_{154}$ &    -0.9997838456 &    -0.0010195948 &     0.2996100208 &     0.2297555256 &   191.7219761952\\
I.B$^{i.c.}_{155}$ &    -1.0000308708 &     0.0002452872 &     0.3310816665 &     0.1350483602 &   192.5848904710\\
II.A$^{i.c.}_{1}$ &    -0.9947533301 &    -0.0073626081 &     0.4213901379 &     0.4934387939 &   120.7898940623\\
II.A$^{i.c.}_{2}$ &    -1.0000152072 &     0.0000173254 &     0.4210984475 &     0.4930872974 &   135.8658689640\\
II.A$^{i.c.}_{3}$ &    -0.9951268410 &    -0.0129271249 &     0.3170127745 &     0.3076871295 &   144.2889766691\\
II.B$^{i.c.}_{3}$ &    -0.9993790431 &    -0.0082364431 &     0.5083086269 &     0.1185218584 &   120.7766641833\\
II.B$^{i.c.}_{4}$ &    -0.9988597232 &    -0.0012136651 &     0.2064776216 &     0.5777876235 &   182.9481851424\\
II.B$^{i.c.}_{5}$ &    -0.9999620559 &    -0.0002319517 &     0.2465359292 &     0.2580416156 &   111.2560502029\\
\hline
\end{tabular*}
{\rule{\temptablewidth}{1pt}}
\end{center}
\end{table*}

\begin{table*}
\tabcolsep 0pt \caption{Initial conditions and periods $T_0$  of the additional periodic three-body orbits for class I.B , II.A and II.B in case of the isosceles collinear configurations:  $\bm{r}_1(0)=(x_1,x_2)=-\bm{r}_2(0)$,  $\dot{\bm{r}}_1(0)=(v_1,v_2)=\dot{\bm{r}}_2(0)$ and $\bm{r}_3(0)=(0,0)$, $\dot{\bm{r}}_3(0)=(-2v_1, -2v_2)$ when $G=1$ and $m_1=m_2=m_3=1$ by means of the search grid $4000 \times 4000$ in the interval $T_0\in[0,200]$. Here, the superscript {\em i.c.} indicates the case of the initial conditions with {\em isosceles collinear}  configuration, due to the fact that there exist periodic orbits in many other cases.} \label{table-S3} \vspace*{-12pt}
\begin{center}
\def\temptablewidth{1\textwidth}
{\rule{\temptablewidth}{1pt}}
\begin{tabular*}{\temptablewidth}{@{\extracolsep{\fill}}lccccc}
\hline
Class and number & $x_1$ & $x_2$ & $v_1$ & $v_2$  & $T_0$\\
\hline
I.B$^{i.c.}_{156}$ &    -0.9999877856 &    -0.0005814781 &     0.3977486296 &     0.5041248838 &    82.2478390612\\
I.B$^{i.c.}_{157}$ &    -0.9999831765 &    -0.0000995165 &     0.2952335680 &     0.5534931232 &    97.3475249982\\
I.B$^{i.c.}_{158}$ &    -1.0020247786 &     0.0045986505 &     0.4812311776 &     0.4086868668 &   106.8815187510\\
I.B$^{i.c.}_{159}$ &    -0.9987986134 &    -0.0021552055 &     0.4132839850 &     0.3448189244 &   110.8560617934\\
I.B$^{i.c.}_{160}$ &    -1.0006864935 &     0.0017999067 &     0.3688795678 &     0.3990936658 &   117.0527690312\\
I.B$^{i.c.}_{161}$ &    -1.0001480668 &     0.0003414076 &     0.4647459174 &     0.3881587066 &   149.5193645034\\
I.B$^{i.c.}_{162}$ &    -1.0007082297 &     0.0006320112 &     0.4032649690 &     0.4613151212 &   156.1597649126\\
I.B$^{i.c.}_{163}$ &    -0.9999144921 &    -0.0002725652 &     0.4237854322 &     0.4720475669 &   170.6877509743\\
I.B$^{i.c.}_{164}$ &    -0.9994975635 &    -0.0006847574 &     0.4205764510 &     0.4491433227 &   176.0861023461\\
I.B$^{i.c.}_{165}$ &    -1.0006500507 &     0.0020683553 &     0.4561197148 &     0.4076973098 &   176.5980377684\\
I.B$^{i.c.}_{166}$ &    -1.0000695286 &     0.0015443855 &     0.5092287565 &     0.1057790450 &   138.5149518136\\
I.B$^{i.c.}_{167}$ &    -0.9999362907 &    -0.0006678556 &     0.3987762896 &     0.1831075465 &   125.1459851583\\
I.B$^{i.c.}_{168}$ &    -1.0002466667 &     0.0007834015 &     0.4108947045 &     0.2574916947 &   138.4155924232\\
I.B$^{i.c.}_{169}$ &    -1.0004087874 &     0.0021091502 &     0.5116991389 &     0.1510204034 &   172.2214478000\\
I.B$^{i.c.}_{170}$ &    -1.0002203056 &     0.0006405360 &     0.4089433249 &     0.2469756636 &   152.3917721970\\
I.B$^{i.c.}_{171}$ &    -1.0002095177 &     0.0001707503 &     0.3611700432 &     0.3726074018 &   171.3495314893\\
I.B$^{i.c.}_{172}$ &    -0.9998978817 &    -0.0003607301 &     0.2815198551 &     0.1834115855 &   120.7024583397\\
I.B$^{i.c.}_{173}$ &    -1.0001637291 &     0.0007797193 &     0.3952704790 &     0.1494364180 &   139.3700881635\\
I.B$^{i.c.}_{174}$ &    -0.9999059812 &    -0.0002102782 &     0.4157780204 &     0.3473362213 &   193.3229481677\\
I.B$^{i.c.}_{175}$ &    -0.9998860717 &    -0.0011127305 &     0.4134226210 &     0.2720511536 &   179.6762837838\\
I.B$^{i.c.}_{176}$ &    -0.9998070036 &    -0.0004408509 &     0.4153042995 &     0.2959416679 &   190.6770137480\\
I.B$^{i.c.}_{177}$ &    -1.0004885274 &     0.0020285549 &     0.3995901276 &     0.1852460454 &   166.9692629914\\
I.B$^{i.c.}_{178}$ &    -1.0023524869 &     0.0063300001 &     0.4151172651 &     0.2834533744 &   192.6591260323\\
I.B$^{i.c.}_{179}$ &    -1.0004655097 &     0.0022138318 &     0.3970869351 &     0.1647105126 &   167.1588555241\\
I.B$^{i.c.}_{180}$ &    -1.0003519850 &     0.0012223132 &     0.4036395425 &     0.2142764848 &   180.5459604732\\
I.B$^{i.c.}_{181}$ &    -0.9998883900 &    -0.0003760745 &     0.4092850740 &     0.2503793652 &   193.8098850814\\
I.B$^{i.c.}_{182}$ &    -1.0004489291 &     0.0019157114 &     0.4446265683 &     0.1840121567 &   195.1106076577\\
I.B$^{i.c.}_{183}$ &    -0.9996327583 &    -0.0038817861 &     0.3946193172 &     0.0770908142 &   168.8729666850\\
I.B$^{i.c.}_{184}$ &    -0.9987777148 &    -0.0060204525 &     0.3512447995 &     0.1758870428 &   169.2218794279\\
I.B$^{i.c.}_{185}$ &    -1.0007336752 &     0.0027681588 &     0.3294190402 &     0.2356122228 &   173.2836318829\\
I.B$^{i.c.}_{186}$ &    -1.0001128465 &     0.0008015762 &     0.3692044602 &     0.1227509788 &   175.4157416773\\
I.B$^{i.c.}_{187}$ &    -0.9999424536 &    -0.0003920874 &     0.3945062919 &     0.0777433860 &   182.9079846060\\
I.B$^{i.c.}_{188}$ &    -0.9994592107 &    -0.0028466487 &     0.3300070176 &     0.1762940372 &   177.4881409911\\
I.B$^{i.c.}_{189}$ &    -0.9993848724 &    -0.0058921104 &     0.3939316576 &     0.0816178085 &   196.7177605118\\
I.B$^{i.c.}_{190}$ &    -1.0000184961 &    -0.0000127634 &     0.3165064956 &     0.1757257867 &   187.3274416387\\
I.B$^{i.c.}_{191}$ &    -0.9998781064 &     0.0004126731 &     0.3135571349 &     0.2018201864 &   197.0646973632\\
II.A$^{i.c.}_{4}$ &    -0.9998795020 &     0.0000320054 &     0.0262269748 &     0.6907879303 &    46.5682885545\\
II.B$^{i.c.}_{6}$ &    -0.9997601327 &     0.0044033483 &     0.0726543370 &     0.6187791387 &    95.2555056523\\
II.B$^{i.c.}_{7}$ &    -1.0010623623 &     0.0011171795 &     0.3960131078 &     0.5090592509 &   110.0499076916\\
II.B$^{i.c.}_{8}$ &    -1.0001494432 &     0.0003104221 &     0.3066286068 &     0.5511835901 &   111.2182726866\\
II.B$^{i.c.}_{9}$ &    -1.0000083032 &     0.0001418212 &     0.5122310700 &     0.1632796412 &   155.0288369362\\
II.B$^{i.c.}_{10}$ &    -1.0000043765 &    -0.0001456734 &     0.5152419669 &     0.2342893192 &   189.7706914768\\

\hline
\end{tabular*}
{\rule{\temptablewidth}{1pt}}
\end{center}
\end{table*}

\begin{table*}
\tabcolsep 0pt \caption{Initial conditions and periods $T_0$ of the periodic  three-body orbits for class II.C in case of the isosceles collinear configurations:  $\bm{r}_1(0)=(x_1,x_2)=-\bm{r}_2(0)$,  $\dot{\bm{r}}_1(0)=(v_1,v_2)=\dot{\bm{r}}_2(0)$ and $\bm{r}_3(0)=(0,0)$, $\dot{\bm{r}}_3(0)=(-2v_1, -2v_2)$ when $G=1$ and $m_1=m_2=m_3=1$ by means of the search grid $1000 \times 1000$  in the interval $T_0 \in [0,100]$.  } \label{table-S4} \vspace*{-12pt}
\begin{center}
\def\temptablewidth{1\textwidth}
{\rule{\temptablewidth}{1pt}}
\begin{tabular*}{\temptablewidth}{@{\extracolsep{\fill}}lccccc}
\hline
Class and number & $x_1$ & $x_2$ & $v_1$ & $v_2$  & $T_0$\\
\hline
II.C$^{i.c.}_{1}$ &    -0.9826146484 &    -0.0411837391 &     0.2710001824 &     0.3415940623 &    10.6927139709\\
II.C$^{i.c.}_{2}$ &    -1.0016879974 &     0.0026309175 &     0.5656359883 &     0.5349017518 &    54.8891607912\\
II.C$^{i.c.}_{3}$ &    -1.0010192913 &     0.0028805303 &     0.5229031501 &     0.2251886482 &    26.3597995195\\
II.C$^{i.c.}_{4}$ &    -0.9979319887 &    -0.0029629112 &     0.2006675819 &     0.4109167331 &    20.9554821689\\
II.C$^{i.c.}_{5}$ &    -0.9961434009 &    -0.0050071036 &     0.5495563197 &     0.5543125804 &    86.4408761911\\
II.C$^{i.c.}_{6}$ &    -1.0007357274 &     0.0020118982 &     0.5238826063 &     0.3409361213 &    44.9799754767\\
II.C$^{i.c.}_{7}$ &    -1.0000747816 &     0.0003915745 &     0.4738937949 &     0.4310750400 &    46.8999197399\\
II.C$^{i.c.}_{8}$ &    -1.0001704251 &    -0.0022610973 &     0.2427965934 &     0.2510445612 &    22.3391739136\\
II.C$^{i.c.}_{9}$ &    -1.0006221621 &     0.0012247344 &     0.1740037507 &     0.1138359814 &    21.7738630579\\
II.C$^{i.c.}_{10}$ &    -0.9975148730 &     0.0001297649 &     0.1450629144 &     0.5435551180 &    37.9212316240\\
II.C$^{i.c.}_{11}$ &    -0.9943383314 &    -0.0057812831 &     0.1331341835 &     0.5429944488 &    37.2103034833\\
II.C$^{i.c.}_{12}$ &    -1.0020231439 &     0.0037713135 &     0.5551632149 &     0.2698044462 &    53.1094005605\\
II.C$^{i.c.}_{13}$ &    -1.0032605106 &     0.0045167438 &     0.4472991527 &     0.4896750344 &    62.8530057118\\
II.C$^{i.c.}_{14}$ &    -0.9985578294 &    -0.0027678151 &     0.5362144152 &     0.3723306084 &    60.4301224826\\
II.C$^{i.c.}_{15}$ &    -1.0029500764 &     0.0070627083 &     0.5213493496 &     0.3491337692 &    58.2492464733\\
II.C$^{i.c.}_{16}$ &    -1.0031391560 &     0.0039466348 &     0.2468215905 &     0.5672081394 &    53.1896103937\\
II.C$^{i.c.}_{17}$ &    -1.0000862446 &     0.0003055319 &     0.5164852735 &     0.1636660959 &    43.4802911792\\
II.C$^{i.c.}_{18}$ &    -1.0008489261 &     0.0013290610 &     0.4057627077 &     0.4954415922 &    62.3794063369\\
II.C$^{i.c.}_{19}$ &    -0.9998645606 &     0.0006210543 &     0.1869988183 &     0.2043525957 &    27.7197406792\\
II.C$^{i.c.}_{20}$ &    -0.9986419306 &     0.0064065752 &     0.3051083709 &     0.3559741487 &    37.1024009040\\
II.C$^{i.c.}_{21}$ &    -1.0007988437 &     0.0014317851 &     0.5048230051 &     0.3964978849 &    70.9198301370\\
II.C$^{i.c.}_{22}$ &    -0.9998260895 &    -0.0001772606 &     0.2660412682 &     0.5521309238 &    62.9700624858\\
II.C$^{i.c.}_{23}$ &    -1.0019501117 &     0.0020107888 &     0.2874388693 &     0.5596327476 &    67.7844044591\\
II.C$^{i.c.}_{24}$ &    -1.0012529935 &     0.0028881095 &     0.0892701310 &     0.3610325208 &    35.8845749685\\
II.C$^{i.c.}_{25}$ &    -1.0027353913 &     0.0005157812 &     0.5153175411 &     0.1440310535 &    52.4270825010\\
II.C$^{i.c.}_{26}$ &    -1.0044088408 &     0.0008204369 &     0.3965893119 &     0.4982442012 &    74.8967977689\\
II.C$^{i.c.}_{27}$ &    -0.9974591829 &     0.0014434640 &     0.4178305701 &     0.3301492680 &    55.5769080376\\
II.C$^{i.c.}_{28}$ &    -0.9996200271 &    -0.0008020030 &     0.2661357628 &     0.3362393813 &    43.5093590242\\
II.C$^{i.c.}_{29}$ &    -0.9999611432 &     0.0017175644 &     0.4299740470 &     0.3730083098 &    64.8696729632\\
II.C$^{i.c.}_{30}$ &    -1.0002778408 &    -0.0007336310 &     0.2540745616 &     0.2155375716 &    40.6100821848\\
II.C$^{i.c.}_{31}$ &    -1.0043264234 &     0.0085351855 &     0.5081224548 &     0.3842496864 &    86.9841516452\\
II.C$^{i.c.}_{32}$ &    -1.0022103816 &     0.0102733510 &     0.4147469694 &     0.4067682790 &    73.1187153646\\
II.C$^{i.c.}_{33}$ &    -0.9883083190 &    -0.0266620699 &     0.4097218737 &     0.3197223480 &    59.1311936461\\
II.C$^{i.c.}_{34}$ &    -0.9982384101 &    -0.0026965767 &     0.4471973476 &     0.4612592418 &    96.5838877286\\
II.C$^{i.c.}_{35}$ &    -0.9951304413 &    -0.0113745372 &     0.4152046229 &     0.3089526360 &    66.2706254945\\
II.C$^{i.c.}_{36}$ &    -0.9997799146 &    -0.0000871689 &     0.2709972127 &     0.5711878526 &    93.0111567688\\
II.C$^{i.c.}_{37}$ &    -1.0046985762 &     0.0162831677 &     0.3530834895 &     0.2445162178 &    57.2474723960\\
II.C$^{i.c.}_{38}$ &    -1.0008608869 &     0.0011280380 &     0.3788005974 &     0.5094274206 &    98.7297713440\\
II.C$^{i.c.}_{39}$ &    -1.0012100852 &     0.0028099578 &     0.4277768060 &     0.3389118542 &    80.1840191342\\
II.C$^{i.c.}_{40}$ &    -1.0178257900 &     0.0160007544 &     0.1778768589 &     0.5724956206 &    96.7051205494\\
II.C$^{i.c.}_{41}$ &    -0.9999159782 &    -0.0012779406 &     0.4758965652 &     0.2693288357 &    85.3853381329\\
II.C$^{i.c.}_{42}$ &    -1.0005146883 &     0.0178470909 &     0.4338014225 &     0.1146294454 &    66.8767469903\\
II.C$^{i.c.}_{43}$ &    -0.9969262199 &    -0.0055327948 &     0.4135974680 &     0.3039779928 &    78.9201147757\\
II.C$^{i.c.}_{44}$ &    -0.9997689776 &    -0.0006983641 &     0.2708168388 &     0.3123404924 &    63.3716725823\\
II.C$^{i.c.}_{45}$ &    -1.0058335497 &     0.0141374874 &     0.4263183336 &     0.2920871571 &    88.0387952421\\
II.C$^{i.c.}_{46}$ &    -1.0004035587 &     0.0031565296 &     0.3287145259 &     0.1543359704 &    63.6946794825\\
II.C$^{i.c.}_{47}$ &    -1.0001981436 &    -0.0005327229 &     0.2949051506 &     0.2388052958 &    65.2492675927\\
II.C$^{i.c.}_{48}$ &    -1.0025031994 &     0.0052121421 &     0.4082798530 &     0.3518144302 &    95.1326222384\\
II.C$^{i.c.}_{49}$ &    -1.0083313286 &    -0.0286778328 &     0.4077992979 &     0.3099009593 &    94.2024974068\\
II.C$^{i.c.}_{50}$ &    -0.9991000611 &    -0.0029159692 &     0.2874417121 &     0.2789397413 &    74.2602225931\\
II.C$^{i.c.}_{51}$ &    -1.0019070615 &     0.0078677169 &     0.2848590503 &     0.2308658034 &    70.7113437472\\
II.C$^{i.c.}_{52}$ &    -0.9996332993 &    -0.0028910173 &     0.4471116841 &     0.1069668572 &    83.9891019124\\
II.C$^{i.c.}_{53}$ &    -0.9947326912 &    -0.0128306041 &     0.4093239532 &     0.3063748779 &    97.6687075761\\
II.C$^{i.c.}_{54}$ &    -0.9997446546 &    -0.0003123011 &     0.3351127698 &     0.3541842111 &    96.8684615916\\
II.C$^{i.c.}_{55}$ &    -0.9993349134 &    -0.0032458410 &     0.2754489605 &     0.2297037522 &    76.0001131456\\
II.C$^{i.c.}_{56}$ &    -1.0003357519 &     0.0051959956 &     0.4507218273 &     0.0976302349 &    92.2314398981\\
II.C$^{i.c.}_{57}$ &    -1.0019558752 &     0.0070044294 &     0.2809143305 &     0.2689766307 &    80.0099426092\\
II.C$^{i.c.}_{58}$ &    -1.0015893090 &     0.0071301440 &     0.2737543421 &     0.2225574422 &    82.0867499689\\
II.C$^{i.c.}_{59}$ &    -0.9989466432 &    -0.0050002701 &     0.2716591775 &     0.2221618499 &    87.8794229689\\
II.C$^{i.c.}_{60}$ &    -0.9998090862 &    -0.0001193521 &     0.2769776134 &     0.2163748880 &    94.4991196762\\
\hline
\end{tabular*}
{\rule{\temptablewidth}{1pt}}
\end{center}
\end{table*}

\begin{table*}
\tabcolsep 0pt \caption{Initial conditions and periods $T_0$ of the additional periodic  three-body orbits for class II.C in case of the isosceles collinear configurations:  $\bm{r}_1(0)=(x_1,x_2)=-\bm{r}_2(0)$,  $\dot{\bm{r}}_1(0)=(v_1,v_2)=\dot{\bm{r}}_2(0)$ and $\bm{r}_3(0)=(0,0)$, $\dot{\bm{r}}_3(0)=(-2v_1, -2v_2)$ when $G=1$ and $m_1=m_2=m_3=1$ by means of the search grid $2000\times 2000$ in the interval $T_0 \in [0,200]$. Here, the superscript {\em i.c.} indicates the case of the initial conditions with {\em isosceles collinear}  configuration, due to the fact that there exist periodic orbits in many other cases. } \label{table-S4} \vspace*{-12pt}
\begin{center}
\def\temptablewidth{1\textwidth}
{\rule{\temptablewidth}{1pt}}
\begin{tabular*}{\temptablewidth}{@{\extracolsep{\fill}}lccccc}
\hline
Class and number & $x_1$ & $x_2$ & $v_1$ & $v_2$  & $T_0$\\
\hline
II.C$^{i.c.}_{61}$ &    -1.0052419067 &     0.0069932577 &     0.6712873807 &     0.4803461918 &   112.5568596820\\
II.C$^{i.c.}_{62}$ &    -0.9993020486 &    -0.0012318466 &     0.1465770165 &     0.1990740813 &    15.5689081167\\
II.C$^{i.c.}_{63}$ &    -0.9987921710 &    -0.0000376168 &     0.2981862979 &     0.6489381132 &    63.8148570065\\
II.C$^{i.c.}_{64}$ &    -0.9997342208 &    -0.0003340733 &     0.1065412613 &     0.2370411564 &    21.4507790042\\
II.C$^{i.c.}_{65}$ &    -0.9986440111 &    -0.0016444377 &     0.1181627019 &     0.5392186294 &    41.3525424135\\
II.C$^{i.c.}_{66}$ &    -0.9995844179 &    -0.0007608806 &     0.1348802539 &     0.1389474133 &    26.9005921406\\
II.C$^{i.c.}_{67}$ &    -0.9999689110 &     0.0000092431 &     0.0191038782 &     0.7881103514 &   174.4656744353\\
II.C$^{i.c.}_{68}$ &    -1.0002658777 &     0.0004737446 &     0.5314839122 &     0.3437559115 &    72.9208664122\\
II.C$^{i.c.}_{69}$ &    -0.9997112914 &     0.0023860970 &     0.1986151928 &     0.1079063889 &    33.1436831407\\
II.C$^{i.c.}_{70}$ &    -0.9986215413 &    -0.0015062085 &     0.3956424444 &     0.5085462937 &    82.7383275739\\
II.C$^{i.c.}_{71}$ &    -0.9994172355 &    -0.0004868193 &     0.3889596756 &     0.4367293130 &    67.5342475591\\
II.C$^{i.c.}_{72}$ &    -0.9967406301 &    -0.0042695019 &     0.4649642933 &     0.4808871151 &    97.3754612603\\
II.C$^{i.c.}_{73}$ &    -1.0006801652 &     0.0006105034 &     0.3914670389 &     0.5079797815 &    88.6055743961\\
II.C$^{i.c.}_{74}$ &    -1.0063577238 &    -0.0007899868 &     0.0469539964 &     0.5629093243 &    73.0920946244\\
II.C$^{i.c.}_{75}$ &    -1.0001323147 &     0.0001097191 &     0.2744638308 &     0.5579447167 &    89.1893629816\\
II.C$^{i.c.}_{76}$ &    -1.0002837014 &     0.0006003160 &     0.3633489139 &     0.4098810239 &    72.5678295850\\
II.C$^{i.c.}_{77}$ &    -1.0000824509 &     0.0026141957 &     0.2869528423 &     0.5549137288 &    90.2783841725\\
II.C$^{i.c.}_{78}$ &    -0.9996237569 &    -0.0007436637 &     0.2574689818 &     0.2094494453 &    46.7295611529\\
II.C$^{i.c.}_{79}$ &    -1.0006721678 &     0.0011219348 &     0.2831195994 &     0.5569547929 &    96.4949377862\\
II.C$^{i.c.}_{80}$ &    -0.9992205369 &    -0.0017928162 &     0.4612684850 &     0.3859118336 &   101.3746560458\\
II.C$^{i.c.}_{81}$ &    -0.9997738485 &    -0.0005457907 &     0.4455091949 &     0.3437749174 &    89.2418884626\\
II.C$^{i.c.}_{82}$ &    -1.0000924456 &     0.0000952707 &     0.3379818190 &     0.5473980161 &   117.6503619533\\
II.C$^{i.c.}_{83}$ &    -1.0000577962 &     0.0001145427 &     0.3129702933 &     0.5484224438 &   111.5323785796\\
II.C$^{i.c.}_{84}$ &    -1.0003199257 &     0.0002461198 &     0.2969756431 &     0.5438038303 &   106.1473549083\\
II.C$^{i.c.}_{85}$ &    -1.0004826776 &    -0.0004828304 &     0.2509154009 &     0.5668973521 &   106.1918301424\\
II.C$^{i.c.}_{86}$ &    -0.9993126562 &    -0.0007439249 &     0.3990385411 &     0.5099682634 &   124.1160017297\\
II.C$^{i.c.}_{87}$ &    -1.0001868367 &     0.0001926681 &     0.2254545025 &     0.3468678048 &    66.3708668659\\
II.C$^{i.c.}_{88}$ &    -1.0001466147 &     0.0002279102 &     0.5235135756 &     0.3428681229 &   122.7063150460\\
II.C$^{i.c.}_{89}$ &    -1.0006742054 &     0.0007036192 &     0.2648339916 &     0.5645711256 &   113.8940933586\\
II.C$^{i.c.}_{90}$ &    -1.0001562065 &     0.0002008968 &     0.3754694996 &     0.5198808856 &   127.3757065957\\
II.C$^{i.c.}_{91}$ &    -1.0019052297 &     0.0010913146 &     0.3736441570 &     0.5174825231 &   126.2975030884\\
II.C$^{i.c.}_{92}$ &    -1.0003798265 &     0.0006884069 &     0.4245613810 &     0.4896712222 &   135.8044318822\\
II.C$^{i.c.}_{93}$ &    -1.0002500671 &     0.0005219493 &     0.4780256232 &     0.4359311704 &   138.2492171687\\
II.C$^{i.c.}_{94}$ &    -1.0002459828 &     0.0004903903 &     0.4480253630 &     0.4240037755 &   121.1918921896\\
II.C$^{i.c.}_{95}$ &    -1.0008130645 &     0.0009030901 &     0.3683609414 &     0.5194758930 &   131.4440221037\\
II.C$^{i.c.}_{96}$ &    -1.0003357030 &     0.0004210132 &     0.4440770971 &     0.4217980998 &   122.5879353267\\
II.C$^{i.c.}_{97}$ &    -1.0003144961 &     0.0002566797 &     0.5050088314 &     0.4001484047 &   143.2104794611\\
II.C$^{i.c.}_{98}$ &    -1.0007497858 &     0.0009535509 &     0.3623244947 &     0.5155126495 &   131.5097060260\\
II.C$^{i.c.}_{99}$ &    -0.9996516166 &    -0.0009118649 &     0.4961624872 &     0.2755087892 &   111.7780338451\\
II.C$^{i.c.}_{100}$ &    -0.9975087774 &    -0.0006927134 &     0.1598422080 &     0.5839361975 &   121.3004546165\\
II.C$^{i.c.}_{101}$ &    -1.0011423223 &     0.0024436106 &     0.5108696475 &     0.3561396647 &   145.0362346455\\
II.C$^{i.c.}_{102}$ &    -0.9979476547 &    -0.0053913440 &     0.5023411523 &     0.2715293365 &   123.2578702276\\
II.C$^{i.c.}_{103}$ &    -0.9997027246 &    -0.0019259819 &     0.4825165193 &     0.3439517689 &   132.9451699407\\
II.C$^{i.c.}_{104}$ &    -0.9998167789 &    -0.0002616082 &     0.4600260151 &     0.4391474283 &   154.2830105938\\
II.C$^{i.c.}_{105}$ &    -0.9983176481 &    -0.0050269121 &     0.4123596897 &     0.3022004565 &   104.9731327132\\
II.C$^{i.c.}_{106}$ &    -1.0097132550 &     0.0200432827 &     0.4348329295 &     0.3451642094 &   122.0152451676\\
II.C$^{i.c.}_{107}$ &    -0.9985617493 &    -0.0019016284 &     0.4122482947 &     0.4880002442 &   159.0159956749\\
II.C$^{i.c.}_{108}$ &    -0.9994327398 &    -0.0009835969 &     0.2333863975 &     0.4714882786 &   107.5991420560\\
II.C$^{i.c.}_{109}$ &    -0.9996354196 &    -0.0005084127 &     0.5020029077 &     0.4268703112 &   178.8233665414\\
II.C$^{i.c.}_{110}$ &    -1.0000447633 &     0.0000248229 &     0.2164962932 &     0.5799688472 &   146.6269775104\\

\hline
\end{tabular*}
{\rule{\temptablewidth}{1pt}}
\end{center}
\end{table*}

\begin{table*}
\tabcolsep 0pt \caption{Initial conditions and periods $T_0$ of the additional periodic  three-body orbits for class II.C in case of the isosceles collinear configurations:  $\bm{r}_1(0)=(x_1,x_2)=-\bm{r}_2(0)$,  $\dot{\bm{r}}_1(0)=(v_1,v_2)=\dot{\bm{r}}_2(0)$ and $\bm{r}_3(0)=(0,0)$, $\dot{\bm{r}}_3(0)=(-2v_1, -2v_2)$ when $G=1$ and $m_1=m_2=m_3=1$ by means of the search grid $2000\times 2000$ in the interval $T_0 \in [0,200]$.  Here, the superscript {\em i.c.} indicates the case of the initial conditions with {\em isosceles collinear}  configuration, due to the fact that there exist periodic orbits in many other cases.} \label{table-S4} \vspace*{-12pt}
\begin{center}
\def\temptablewidth{1\textwidth}
{\rule{\temptablewidth}{1pt}}
\begin{tabular*}{\temptablewidth}{@{\extracolsep{\fill}}lccccc}
\hline
Class and number & $x_1$ & $x_2$ & $v_1$ & $v_2$  & $T_0$\\
\hline
II.C$^{i.c.}_{111}$ &    -1.0001878738 &     0.0001976041 &     0.3580246964 &     0.3620240468 &   108.4692501902\\
II.C$^{i.c.}_{112}$ &    -0.9999338757 &    -0.0000662093 &     0.4030942247 &     0.5021397842 &   165.8163230030\\
II.C$^{i.c.}_{113}$ &    -0.9973564264 &    -0.0055480022 &     0.4467609420 &     0.3632463403 &   131.8173163660\\
II.C$^{i.c.}_{114}$ &    -0.9998705005 &    -0.0002412936 &     0.5500166603 &     0.3816208882 &   196.2030131227\\
II.C$^{i.c.}_{115}$ &    -0.9996408082 &    -0.0034100425 &     0.3544729151 &     0.5263053524 &   161.0933835172\\
II.C$^{i.c.}_{116}$ &    -1.0016639152 &     0.0027369839 &     0.4124524496 &     0.3418039858 &   120.3537531454\\
II.C$^{i.c.}_{117}$ &    -0.9971942551 &    -0.0075528823 &     0.4786726614 &     0.2911739528 &   128.4789393448\\
II.C$^{i.c.}_{118}$ &    -1.0008736944 &     0.0008342668 &     0.5372631027 &     0.3978725791 &   195.9255232918\\
II.C$^{i.c.}_{119}$ &    -1.0000041304 &    -0.0002979194 &     0.2725026918 &     0.2645153395 &    85.2275180018\\
II.C$^{i.c.}_{120}$ &    -0.9992947526 &    -0.0014626312 &     0.4495699503 &     0.4220694808 &   157.1326513904\\
II.C$^{i.c.}_{121}$ &    -1.0013880192 &     0.0026662578 &     0.4446910037 &     0.4211013518 &   158.8738763042\\
II.C$^{i.c.}_{122}$ &    -1.0069628973 &     0.0045846719 &     0.0954822369 &     0.5729817986 &   139.2280448192\\
II.C$^{i.c.}_{123}$ &    -0.9993665091 &    -0.0012497302 &     0.4172328141 &     0.3437263248 &   128.6251230999\\
II.C$^{i.c.}_{124}$ &    -1.0015537301 &     0.0020229318 &     0.3607922062 &     0.5210463878 &   174.0275598937\\
II.C$^{i.c.}_{125}$ &    -0.9993019515 &    -0.0041030158 &     0.3793173882 &     0.3517942721 &   126.4851896900\\
II.C$^{i.c.}_{126}$ &    -1.0003617169 &     0.0006249003 &     0.3839326082 &     0.3566474946 &   129.1561627606\\
II.C$^{i.c.}_{127}$ &    -1.0000882275 &     0.0004083953 &     0.0718196326 &     0.5614066244 &   139.6924992835\\
II.C$^{i.c.}_{128}$ &    -0.9997939910 &    -0.0001974850 &     0.5084636066 &     0.3972587492 &   198.0966564176\\
II.C$^{i.c.}_{129}$ &    -1.0000059859 &     0.0000511211 &     0.5129973593 &     0.1994916873 &   146.1319840563\\
II.C$^{i.c.}_{130}$ &    -1.0000595249 &     0.0000721456 &     0.2239773950 &     0.5629479931 &   166.9759210601\\
II.C$^{i.c.}_{131}$ &    -0.9998033008 &    -0.0020199079 &     0.3641720646 &     0.2333768494 &   110.4901291852\\
II.C$^{i.c.}_{132}$ &    -0.9996158726 &    -0.0008571195 &     0.3756344279 &     0.3718126856 &   139.3960578613\\
II.C$^{i.c.}_{133}$ &    -1.0031796856 &     0.0116849810 &     0.4113128286 &     0.3414653995 &   143.6544337992\\
II.C$^{i.c.}_{134}$ &    -0.9970069552 &    -0.0071385299 &     0.2564915763 &     0.3009516148 &   103.8548828437\\
II.C$^{i.c.}_{135}$ &    -0.9997726615 &     0.0136501602 &     0.2654701477 &     0.2974883536 &   108.3401858875\\
II.C$^{i.c.}_{136}$ &    -0.9974381161 &    -0.0048082175 &     0.3942274845 &     0.3866729863 &   153.5487105246\\
II.C$^{i.c.}_{137}$ &    -1.0000722688 &    -0.0001301071 &     0.3454904870 &     0.3274211355 &   126.5668880037\\
II.C$^{i.c.}_{138}$ &    -1.0001417901 &     0.0012105130 &     0.3388352388 &     0.2310988162 &   112.9726097141\\
II.C$^{i.c.}_{139}$ &    -1.0002283724 &     0.0007510855 &     0.3579565048 &     0.1827031638 &   111.4365930773\\
II.C$^{i.c.}_{140}$ &    -0.9998831787 &     0.0004530519 &     0.2663540278 &     0.2616234253 &   107.4950725501\\
II.C$^{i.c.}_{141}$ &    -0.9985360463 &    -0.0028128349 &     0.4449720238 &     0.4234720627 &   194.7437039718\\
II.C$^{i.c.}_{142}$ &    -0.9986085844 &    -0.0026634557 &     0.4154551424 &     0.3592708320 &   157.9125298899\\
II.C$^{i.c.}_{143}$ &    -1.0001176968 &     0.0007557802 &     0.4106804220 &     0.3811624133 &   166.1880135854\\
II.C$^{i.c.}_{144}$ &    -0.9994752681 &     0.0006281915 &     0.0697668516 &     0.6034792439 &   184.2941396316\\
II.C$^{i.c.}_{145}$ &    -0.9980925280 &    -0.0079164176 &     0.4721914878 &     0.2352875371 &   156.5421248415\\
II.C$^{i.c.}_{146}$ &    -1.0000434585 &     0.0000444687 &     0.4799900474 &     0.2624524953 &   166.2108820260\\
II.C$^{i.c.}_{147}$ &    -1.0018301183 &     0.0044292283 &     0.4645435057 &     0.2835787142 &   165.0568233778\\
II.C$^{i.c.}_{148}$ &    -0.9998863289 &    -0.0006362756 &     0.5205233526 &     0.2387134015 &   183.2784602197\\
II.C$^{i.c.}_{149}$ &    -1.0010305795 &     0.0016795676 &     0.4179077416 &     0.3237817307 &   159.2989795739\\
II.C$^{i.c.}_{150}$ &    -1.0002299017 &     0.0010585324 &     0.5094591003 &     0.1585266374 &   162.7149597025\\
\hline
\end{tabular*}
{\rule{\temptablewidth}{1pt}}
\end{center}
\end{table*}

\begin{table*}
\tabcolsep 0pt \caption{Initial conditions and periods $T_0$ of the additional periodic  three-body orbits for class II.C in case of the isosceles collinear configurations:  $\bm{r}_1(0)=(x_1,x_2)=-\bm{r}_2(0)$,  $\dot{\bm{r}}_1(0)=(v_1,v_2)=\dot{\bm{r}}_2(0)$ and $\bm{r}_3(0)=(0,0)$, $\dot{\bm{r}}_3(0)=(-2v_1, -2v_2)$ when $G=1$ and $m_1=m_2=m_3=1$ by means of the search grid $2000\times 2000$ in the interval $T_0 \in [0,200]$. Here, the superscript {\em i.c.} indicates the case of the initial conditions with {\em isosceles collinear}  configuration, due to the fact that there exist periodic orbits in many other cases.} \label{table-S4} \vspace*{-12pt}
\begin{center}
\def\temptablewidth{1\textwidth}
{\rule{\temptablewidth}{1pt}}
\begin{tabular*}{\temptablewidth}{@{\extracolsep{\fill}}lccccc}
\hline
Class and number & $x_1$ & $x_2$ & $v_1$ & $v_2$  & $T_0$\\
\hline
II.C$^{i.c.}_{151}$ &    -1.0003579297 &     0.0001386626 &     0.0684520890 &     0.5832514468 &   181.5402088564\\
II.C$^{i.c.}_{152}$ &    -1.0000062160 &     0.0000137970 &     0.4639984735 &     0.3539914399 &   196.2979767201\\
II.C$^{i.c.}_{153}$ &    -1.0004579537 &    -0.0003040420 &     0.3516419315 &     0.3640172911 &   155.5273430215\\
II.C$^{i.c.}_{154}$ &    -0.9993744752 &    -0.0011606107 &     0.3151515352 &     0.2974388166 &   135.8847384753\\
II.C$^{i.c.}_{155}$ &    -1.0009338406 &     0.0025136409 &     0.4107220506 &     0.3694016013 &   184.7047155440\\
II.C$^{i.c.}_{156}$ &    -0.9998947485 &    -0.0004613020 &     0.3230582846 &     0.3280798899 &   145.3569669177\\
II.C$^{i.c.}_{157}$ &    -0.9988305188 &    -0.0071836995 &     0.4761411271 &     0.1836551378 &   167.4336860617\\
II.C$^{i.c.}_{158}$ &    -1.0012344570 &     0.0045607417 &     0.2983656665 &     0.2528048642 &   129.8125656142\\
II.C$^{i.c.}_{159}$ &    -0.9996815616 &    -0.0014949294 &     0.3132461447 &     0.1845645006 &   124.0109852724\\
II.C$^{i.c.}_{160}$ &    -1.0001427298 &     0.0001039303 &     0.4160683467 &     0.3349307497 &   182.5831598106\\
II.C$^{i.c.}_{161}$ &    -1.0029684695 &     0.0098848810 &     0.3391135928 &     0.3842089282 &   173.5970900472\\
II.C$^{i.c.}_{162}$ &    -0.9957240222 &    -0.0145517994 &     0.4849091125 &     0.2174783545 &   185.4070301506\\
II.C$^{i.c.}_{163}$ &    -1.0000763029 &     0.0003544606 &     0.2970550726 &     0.2220117477 &   132.0367794399\\
II.C$^{i.c.}_{164}$ &    -0.9998630405 &    -0.0002625686 &     0.4155954294 &     0.3580933511 &   199.4000133450\\
II.C$^{i.c.}_{165}$ &    -0.9998043545 &    -0.0005831287 &     0.4871434830 &     0.1664499354 &   187.7250828374\\
II.C$^{i.c.}_{166}$ &    -0.9990786651 &    -0.0017944247 &     0.3633486143 &     0.3778437601 &   191.4082770477\\
II.C$^{i.c.}_{167}$ &    -0.9995072442 &    -0.0011060861 &     0.2982624015 &     0.3186046105 &   160.0960194781\\
II.C$^{i.c.}_{168}$ &    -1.0001341807 &     0.0016758159 &     0.3107810092 &     0.3272211576 &   165.3545618995\\
II.C$^{i.c.}_{169}$ &    -1.0005876572 &     0.0014169677 &     0.3102735439 &     0.3279657982 &   165.5714876065\\
II.C$^{i.c.}_{170}$ &    -0.9989490140 &    -0.0049306902 &     0.2756587302 &     0.2462901387 &   144.5568341268\\
II.C$^{i.c.}_{171}$ &    -1.0002776965 &     0.0009891203 &     0.4390360091 &     0.2367245735 &   190.5169699427\\
II.C$^{i.c.}_{172}$ &    -0.9999004074 &    -0.0001909213 &     0.3625268037 &     0.3645943176 &   195.5861162535\\
II.C$^{i.c.}_{173}$ &    -0.9989831540 &     0.0044951487 &     0.4291446608 &     0.2622089189 &   191.7314605155\\
II.C$^{i.c.}_{174}$ &    -1.0000218503 &     0.0000970444 &     0.3729890521 &     0.2249735407 &   165.0144539545\\
II.C$^{i.c.}_{175}$ &    -1.0002415919 &     0.0000875308 &     0.2888230696 &     0.2160090416 &   149.5190011938\\
II.C$^{i.c.}_{176}$ &    -1.0006287306 &     0.0013765383 &     0.3667898158 &     0.1910813836 &   165.5341361180\\
II.C$^{i.c.}_{177}$ &    -1.0000437442 &     0.0001069776 &     0.2884795245 &     0.1819433227 &   151.5959091689\\
II.C$^{i.c.}_{178}$ &    -1.0010470776 &     0.0061711582 &     0.3756890358 &     0.1450913247 &   169.4799994831\\
II.C$^{i.c.}_{179}$ &    -1.0004586234 &     0.0008217659 &     0.3019673227 &     0.3276357315 &   186.0057961528\\
II.C$^{i.c.}_{180}$ &    -1.0000180537 &    -0.0002026011 &     0.2920408658 &     0.2114136875 &   162.4448908003\\
II.C$^{i.c.}_{181}$ &    -1.0005966257 &     0.0027068678 &     0.3367428951 &     0.1921541610 &   171.6538930140\\
II.C$^{i.c.}_{182}$ &    -0.9999667271 &    -0.0000131797 &     0.2965219810 &     0.3180219666 &   192.7617594456\\
II.C$^{i.c.}_{183}$ &    -0.9992491629 &    -0.0034871098 &     0.3258429673 &     0.1940315317 &   175.5129086549\\
II.C$^{i.c.}_{184}$ &    -0.9999718046 &    -0.0005157440 &     0.3065429253 &     0.2615287991 &   189.5799746897\\
II.C$^{i.c.}_{185}$ &    -0.9999108973 &    -0.0005680567 &     0.2991509810 &     0.1592369292 &   173.1650000611\\
II.C$^{i.c.}_{186}$ &    -1.0005529897 &     0.0027084473 &     0.3122232581 &     0.1907334790 &   185.3134789100\\
II.C$^{i.c.}_{187}$ &    -1.0000232543 &     0.0004999932 &     0.3034790834 &     0.2174643065 &   193.8838763097\\
II.C$^{i.c.}_{188}$ &    -1.0005361761 &     0.0025270970 &     0.2609453389 &     0.2249677059 &   196.8070490286\\
\hline
\end{tabular*}
{\rule{\temptablewidth}{1pt}}
\end{center}
\end{table*}

\begin{table*}
\tabcolsep 0pt \caption{Initial conditions and periods $T_0$ of the additional periodic  three-body orbits for class II.C in case of the isosceles collinear configurations:  $\bm{r}_1(0)=(x_1,x_2)=-\bm{r}_2(0)$,  $\dot{\bm{r}}_1(0)=(v_1,v_2)=\dot{\bm{r}}_2(0)$ and $\bm{r}_3(0)=(0,0)$, $\dot{\bm{r}}_3(0)=(-2v_1, -2v_2)$ when $G=1$ and $m_1=m_2=m_3=1$ by means of the search grid $4000\times 4000$ in the interval $T_0 \in [0,200]$. } \label{table-S4} \vspace*{-12pt}
\begin{center}
\def\temptablewidth{1\textwidth}
{\rule{\temptablewidth}{1pt}}
\begin{tabular*}{\temptablewidth}{@{\extracolsep{\fill}}lccccc}
\hline
Class and number & $x_1$ & $x_2$ & $v_1$ & $v_2$  & $T_0$\\
\hline
II.C$^{i.c.}_{189}$ &    -0.9987156003 &    -0.0001808773 &     0.0667228965 &     0.2914860882 &    13.3439615814\\
II.C$^{i.c.}_{190}$ &    -1.0004056726 &     0.0004036254 &     0.0541196155 &     0.3916581165 &    15.2889216727\\
II.C$^{i.c.}_{191}$ &    -0.9962117149 &    -0.0104165391 &     0.4213247584 &     0.6338536605 &    58.5664322507\\
II.C$^{i.c.}_{192}$ &    -0.9981975513 &    -0.0043835873 &     0.1458583114 &     0.0441379641 &    16.7744994586\\
II.C$^{i.c.}_{193}$ &    -0.9997643579 &     0.0000791554 &     0.0734663749 &     0.2957075354 &    22.3933998449\\
II.C$^{i.c.}_{194}$ &    -0.9977672957 &    -0.0041772092 &     0.5544350298 &     0.4242879511 &    74.8547183310\\
II.C$^{i.c.}_{195}$ &    -1.0009485148 &     0.0015951145 &     0.1880792732 &     0.4824111698 &    40.3627303824\\
II.C$^{i.c.}_{196}$ &    -0.9994746149 &    -0.0004509771 &     0.3267558947 &     0.4442573138 &    46.4115156704\\
II.C$^{i.c.}_{197}$ &    -0.9998362308 &    -0.0001184829 &     0.0772502870 &     0.0963105429 &    25.8332940780\\
II.C$^{i.c.}_{198}$ &    -1.0001352357 &     0.0000728370 &     0.4805176176 &     0.5831724751 &   131.5979482665\\
II.C$^{i.c.}_{199}$ &    -0.9999974677 &     0.0000048132 &     0.0759952308 &     0.2997588380 &    31.5190459098\\
II.C$^{i.c.}_{200}$ &    -0.9999568995 &    -0.0001864658 &     0.2512828957 &     0.3012941946 &    37.4190007658\\
II.C$^{i.c.}_{201}$ &    -0.9998989469 &    -0.0004479285 &     0.2223468570 &     0.3220102801 &    38.7852937663\\
II.C$^{i.c.}_{202}$ &    -1.0000351771 &    -0.0000300698 &     0.1727280013 &     0.3365000175 &    39.3465006036\\
II.C$^{i.c.}_{203}$ &    -1.0019177030 &     0.0026394585 &     0.3459003198 &     0.4904903533 &    72.8641921756\\
II.C$^{i.c.}_{204}$ &    -0.9996352172 &    -0.0001451789 &     0.3984906406 &     0.4529025301 &    76.7247467269\\
II.C$^{i.c.}_{205}$ &    -0.9998558979 &    -0.0001142152 &     0.3107629111 &     0.4360285084 &    67.9203149209\\
II.C$^{i.c.}_{206}$ &    -0.9999867179 &    -0.0000045996 &     0.0995065550 &     0.3957741246 &    51.8946950504\\
II.C$^{i.c.}_{207}$ &    -0.9953690728 &    -0.0011563935 &     0.2712213964 &     0.5783454991 &    94.5291576683\\
II.C$^{i.c.}_{208}$ &    -0.9999881008 &     0.0001490234 &     0.3791820867 &     0.5142843068 &   100.3132215843\\
II.C$^{i.c.}_{209}$ &    -1.0000873157 &     0.0000620007 &     0.2816591107 &     0.3514427663 &    60.0455107428\\
II.C$^{i.c.}_{210}$ &    -0.9999740133 &    -0.0000063259 &     0.3851703094 &     0.5086505769 &   106.2793970063\\
II.C$^{i.c.}_{211}$ &    -1.0004773198 &     0.0006990473 &     0.4038571091 &     0.5024819268 &   109.7067469940\\
II.C$^{i.c.}_{212}$ &    -0.9997285801 &    -0.0002552208 &     0.3010549515 &     0.5488680845 &   102.7966886504\\
II.C$^{i.c.}_{213}$ &    -1.0006099242 &     0.0005401982 &     0.3084374588 &     0.5513227429 &   105.6300556053\\
II.C$^{i.c.}_{214}$ &    -1.0007040828 &    -0.0043204430 &     0.5190849396 &     0.2144998198 &    87.3795069772\\
II.C$^{i.c.}_{215}$ &    -1.0000626188 &     0.0000295817 &     0.2512399360 &     0.3272223702 &    61.4049734638\\
II.C$^{i.c.}_{216}$ &    -1.0000279578 &    -0.0000053618 &     0.3800152197 &     0.5162309497 &   114.2190811967\\
II.C$^{i.c.}_{217}$ &    -0.9999453927 &    -0.0000525535 &     0.3975071879 &     0.4882801093 &   112.4719632569\\
II.C$^{i.c.}_{218}$ &    -0.9999884253 &     0.0001286863 &     0.1829866856 &     0.2103272454 &    55.5601441746\\
II.C$^{i.c.}_{219}$ &    -0.9971285533 &    -0.0025848850 &     0.3075439756 &     0.5536254588 &   117.7057392712\\
II.C$^{i.c.}_{220}$ &    -0.9999635256 &     0.0000272355 &     0.4492456452 &     0.4587792545 &   127.0597565659\\
II.C$^{i.c.}_{221}$ &    -0.9999420692 &    -0.0000020503 &     0.3837333109 &     0.5140293567 &   127.3976992403\\
II.C$^{i.c.}_{222}$ &    -1.0006615015 &     0.0012450475 &     0.3437500104 &     0.3426947718 &    80.2375649425\\
II.C$^{i.c.}_{223}$ &    -1.0002577628 &     0.0002833695 &     0.3086887635 &     0.5448275268 &   121.2449173210\\
II.C$^{i.c.}_{224}$ &    -1.0000006686 &     0.0000370596 &     0.5144905585 &     0.2857462788 &   117.3489868466\\
II.C$^{i.c.}_{225}$ &    -1.0004622415 &     0.0004509313 &     0.2874162072 &     0.5599798038 &   129.0094022535\\
II.C$^{i.c.}_{226}$ &    -1.0000737409 &     0.0000685736 &     0.3934962614 &     0.4901913603 &   130.4066658899\\
II.C$^{i.c.}_{227}$ &    -0.9996589411 &    -0.0004650065 &     0.5185413710 &     0.4545088680 &   178.5446004292\\
II.C$^{i.c.}_{228}$ &    -1.0003821718 &     0.0002866767 &     0.3525549226 &     0.4503899441 &   110.5646092212\\
II.C$^{i.c.}_{229}$ &    -0.9969096477 &    -0.0037112567 &     0.3683870820 &     0.5227999790 &   137.9240293649\\
II.C$^{i.c.}_{230}$ &    -0.9996506355 &    -0.0004974086 &     0.1741471223 &     0.5734799650 &   119.2783925080\\
II.C$^{i.c.}_{231}$ &    -1.0000418944 &     0.0000309536 &     0.2507445323 &     0.5609705084 &   132.5909859053\\
II.C$^{i.c.}_{232}$ &    -0.9999993626 &     0.0000175831 &     0.3102635877 &     0.3782403600 &    98.4449177453\\
II.C$^{i.c.}_{233}$ &    -0.9995713012 &     0.0001457361 &     0.3165695623 &     0.3698366478 &   100.0503333445\\
II.C$^{i.c.}_{234}$ &    -1.0000322260 &     0.0000882641 &     0.5017329443 &     0.3989758208 &   164.4327683942\\
II.C$^{i.c.}_{235}$ &    -1.0000162983 &     0.0000319933 &     0.4964924947 &     0.4057555444 &   164.5550366898\\
II.C$^{i.c.}_{236}$ &    -0.9998253224 &     0.0000157250 &     0.4409388563 &     0.4698571723 &   167.8254115805\\
II.C$^{i.c.}_{237}$ &    -1.0000045664 &     0.0001083019 &     0.2462147856 &     0.6145115166 &   178.2102563595\\
II.C$^{i.c.}_{238}$ &    -0.9996121801 &    -0.0007145032 &     0.4982828137 &     0.3333895803 &   148.2688763644\\
II.C$^{i.c.}_{239}$ &    -0.9999745932 &    -0.0000680219 &     0.3699940868 &     0.5202873176 &   164.0681520596\\
II.C$^{i.c.}_{240}$ &    -0.9991569403 &    -0.0014204011 &     0.0752963046 &     0.5599873395 &   124.5823286975\\
II.C$^{i.c.}_{241}$ &    -0.9999488227 &    -0.0000023867 &     0.2629902490 &     0.5682981139 &   157.2222631073\\
II.C$^{i.c.}_{242}$ &    -1.0000672764 &     0.0000778419 &     0.4412337213 &     0.4674745244 &   173.3014324846\\
II.C$^{i.c.}_{243}$ &    -0.9991553462 &    -0.0025270317 &     0.2816635321 &     0.3056086144 &    93.8195796981\\
II.C$^{i.c.}_{244}$ &    -0.9999751274 &    -0.0001288570 &     0.4148348037 &     0.4294973613 &   152.3950327025\\
II.C$^{i.c.}_{245}$ &    -1.0001978574 &     0.0002371804 &     0.2591616337 &     0.5674280924 &   161.7537071941\\
II.C$^{i.c.}_{246}$ &    -0.9990481655 &    -0.0016568759 &     0.4121147162 &     0.4820443096 &   172.4809267651\\
II.C$^{i.c.}_{247}$ &    -0.9995316642 &    -0.0011980010 &     0.2981739309 &     0.3116833545 &    99.8938774737\\

\hline
\end{tabular*}
{\rule{\temptablewidth}{1pt}}
\end{center}
\end{table*}

\begin{table*}
\tabcolsep 0pt \caption{Initial conditions and periods $T_0$ of the additional periodic  three-body orbits for class II.C in case of the isosceles collinear configurations:  $\bm{r}_1(0)=(x_1,x_2)=-\bm{r}_2(0)$,  $\dot{\bm{r}}_1(0)=(v_1,v_2)=\dot{\bm{r}}_2(0)$ and $\bm{r}_3(0)=(0,0)$, $\dot{\bm{r}}_3(0)=(-2v_1, -2v_2)$ when $G=1$ and $m_1=m_2=m_3=1$ by means of the search grid $4000\times 4000$ in the interval $T_0 \in [0,200]$.  Here, the superscript {\em i.c.} indicates the case of the initial conditions with {\em isosceles collinear}  configuration, due to the fact that there exist periodic orbits in many other cases.} \label{table-S4} \vspace*{-12pt}
\begin{center}
\def\temptablewidth{1\textwidth}
{\rule{\temptablewidth}{1pt}}
\begin{tabular*}{\temptablewidth}{@{\extracolsep{\fill}}lccccc}
\hline
Class and number & $x_1$ & $x_2$ & $v_1$ & $v_2$  & $T_0$\\
\hline
II.C$^{i.c.}_{248}$ &    -0.9998725873 &    -0.0012619861 &     0.4547945347 &     0.0902252115 &   108.4318363374\\
II.C$^{i.c.}_{249}$ &    -1.0001338896 &     0.0003895034 &     0.2666839601 &     0.3021242917 &   102.2029657706\\
II.C$^{i.c.}_{250}$ &    -1.0016442384 &     0.0019831858 &     0.4847429580 &     0.3723593286 &   177.7896653009\\
II.C$^{i.c.}_{251}$ &    -1.0000135544 &    -0.0002722145 &     0.3410872967 &     0.3826583264 &   131.0549830050\\
II.C$^{i.c.}_{252}$ &    -1.0002665933 &     0.0002187864 &     0.0587511984 &     0.5596235802 &   143.0177124983\\
II.C$^{i.c.}_{253}$ &    -0.9997827278 &    -0.0003313134 &     0.3803067198 &     0.4573989071 &   169.3150964920\\
II.C$^{i.c.}_{254}$ &    -0.9999261907 &    -0.0002834034 &     0.5037613707 &     0.1748930395 &   143.5829989597\\
II.C$^{i.c.}_{255}$ &    -0.9990477767 &     0.0012590880 &     0.2569557640 &     0.5607591530 &   179.9010137902\\
II.C$^{i.c.}_{256}$ &    -1.0000434033 &     0.0000190077 &     0.2837527401 &     0.5614788137 &   189.9391680740\\
II.C$^{i.c.}_{257}$ &    -1.0001594223 &     0.0006217366 &     0.5132795779 &     0.2599206266 &   168.6313690226\\
II.C$^{i.c.}_{258}$ &    -0.9995501395 &    -0.0017130408 &     0.4643508516 &     0.1962106112 &   147.5476763432\\
II.C$^{i.c.}_{259}$ &    -1.0000137217 &     0.0001032477 &     0.2917566760 &     0.3452216814 &   130.3556831256\\
II.C$^{i.c.}_{260}$ &    -0.9999758464 &    -0.0001233067 &     0.3112649021 &     0.2267781934 &   121.5980409635\\
II.C$^{i.c.}_{261}$ &    -1.0001407442 &    -0.0000788778 &     0.2454576748 &     0.2540970400 &   117.1373819005\\
II.C$^{i.c.}_{262}$ &    -0.9999905041 &    -0.0000356075 &     0.0287660408 &     0.5707570417 &   181.4562603011\\
II.C$^{i.c.}_{263}$ &    -0.9999996482 &     0.0002578254 &     0.2562868409 &     0.2837424323 &   125.6569499608\\
II.C$^{i.c.}_{264}$ &    -1.0000189688 &    -0.0000158115 &     0.4169946169 &     0.3102298659 &   168.4699848812\\
II.C$^{i.c.}_{265}$ &    -0.9987788922 &    -0.0011875959 &     0.3142668032 &     0.3390312003 &   147.2570456988\\
II.C$^{i.c.}_{266}$ &    -0.9997894125 &    -0.0002865976 &     0.4918287020 &     0.2456593661 &   187.3950204843\\
II.C$^{i.c.}_{267}$ &    -1.0001427704 &     0.0000915041 &     0.3761863814 &     0.3531462897 &   170.5503382220\\
II.C$^{i.c.}_{268}$ &    -0.9994954389 &    -0.0025332078 &     0.3052493724 &     0.1844460980 &   129.1199736931\\
II.C$^{i.c.}_{269}$ &    -1.0002399197 &     0.0003816371 &     0.4961292009 &     0.2383267110 &   197.9343910833\\
II.C$^{i.c.}_{270}$ &    -1.0011384426 &     0.0017612350 &     0.4107828747 &     0.3436504850 &   192.8739022470\\
II.C$^{i.c.}_{271}$ &    -0.9999713559 &    -0.0005081604 &     0.2465996681 &     0.2776393934 &   136.6852840505\\
II.C$^{i.c.}_{272}$ &    -0.9999309348 &    -0.0003572874 &     0.4575130947 &     0.1344272394 &   170.6883836240\\
II.C$^{i.c.}_{273}$ &    -0.9992867062 &    -0.0019384049 &     0.4167884076 &     0.3002718583 &   191.9211595245\\
II.C$^{i.c.}_{274}$ &    -0.9998356367 &    -0.0008239274 &     0.2948453667 &     0.1829736798 &   140.1021980522\\
II.C$^{i.c.}_{275}$ &    -1.0000804524 &     0.0004189259 &     0.2856834408 &     0.2658866170 &   149.6772537639\\
II.C$^{i.c.}_{276}$ &    -0.9993980509 &    -0.0017679740 &     0.3599600569 &     0.2716460452 &   177.4119922420\\
II.C$^{i.c.}_{277}$ &    -1.0000035902 &     0.0001339393 &     0.2507649288 &     0.2254700662 &   149.0898631488\\
II.C$^{i.c.}_{278}$ &    -1.0006654019 &     0.0015511428 &     0.3633059667 &     0.3155936500 &   195.5239398101\\
II.C$^{i.c.}_{279}$ &    -0.9998387459 &    -0.0007317615 &     0.3488179941 &     0.1932276404 &   167.8110014841\\
II.C$^{i.c.}_{280}$ &    -1.0000138576 &     0.0000912128 &     0.3367373516 &     0.2244889746 &   169.5324339488\\
II.C$^{i.c.}_{281}$ &    -0.9997900749 &    -0.0009843648 &     0.3343414299 &     0.2494882433 &   173.2442087382\\
II.C$^{i.c.}_{282}$ &    -1.0000971420 &     0.0003674499 &     0.2661903172 &     0.2371482182 &   161.3439815607\\
II.C$^{i.c.}_{283}$ &    -0.9999465107 &    -0.0003739149 &     0.3262703600 &     0.2233279843 &   173.6099578753\\
II.C$^{i.c.}_{284}$ &    -1.0000081675 &     0.0001861087 &     0.2692299144 &     0.2735207886 &   174.8769317640\\
II.C$^{i.c.}_{285}$ &    -1.0001629243 &     0.0015686838 &     0.3871981754 &     0.0809189184 &   180.8747987036\\
II.C$^{i.c.}_{286}$ &    -0.9996999528 &    -0.0009522718 &     0.3158753573 &     0.2640693312 &   185.4438380975\\
II.C$^{i.c.}_{287}$ &    -1.0007310282 &     0.0015516783 &     0.2763871353 &     0.3026990853 &   186.1998273962\\
II.C$^{i.c.}_{288}$ &    -1.0000321666 &    -0.0000515066 &     0.2727347149 &     0.2677280585 &   181.1658791593\\
II.C$^{i.c.}_{289}$ &    -1.0000238493 &    -0.0000367018 &     0.2622640785 &     0.2330143141 &   172.7886677901\\
II.C$^{i.c.}_{290}$ &    -1.0001107093 &     0.0005574302 &     0.3184424756 &     0.1920983970 &   180.3212998730\\
II.C$^{i.c.}_{291}$ &    -1.0001813181 &     0.0013450542 &     0.3401309720 &     0.1536431824 &   180.8060077523\\
II.C$^{i.c.}_{292}$ &    -0.9998013254 &    -0.0015808931 &     0.3630625715 &     0.0955715471 &   181.3670063858\\
II.C$^{i.c.}_{293}$ &    -1.0012110874 &     0.0011146648 &     0.3066659300 &     0.2422317252 &   186.1816524786\\
II.C$^{i.c.}_{294}$ &    -1.0007170582 &     0.0065592095 &     0.3504155672 &     0.1001305157 &   185.0680274964\\
II.C$^{i.c.}_{295}$ &    -0.9994226525 &    -0.0020297543 &     0.2987828627 &     0.2598745734 &   193.9609882459\\
II.C$^{i.c.}_{296}$ &    -0.9997829514 &    -0.0006793886 &     0.3073462761 &     0.2754690751 &   199.8303917589\\
II.C$^{i.c.}_{297}$ &    -1.0001577839 &     0.0014271273 &     0.3399198897 &     0.1073727897 &   189.0906833247\\
II.C$^{i.c.}_{298}$ &    -0.9999398533 &    -0.0002715230 &     0.2460657703 &     0.2457722296 &   191.0054386580\\
II.C$^{i.c.}_{299}$ &    -1.0000155059 &     0.0006008126 &     0.2471482953 &     0.2435552352 &   197.1915958863\\
II.C$^{i.c.}_{300}$ &    -0.9990460870 &    -0.0076394565 &     0.3252039063 &     0.1170310044 &   198.4164844102\\
\hline
\end{tabular*}
{\rule{\temptablewidth}{1pt}}
\end{center}
\end{table*}

\begin{table*}
\tabcolsep 0pt \caption{Initial conditions and periods $T$ of the periodic three-body orbits for class I.A in case of the isosceles collinear configurations:  $\bm{r}_1(0)=(-1,0)=-\bm{r}_2(0)$,  $\dot{\bm{r}}_1(0)=(v_1,v_2)=\dot{\bm{r}}_2(0)$ and $\bm{r}_3(0)=(0,0)$, $\dot{\bm{r}}_3(0)=(-2v_1, -2v_2)$ when $G=1$ and $m_1=m_2=m_3=1$ by means of the search grid $1000\times 1000$ in the interval $T\in[0,100]$, where $T^*=T |E|^{3/2}$ is its scale-invariant period, $L_f$ is the length of the free group element.  Here, the superscript {\em i.c.} indicates the case of the initial conditions with {\em isosceles collinear}  configuration, due to the fact that there exist periodic orbits in many other cases.} \label{table-S5} \vspace*{-12pt}
\begin{center}
\def\temptablewidth{1\textwidth}
{\rule{\temptablewidth}{1pt}}
\begin{tabular*}{\temptablewidth}{@{\extracolsep{\fill}}lccccc}
\hline
Class and number  & $v_1$ & $v_2$  & $T$  & $T^*$ & $L_f$\\
\hline
I.A$^{i.c.}_{1}$ &     0.3471168881 &     0.5327249454 &     6.3259139829 &      9.238 &  4\\
I.A$^{i.c.}_{2}$ &     0.3068934205 &     0.1255065670 &     6.2346748391 &     19.932 &  8\\
I.A$^{i.c.}_{3}$ &     0.6150407229 &     0.5226158545 &    37.3205235945 &     15.048 &  8\\
I.A$^{i.c.}_{4}$ &     0.5379557207 &     0.3414578545 &    26.9186696160 &     39.075 &  16\\
I.A$^{i.c.}_{5}$ &     0.4112926910 &     0.2607551013 &    20.7490650080 &     49.630 &  20\\
I.A$^{i.c.}_{6}$ &     0.4425908552 &     0.4235138348 &    35.8334644158 &     57.728 &  24\\
I.A$^{i.c.}_{7}$ &     0.1214534165 &     0.1012023800 &    15.7440944954 &     59.455 &  24\\
I.A$^{i.c.}_{8}$ &     0.4094945913 &     0.3628231655 &    33.8677507235 &     68.673 &  28\\
I.A$^{i.c.}_{9}$ &     0.5255769251 &     0.2501253528 &    43.8698538026 &     79.277 &  32\\
I.A$^{i.c.}_{10}$ &     0.4121028725 &     0.2833837497 &    34.2470499687 &     79.255 &  32\\
I.A$^{i.c.}_{11}$ &     0.4364192674 &     0.4457095477 &    49.5954033190 &     76.298 &  32\\
I.A$^{i.c.}_{12}$ &     0.1182009612 &     0.5847367662 &    48.2723717545 &     82.749 &  36\\
I.A$^{i.c.}_{13}$ &     0.4763527642 &     0.3789434497 &    53.4714914343 &     87.484 &  36\\
I.A$^{i.c.}_{14}$ &     0.4027121690 &     0.2100155085 &    34.7120153702 &     89.560 &  36\\
I.A$^{i.c.}_{15}$ &     0.3960494651 &     0.3529413931 &    46.1340735475 &     98.289 &  40\\
I.A$^{i.c.}_{16}$ &     0.4340543928 &     0.4606927255 &    64.1162217970 &     94.824 &  40\\
I.A$^{i.c.}_{17}$ &     0.1842784887 &     0.5871881740 &    63.5343529785 &    101.184 &  44\\
I.A$^{i.c.}_{18}$ &     0.4580378828 &     0.4093753024 &    66.3519736810 &    106.147 &  44\\
I.A$^{i.c.}_{19}$ &     0.4152505707 &     0.2913461808 &    47.9254977196 &    108.869 &  44\\
I.A$^{i.c.}_{20}$ &     0.0969422653 &     0.5615968396 &    53.6899444096 &    101.174 &  44\\
I.A$^{i.c.}_{21}$ &     0.4082108156 &     0.2436851904 &    48.4868513719 &    119.241 &  48\\
I.A$^{i.c.}_{22}$ &     0.4289870678 &     0.4723797648 &    78.3984979261 &    113.331 &  48\\
I.A$^{i.c.}_{23}$ &     0.4138807521 &     0.3477955684 &    61.8447789156 &    127.900 &  52\\
I.A$^{i.c.}_{24}$ &     0.4905050535 &     0.4044215155 &    85.3859524958 &    124.747 &  52\\
I.A$^{i.c.}_{25}$ &     0.1982770999 &     0.5760625510 &    73.2768117405 &    119.634 &  52\\
I.A$^{i.c.}_{26}$ &     0.3991287659 &     0.1847081193 &    48.6673769352 &    129.450 &  52\\
I.A$^{i.c.}_{27}$ &     0.4151691260 &     0.2953409098 &    61.3228288140 &    138.477 &  56\\
I.A$^{i.c.}_{28}$ &     0.0490506729 &     0.5901941115 &    79.1518362736 &    137.886 &  60\\
I.A$^{i.c.}_{29}$ &     0.3447503346 &     0.3930446386 &    67.3877746400 &    146.737 &  60\\
I.A$^{i.c.}_{30}$ &     0.2203123981 &     0.5718227262 &    85.7935924161 &    138.093 &  60\\
I.A$^{i.c.}_{31}$ &     0.4441311511 &     0.3823106072 &    82.3618800834 &    146.756 &  60\\
I.A$^{i.c.}_{32}$ &     0.4041322114 &     0.2191641945 &    62.4468134238 &    159.167 &  64\\
I.A$^{i.c.}_{33}$ &     0.4089912496 &     0.3457125084 &    75.0210999574 &    157.507 &  64\\
I.A$^{i.c.}_{34}$ &     0.4572423635 &     0.4074171858 &    95.8806923191 &    154.561 &  64\\
I.A$^{i.c.}_{35}$ &     0.3972193787 &     0.1691985386 &    62.6275833821 &    169.327 &  68\\
I.A$^{i.c.}_{36}$ &     0.4160674674 &     0.2971499303 &    74.7892110946 &    168.084 &  68\\
I.A$^{i.c.}_{37}$ &     0.1038901209 &     0.5907210858 &    92.2887207815 &    156.291 &  68\\
I.A$^{i.c.}_{38}$ &     0.4135366646 &     0.2710056059 &    76.0285224477 &    178.523 &  72\\
I.A$^{i.c.}_{39}$ &     0.3967429300 &     0.3708809839 &    85.9343829011 &    176.391 &  72\\
I.A$^{i.c.}_{40}$ &     0.4073762182 &     0.2388431491 &    76.2219729678 &    188.847 &  76\\
I.A$^{i.c.}_{41}$ &     0.1334658448 &     0.5838277552 &   102.8155095976 &    174.712 &  76\\
I.A$^{i.c.}_{42}$ &     0.4015585101 &     0.2022664462 &    76.4010838847 &    199.069 &  80\\
I.A$^{i.c.}_{43}$ &     0.4157753114 &     0.2981866614 &    88.0489003051 &    197.691 &  80\\
I.A$^{i.c.}_{44}$ &     0.4188475683 &     0.2770588072 &    90.4212103693 &    208.148 &  84\\
I.A$^{i.c.}_{45}$ &     0.3960577146 &     0.1586009694 &    76.5926988984 &    209.198 &  84\\
I.A$^{i.c.}_{46}$ &     0.4096220179 &     0.2516961122 &    89.9866883856 &    218.506 &  88\\
I.A$^{i.c.}_{47}$ &     0.4046803380 &     0.2225972157 &    90.1813477230 &    228.773 &  92\\

\hline
\end{tabular*}
{\rule{\temptablewidth}{1pt}}
\end{center}
\end{table*}

\begin{table*}
\tabcolsep 0pt \caption{Initial conditions and periods $T$ of the additional periodic three-body orbits for class I.A in case of the isosceles collinear configurations:  $\bm{r}_1(0)=(-1,0)=-\bm{r}_2(0)$,  $\dot{\bm{r}}_1(0)=(v_1,v_2)=\dot{\bm{r}}_2(0)$ and $\bm{r}_3(0)=(0,0)$, $\dot{\bm{r}}_3(0)=(-2v_1, -2v_2)$ when $G=1$ and $m_1=m_2=m_3=1$ by means of the search grid $2000\times 2000$ in the interval $T \in [0,200]$, where $T^*=T |E|^{3/2}$ is its scale-invariant period, $L_f$ is the length of the free group element.  Here, the superscript {\em i.c.} indicates the case of the initial conditions with {\em isosceles collinear}  configuration, due to the fact that there exist periodic orbits in many other cases.} \label{table-S5} \vspace*{-12pt}
\begin{center}
\def\temptablewidth{1\textwidth}
{\rule{\temptablewidth}{1pt}}
\begin{tabular*}{\temptablewidth}{@{\extracolsep{\fill}}lccccc}
\hline
Class and number  & $v_1$ & $v_2$  & $T$  & $T^*$ & $L_f$\\
\hline
I.A$^{i.c.}_{48}$ &     0.2744365569 &     0.3448729405 &    37.0281899697 &     98.299 &  40\\
I.A$^{i.c.}_{49}$ &     0.5196830772 &     0.2059463323 &    61.0651489089 &    119.273 &  48\\
I.A$^{i.c.}_{50}$ &     0.0481112024 &     0.5584284134 &    61.5066511920 &    119.557 &  52\\
I.A$^{i.c.}_{51}$ &     0.3240593189 &     0.3665712251 &    61.6980804867 &    146.748 &  60\\
I.A$^{i.c.}_{52}$ &     0.4392326535 &     0.4490130778 &   107.1704322355 &    161.864 &  68\\
I.A$^{i.c.}_{53}$ &     0.4157392818 &     0.3443683086 &    90.2692922120 &    187.114 &  76\\
I.A$^{i.c.}_{54}$ &     0.2591633491 &     0.5655538324 &   112.9677142326 &    175.025 &  76\\
I.A$^{i.c.}_{55}$ &     0.4372549674 &     0.4568896400 &   121.6765447969 &    180.390 &  76\\
I.A$^{i.c.}_{56}$ &     0.4713347232 &     0.3856148608 &   112.7807985048 &    184.312 &  76\\
I.A$^{i.c.}_{57}$ &     0.1538860573 &     0.5805527776 &   114.4056830978 &    193.146 &  84\\
I.A$^{i.c.}_{58}$ &     0.4162411626 &     0.4666456940 &   130.1232036082 &    198.902 &  84\\
I.A$^{i.c.}_{59}$ &     0.4602603277 &     0.4047836526 &   126.1735921874 &    202.975 &  84\\
I.A$^{i.c.}_{60}$ &     0.4260682804 &     0.4689275280 &   134.8572983004 &    198.903 &  84\\
I.A$^{i.c.}_{61}$ &     0.4115604103 &     0.3439515057 &   103.4777852547 &    216.720 &  88\\
I.A$^{i.c.}_{62}$ &     0.4538585041 &     0.4151335211 &   138.9526711783 &    221.605 &  92\\
I.A$^{i.c.}_{63}$ &     0.1807391308 &     0.5804273075 &   128.9311229083 &    211.590 &  92\\
I.A$^{i.c.}_{64}$ &     0.0616295799 &     0.5887684684 &   121.2588070481 &    211.428 &  92\\
I.A$^{i.c.}_{65}$ &     0.4164511851 &     0.2985460701 &   101.4417348354 &    227.296 &  92\\
I.A$^{i.c.}_{66}$ &     0.4755802612 &     0.3834289883 &   143.4920831622 &    232.725 &  96\\
I.A$^{i.c.}_{67}$ &     0.4189618506 &     0.3575723235 &   117.5497551541 &    235.641 &  96\\
I.A$^{i.c.}_{68}$ &     0.3998065241 &     0.1898070007 &    90.3562840799 &    238.958 &  96\\
I.A$^{i.c.}_{69}$ &     0.1938115742 &     0.5776286093 &   140.9274473884 &    230.040 &  100\\
I.A$^{i.c.}_{70}$ &     0.4311833725 &     0.4733759409 &   164.8353415245 &    235.912 &  100\\
I.A$^{i.c.}_{71}$ &     0.0960540551 &     0.5889671026 &   134.1667723286 &    229.832 &  100\\
I.A$^{i.c.}_{72}$ &     0.3952890968 &     0.1508523182 &    90.5619718006 &    249.067 &  100\\
I.A$^{i.c.}_{73}$ &     0.4395870869 &     0.4502293684 &   164.6132228870 &    247.429 &  104\\
I.A$^{i.c.}_{74}$ &     0.3674149486 &     0.3838228297 &   119.7236640531 &    254.456 &  104\\
I.A$^{i.c.}_{75}$ &     0.3939961214 &     0.3865732823 &   127.4002458975 &    254.459 &  104\\
I.A$^{i.c.}_{76}$ &     0.3430511121 &     0.3718832371 &   111.6289843304 &    254.460 &  104\\
I.A$^{i.c.}_{77}$ &     0.4159559963 &     0.2988672319 &   114.5882843578 &    256.902 &  104\\
I.A$^{i.c.}_{78}$ &     0.4069867693 &     0.2365548555 &   103.9566166835 &    258.453 &  104\\
I.A$^{i.c.}_{79}$ &     0.4589974678 &     0.4050034288 &   155.7660607190 &    251.388 &  104\\
I.A$^{i.c.}_{80}$ &     0.4139140441 &     0.2867091579 &   116.5535148563 &    267.382 &  108\\
I.A$^{i.c.}_{81}$ &     0.4379684294 &     0.4556963530 &   179.1036444528 &    265.955 &  112\\
I.A$^{i.c.}_{82}$ &     0.4126145620 &     0.3435200139 &   131.9570363471 &    275.931 &  112\\
I.A$^{i.c.}_{83}$ &     0.4561844518 &     0.4117892184 &   168.9508799790 &    270.024 &  112\\
I.A$^{i.c.}_{84}$ &     0.3985475384 &     0.1801838327 &   104.3138748764 &    278.839 &  112\\
I.A$^{i.c.}_{85}$ &     0.4126594494 &     0.2674571676 &   117.5120633437 &    277.786 &  112\\
I.A$^{i.c.}_{86}$ &     0.3921967782 &     0.3732670713 &   137.7155574878 &    284.106 &  116\\
I.A$^{i.c.}_{87}$ &     0.4180525186 &     0.2985930530 &   128.3273660780 &    286.508 &  116\\
I.A$^{i.c.}_{88}$ &     0.1382080826 &     0.5834341742 &   157.3466528910 &    266.675 &  116\\
I.A$^{i.c.}_{89}$ &     0.3974576989 &     0.2491169363 &   115.4448017793 &    288.116 &  116\\
I.A$^{i.c.}_{90}$ &     0.0188646524 &     0.5900765061 &   151.9903843649 &    266.578 &  116\\
I.A$^{i.c.}_{91}$ &     0.3947500473 &     0.1449169673 &   104.5346721644 &    288.934 &  116\\
I.A$^{i.c.}_{92}$ &     0.4146876683 &     0.2893277049 &   130.1851727452 &    296.995 &  120\\
I.A$^{i.c.}_{93}$ &     0.4049707755 &     0.2243966352 &   117.9157508493 &    298.378 &  120\\
I.A$^{i.c.}_{94}$ &     0.4948547469 &     0.3816681253 &   195.8237684806 &    299.800 &  124\\
I.A$^{i.c.}_{95}$ &     0.4098282065 &     0.4859736518 &   199.4325323563 &    291.389 &  124\\
I.A$^{i.c.}_{96}$ &     0.3557656112 &     0.2101434426 &   110.1033827881 &    308.577 &  124\\
I.A$^{i.c.}_{97}$ &     0.1458080611 &     0.5760627737 &   164.8777227366 &    285.109 &  124\\
I.A$^{i.c.}_{98}$ &     0.4373040972 &     0.1890148898 &   125.7688001626 &    308.577 &  124\\
I.A$^{i.c.}_{99}$ &     0.3693848766 &     0.2787954099 &   121.4331832032 &    307.415 &  124\\
I.A$^{i.c.}_{100}$ &     0.0670760777 &     0.5889627892 &   163.9101889958 &    284.971 &  124\\
\hline
\end{tabular*}
{\rule{\temptablewidth}{1pt}}
\end{center}
\end{table*}

\begin{table*}
\tabcolsep 0pt \caption{Initial conditions and periods $T$ of the additional periodic three-body orbits for class I.A in case of the isosceles collinear configurations:  $\bm{r}_1(0)=(-1,0)=-\bm{r}_2(0)$,  $\dot{\bm{r}}_1(0)=(v_1,v_2)=\dot{\bm{r}}_2(0)$ and $\bm{r}_3(0)=(0,0)$, $\dot{\bm{r}}_3(0)=(-2v_1, -2v_2)$ when $G=1$ and $m_1=m_2=m_3=1$ by means of the search grid $2000\times 2000$ in the interval $T \in [0,200]$, where $T^*=T |E|^{3/2}$ is its scale-invariant period, $L_f$ is the length of the free group element.  Here, the superscript {\em i.c.} indicates the case of the initial conditions with {\em isosceles collinear}  configuration, due to the fact that there exist periodic orbits in many other cases.} \label{table-S5} \vspace*{-12pt}
\begin{center}
\def\temptablewidth{1\textwidth}
{\rule{\temptablewidth}{1pt}}
\begin{tabular*}{\temptablewidth}{@{\extracolsep{\fill}}lccccc}
\hline
Class and number  & $v_1$ & $v_2$  & $T$  & $T^*$ & $L_f$\\
\hline
I.A$^{i.c.}_{101}$ &     0.4024084539 &     0.3670928188 &   153.5887772816 &    313.740 &  128\\
I.A$^{i.c.}_{102}$ &     0.4101448519 &     0.2545904027 &   131.4852184733 &    317.767 &  128\\
I.A$^{i.c.}_{103}$ &     0.3976053675 &     0.1724965035 &   118.2740940483 &    318.716 &  128\\
I.A$^{i.c.}_{104}$ &     0.4070387234 &     0.3507052842 &   155.3531711573 &    324.480 &  132\\
I.A$^{i.c.}_{105}$ &     0.3654211421 &     0.1497581124 &   113.5033068728 &    328.800 &  132\\
I.A$^{i.c.}_{106}$ &     0.3288064734 &     0.2480300294 &   116.7651488382 &    328.060 &  132\\
I.A$^{i.c.}_{107}$ &     0.0921822548 &     0.5888237645 &   176.5993385200 &    303.375 &  132\\
I.A$^{i.c.}_{108}$ &     0.3943552252 &     0.1402131694 &   118.5101674005 &    328.800 &  132\\
I.A$^{i.c.}_{109}$ &     0.4164334853 &     0.2763211125 &   145.5011286545 &    337.039 &  136\\
I.A$^{i.c.}_{110}$ &     0.4033838433 &     0.2143907433 &   131.8709942386 &    338.289 &  136\\
I.A$^{i.c.}_{111}$ &     0.1818464802 &     0.5795638348 &   195.8266870843 &    321.997 &  140\\
I.A$^{i.c.}_{112}$ &     0.4109222799 &     0.3490743143 &   170.4845107382 &    354.091 &  144\\
I.A$^{i.c.}_{113}$ &     0.4159829843 &     0.2928634271 &   157.4236701613 &    356.215 &  144\\
I.A$^{i.c.}_{114}$ &     0.3968779003 &     0.1661961334 &   132.2368607224 &    358.590 &  144\\
I.A$^{i.c.}_{115}$ &     0.3369172422 &     0.2901238678 &   139.2379425543 &    366.661 &  148\\
I.A$^{i.c.}_{116}$ &     0.4786094582 &     0.2027583452 &   167.5764021354 &    367.991 &  148\\
I.A$^{i.c.}_{117}$ &     0.1262212709 &     0.5844212753 &   199.4597414280 &    340.210 &  148\\
I.A$^{i.c.}_{118}$ &     0.3596345346 &     0.3768544947 &   165.4459242130 &    362.173 &  148\\
I.A$^{i.c.}_{119}$ &     0.3735693647 &     0.3814544097 &   171.6871768580 &    362.171 &  148\\
I.A$^{i.c.}_{120}$ &     0.3940559464 &     0.1363870337 &   132.4879124177 &    368.666 &  148\\
I.A$^{i.c.}_{121}$ &     0.0398375597 &     0.5900216545 &   194.6214443391 &    340.118 &  148\\
I.A$^{i.c.}_{122}$ &     0.3366050958 &     0.1545246545 &   122.1507473925 &    368.667 &  148\\
I.A$^{i.c.}_{123}$ &     0.4148912516 &     0.3594943116 &   185.0089900186 &    372.989 &  152\\
I.A$^{i.c.}_{124}$ &     0.4021066869 &     0.2059880758 &   145.8250755871 &    378.190 &  152\\
I.A$^{i.c.}_{125}$ &     0.2984912721 &     0.2642719294 &   134.6157747992 &    387.392 &  156\\
I.A$^{i.c.}_{126}$ &     0.3151887676 &     0.2026092096 &   129.5701866405 &    388.355 &  156\\
I.A$^{i.c.}_{127}$ &     0.3682819786 &     0.0540125357 &   129.1627539603 &    388.682 &  156\\
I.A$^{i.c.}_{128}$ &     0.4094064944 &     0.2504917575 &   159.2238172130 &    387.380 &  156\\
I.A$^{i.c.}_{129}$ &     0.3963021063 &     0.1609272065 &   146.2019999818 &    398.462 &  160\\
I.A$^{i.c.}_{130}$ &     0.4858230168 &     0.1897852046 &   186.6816128151 &    407.911 &  164\\
I.A$^{i.c.}_{131}$ &     0.3938225013 &     0.1332100629 &   146.4674258412 &    408.531 &  164\\
I.A$^{i.c.}_{132}$ &     0.2844202818 &     0.2576751665 &   144.7361096869 &    427.351 &  172\\
I.A$^{i.c.}_{133}$ &     0.3613650149 &     0.0662137880 &   141.3166709722 &    428.547 &  172\\
I.A$^{i.c.}_{134}$ &     0.4052729644 &     0.2262551267 &   173.3843818847 &    437.587 &  176\\
I.A$^{i.c.}_{135}$ &     0.2792503345 &     0.2309398618 &   146.5213559661 &    447.823 &  180\\
I.A$^{i.c.}_{136}$ &     0.4088863339 &     0.2475714173 &   186.9603460122 &    456.990 &  184\\
I.A$^{i.c.}_{137}$ &     0.2755037979 &     0.2512907499 &   155.4548586490 &    467.296 &  188\\
I.A$^{i.c.}_{138}$ &     0.3000356874 &     0.1954805934 &   152.1599660045 &    468.121 &  188\\
I.A$^{i.c.}_{139}$ &     0.2759217512 &     0.2671023857 &   164.8956499133 &    486.678 &  196\\
I.A$^{i.c.}_{140}$ &     0.3060236507 &     0.1574877923 &   155.4631165813 &    488.269 &  196\\
I.A$^{i.c.}_{141}$ &     0.2738106556 &     0.2266665399 &   157.9002511121 &    487.729 &  196\\
I.A$^{i.c.}_{142}$ &     0.2716635736 &     0.2803602873 &   173.2990630202 &    505.979 &  204\\
I.A$^{i.c.}_{143}$ &     0.2690738475 &     0.2460748830 &   166.5414191753 &    507.230 &  204\\
I.A$^{i.c.}_{144}$ &     0.2813047676 &     0.2931043691 &   185.0094674080 &    525.199 &  212\\
I.A$^{i.c.}_{145}$ &     0.2844033114 &     0.2864772055 &   184.1463025888 &    525.201 &  212\\
I.A$^{i.c.}_{146}$ &     0.3406159593 &     0.0878815592 &   176.4815564147 &    548.142 &  220\\
I.A$^{i.c.}_{147}$ &     0.2745076841 &     0.2185259293 &   182.4778598891 &    567.527 &  228\\
\hline
\end{tabular*}
{\rule{\temptablewidth}{1pt}}
\end{center}
\end{table*}

\begin{table*}
\tabcolsep 0pt \caption{Initial conditions and periods $T$ of the additional periodic three-body orbits for class I.A in case of the isosceles collinear configurations:  $\bm{r}_1(0)=(-1,0)=-\bm{r}_2(0)$,  $\dot{\bm{r}}_1(0)=(v_1,v_2)=\dot{\bm{r}}_2(0)$ and $\bm{r}_3(0)=(0,0)$, $\dot{\bm{r}}_3(0)=(-2v_1, -2v_2)$ when $G=1$ and $m_1=m_2=m_3=1$ by means of the search grid $4000\times 4000$ in the interval $T \in [0,200]$, where $T^*=T |E|^{3/2}$ is its scale-invariant period, $L_f$ is the length of the free group element.  Here, the superscript {\em i.c.} indicates the case of the initial conditions with {\em isosceles collinear}  configuration, due to the fact that there exist periodic orbits in many other cases.} \label{table-S5} \vspace*{-12pt}
\begin{center}
\def\temptablewidth{1\textwidth}
{\rule{\temptablewidth}{1pt}}
\begin{tabular*}{\temptablewidth}{@{\extracolsep{\fill}}lccccc}
\hline
Class and number  & $v_1$ & $v_2$  & $T$  & $T^*$ & $L_f$\\
\hline
I.A$^{i.c.}_{148}$ &     0.5236560165 &     0.2615293843 &    77.5451687834 &    138.512 &  56\\
I.A$^{i.c.}_{149}$ &     0.5252433392 &     0.2395323622 &    97.1790201428 &    178.571 &  72\\
I.A$^{i.c.}_{150}$ &     0.2485711669 &     0.2285602706 &    62.8143378441 &    199.118 &  80\\
I.A$^{i.c.}_{151}$ &     0.4717436769 &     0.4159399383 &   127.4097749092 &    191.768 &  80\\
I.A$^{i.c.}_{152}$ &     0.4244996083 &     0.4906100936 &   136.0765222404 &    187.283 &  80\\
I.A$^{i.c.}_{153}$ &     0.3303010815 &     0.4704859697 &   107.3481952603 &    198.916 &  84\\
I.A$^{i.c.}_{154}$ &     0.2538439618 &     0.5599410925 &   121.1861154175 &    193.495 &  84\\
I.A$^{i.c.}_{155}$ &     0.4824594978 &     0.3739325421 &   131.7096541850 &    214.035 &  88\\
I.A$^{i.c.}_{156}$ &     0.4299480127 &     0.4772693278 &   179.4317919632 &    254.406 &  108\\
I.A$^{i.c.}_{157}$ &     0.2242029652 &     0.5716643145 &   178.2257295831 &    285.417 &  124\\
I.A$^{i.c.}_{158}$ &     0.4721688158 &     0.3825825184 &   189.2718322389 &    310.866 &  128\\
I.A$^{i.c.}_{159}$ &     0.2338958736 &     0.5689557210 &   190.6014315264 &    303.881 &  132\\
I.A$^{i.c.}_{160}$ &     0.4067612870 &     0.2352215102 &   131.6910976773 &    328.058 &  132\\
I.A$^{i.c.}_{161}$ &     0.4664129400 &     0.2183545493 &   147.4412391934 &    328.060 &  132\\
I.A$^{i.c.}_{162}$ &     0.1082811863 &     0.5839631505 &   185.8736775605 &    321.788 &  140\\
I.A$^{i.c.}_{163}$ &     0.3733531828 &     0.0372160210 &   116.4750673801 &    348.816 &  140\\
I.A$^{i.c.}_{164}$ &     0.3289795673 &     0.2068164036 &   118.9835822292 &    348.468 &  140\\
I.A$^{i.c.}_{165}$ &     0.4000579116 &     0.1916547891 &   132.0453999094 &    348.465 &  140\\
I.A$^{i.c.}_{166}$ &     0.4122855598 &     0.2657302751 &   159.0067612686 &    377.047 &  152\\
I.A$^{i.c.}_{167}$ &     0.3897682267 &     0.3744628158 &   189.3994398298 &    391.821 &  160\\
I.A$^{i.c.}_{168}$ &     0.3218496119 &     0.1559752782 &   132.7019091118 &    408.535 &  164\\
I.A$^{i.c.}_{169}$ &     0.2959833070 &     0.2812516245 &   143.8012840531 &    406.693 &  164\\
I.A$^{i.c.}_{170}$ &     0.4132174699 &     0.2698100668 &   172.7929755123 &    406.678 &  164\\
I.A$^{i.c.}_{171}$ &     0.4010600790 &     0.1988153634 &   159.7793954306 &    418.084 &  168\\
I.A$^{i.c.}_{172}$ &     0.3983719278 &     0.1787862095 &   159.9605934536 &    428.228 &  172\\
I.A$^{i.c.}_{173}$ &     0.4134354098 &     0.2854601690 &   185.1176308487 &    425.893 &  172\\
I.A$^{i.c.}_{174}$ &     0.4920112032 &     0.1412368522 &   190.8552755255 &    428.247 &  172\\
I.A$^{i.c.}_{175}$ &     0.3958370784 &     0.1564483425 &   160.1693147940 &    438.332 &  176\\
I.A$^{i.c.}_{176}$ &     0.2816607693 &     0.2752504513 &   153.9063858244 &    446.695 &  180\\
I.A$^{i.c.}_{177}$ &     0.4155517978 &     0.2960847052 &   197.4708451315 &    445.039 &  180\\
I.A$^{i.c.}_{178}$ &     0.3124918736 &     0.1567548343 &   143.9064578456 &    448.402 &  180\\
I.A$^{i.c.}_{179}$ &     0.3936357530 &     0.1305277271 &   160.4482656408 &    448.395 &  180\\
I.A$^{i.c.}_{180}$ &     0.3977453372 &     0.1736695550 &   173.9208149453 &    468.105 &  188\\
I.A$^{i.c.}_{181}$ &     0.4065107464 &     0.2337325768 &   187.1598824391 &    467.267 &  188\\
I.A$^{i.c.}_{182}$ &     0.3934827897 &     0.1282315783 &   174.4300043958 &    488.259 &  196\\
I.A$^{i.c.}_{183}$ &     0.4514746238 &     0.0973655800 &   192.4707838219 &    488.269 &  196\\
I.A$^{i.c.}_{184}$ &     0.3994542144 &     0.1871787471 &   187.6908780809 &    497.857 &  200\\
I.A$^{i.c.}_{185}$ &     0.3466891157 &     0.0826041253 &   164.7857262001 &    508.277 &  204\\
I.A$^{i.c.}_{186}$ &     0.3951370317 &     0.1492265626 &   188.1097039559 &    518.068 &  208\\
I.A$^{i.c.}_{187}$ &     0.2704861637 &     0.2617400496 &   176.1949991668 &    526.644 &  212\\
I.A$^{i.c.}_{188}$ &     0.2701991673 &     0.2977922273 &   183.6255697046 &    525.201 &  212\\
I.A$^{i.c.}_{189}$ &     0.2719181145 &     0.2227169845 &   169.8054736884 &    527.630 &  212\\
I.A$^{i.c.}_{190}$ &     0.3933545122 &     0.1262431781 &   188.4122024989 &    528.123 &  212\\
\hline
\end{tabular*}
{\rule{\temptablewidth}{1pt}}
\end{center}
\end{table*}

\begin{table*}
\tabcolsep 0pt \caption{Initial conditions and periods  $T$ of the periodic three-body orbits for class I.B and II.B  in case of the isosceles collinear configurations:  $\bm{r}_1(0)=(-1,0)=-\bm{r}_2(0)$,  $\dot{\bm{r}}_1(0)=(v_1,v_2)=\dot{\bm{r}}_2(0)$ and $\bm{r}_3(0)=(0,0)$, $\dot{\bm{r}}_3(0)=(-2v_1, -2v_2)$ when $G=1$ and $m_1=m_2=m_3=1$ by means of the search grid $1000\times 1000$ in the interval $T\in[0,100]$, where $T^*=T |E|^{3/2}$ is its scale-invariant period, $L_f$ is the length of the free group element.   Here, the superscript {\em i.c.} indicates the case of the initial conditions with {\em isosceles collinear}  configuration, due to the fact that there exist periodic orbits in many other cases.} \label{table-S6} \vspace*{-12pt}
\begin{center}
\def\temptablewidth{1\textwidth}
{\rule{\temptablewidth}{1pt}}
\begin{tabular*}{\temptablewidth}{@{\extracolsep{\fill}}lccccc}
\hline
Class and number  & $v_1$ & $v_2$  & $T$ & $T^*$ & $L_f$ \\
\hline
I.B$^{i.c.}_{1}$ &     0.4644451728 &     0.3960600146 &    14.8943051743 &     24.206 &  10\\
I.B$^{i.c.}_{2}$ &     0.4059155671 &     0.2301631260 &    13.8671234361 &     34.802 &  14\\
I.B$^{i.c.}_{3}$ &     0.0833000718 &     0.1278892555 &    10.4648495256 &     39.644 &  16\\
I.B$^{i.c.}_{4}$ &     0.0805842255 &     0.5888360898 &    21.2723373956 &     36.771 &  16\\
I.B$^{i.c.}_{5}$ &     0.4391659182 &     0.4529676431 &    28.6692709402 &     42.782 &  18\\
I.B$^{i.c.}_{6}$ &     0.3834435199 &     0.3773636946 &    25.8392363356 &     53.857 &  22\\
I.B$^{i.c.}_{7}$ &     0.1862378160 &     0.5787138661 &    33.6414187604 &     55.204 &  24\\
I.B$^{i.c.}_{8}$ &     0.4290898149 &     0.4753132903 &    42.8299661050 &     61.290 &  26\\
I.B$^{i.c.}_{9}$ &     0.4149129608 &     0.2746187551 &    27.6646471048 &     64.445 &  26\\
I.B$^{i.c.}_{10}$ &     0.3980444335 &     0.1761383334 &    27.8234143343 &     74.695 &  30\\
I.B$^{i.c.}_{11}$ &     0.2302567240 &     0.5696545030 &    46.0668997027 &     73.662 &  32\\
I.B$^{i.c.}_{12}$ &     0.4281877764 &     0.3550351874 &    42.3641397966 &     83.484 &  34\\
I.B$^{i.c.}_{13}$ &     0.3580870041 &     0.4976954491 &    49.6315066172 &     79.780 &  34\\
I.B$^{i.c.}_{14}$ &     0.4097171234 &     0.4392219554 &    54.0572550607 &     91.244 &  38\\
I.B$^{i.c.}_{15}$ &     0.4143481483 &     0.2881031095 &    41.1260553556 &     94.063 &  38\\
I.B$^{i.c.}_{16}$ &     0.2590629830 &     0.5643154612 &    59.1751946306 &     92.130 &  40\\
I.B$^{i.c.}_{17}$ &     0.3956370780 &     0.1544499871 &    41.7884305464 &    114.567 &  46\\
I.B$^{i.c.}_{18}$ &     0.4092519851 &     0.3498343142 &    54.3261092062 &    113.095 &  46\\
I.B$^{i.c.}_{19}$ &     0.2776121408 &     0.5600955198 &    72.2975181708 &    110.601 &  48\\
I.B$^{i.c.}_{20}$ &     0.4174078020 &     0.2932839393 &    54.8606819583 &    123.673 &  50\\
I.B$^{i.c.}_{21}$ &     0.4110550483 &     0.4029989043 &    71.3849893061 &    131.914 &  54\\
I.B$^{i.c.}_{22}$ &     0.4099321826 &     0.2534182814 &    55.3680130042 &    134.068 &  54\\
I.B$^{i.c.}_{23}$ &     0.1287022901 &     0.5842440794 &    75.5738965721 &    128.730 &  56\\
I.B$^{i.c.}_{24}$ &     0.4019178915 &     0.2047154472 &    55.5565256091 &    144.315 &  58\\
I.B$^{i.c.}_{25}$ &     0.4532653563 &     0.4161260139 &    87.6578887763 &    139.667 &  58\\
I.B$^{i.c.}_{26}$ &     0.4156793762 &     0.2964031093 &    68.0656402822 &    153.281 &  62\\
I.B$^{i.c.}_{27}$ &     0.4805889690 &     0.3753996716 &    92.5604403814 &    150.759 &  62\\
I.B$^{i.c.}_{28}$ &     0.4282751070 &     0.4640165419 &    98.6494595687 &    146.864 &  62\\
I.B$^{i.c.}_{29}$ &     0.1551825144 &     0.5792904049 &    86.8760972509 &    147.164 &  64\\
I.B$^{i.c.}_{30}$ &     0.3388749658 &     0.2763529104 &    61.2294979456 &    163.709 &  66\\
I.B$^{i.c.}_{31}$ &     0.4124421756 &     0.2664662522 &    69.1294839452 &    163.708 &  66\\
I.B$^{i.c.}_{32}$ &     0.3751487886 &     0.0241612662 &    54.9741990396 &    164.442 &  66\\
I.B$^{i.c.}_{33}$ &     0.4125928452 &     0.3449403658 &    82.6214161579 &    172.311 &  70\\
I.B$^{i.c.}_{34}$ &     0.4757948431 &     0.1603050376 &    79.9971544120 &    184.207 &  74\\
I.B$^{i.c.}_{35}$ &     0.4141289440 &     0.2980760641 &    81.1505833629 &    182.888 &  74\\
I.B$^{i.c.}_{36}$ &     0.3212269533 &     0.2046903079 &    62.0875306826 &    184.206 &  74\\
I.B$^{i.c.}_{37}$ &     0.4039945439 &     0.3662946692 &    93.7855168318 &    191.207 &  78\\
I.B$^{i.c.}_{38}$ &     0.3939323748 &     0.1347283909 &    69.7387381695 &    194.299 &  78\\
I.B$^{i.c.}_{39}$ &     0.2904226486 &     0.2609195136 &    69.7665857553 &    203.687 &  82\\
I.B$^{i.c.}_{40}$ &     0.4942684807 &     0.2201949355 &    98.6353340741 &    203.687 &  82\\
I.B$^{i.c.}_{41}$ &     0.3650027028 &     0.0605128530 &    67.6506508279 &    204.307 &  82\\
I.B$^{i.c.}_{42}$ &     0.4160188002 &     0.2984267206 &    94.7278887029 &    212.494 &  86\\
I.B$^{i.c.}_{43}$ &     0.4032437210 &     0.2134855934 &    83.2915115056 &    213.925 &  86\\
I.B$^{i.c.}_{44}$ &     0.2995058112 &     0.2924056736 &    80.3687190155 &    222.964 &  90\\
I.B$^{i.c.}_{45}$ &     0.4954934838 &     0.1360770719 &   100.4005086563 &    224.095 &  90\\
I.B$^{i.c.}_{46}$ &     0.3029165156 &     0.1970668629 &    73.2064153230 &    224.090 &  90\\
I.B$^{i.c.}_{47}$ &     0.4105107221 &     0.2565879477 &    96.8662425654 &    233.329 &  94\\
I.B$^{i.c.}_{48}$ &     0.3935556862 &     0.1293368487 &    83.7195247407 &    234.164 &  94\\
I.B$^{i.c.}_{49}$ &     0.3501125857 &     0.0793401820 &    79.4749538995 &    244.172 &  98\\
I.B$^{i.c.}_{50}$ &     0.4013534478 &     0.2008553668 &    97.2456382734 &    253.823 &  102\\
I.B$^{i.c.}_{51}$ &     0.2666293299 &     0.2779915428 &    89.2809106554 &    262.991 &  106\\
I.B$^{i.c.}_{52}$ &     0.2935043434 &     0.1914491983 &    84.8189274623 &    263.969 &  106\\
I.B$^{i.c.}_{53}$ &     0.3969872473 &     0.1671666821 &    97.4321722312 &    263.958 &  106\\
I.B$^{i.c.}_{54}$ &     0.3932974661 &     0.1253456024 &    97.7016625145 &    274.027 &  110\\
I.B$^{i.c.}_{55}$ &     0.4030477277 &     0.0469627166 &    99.9681279479 &    284.035 &  114\\
II.B$^{i.c.}_{1}$ &     0.3962186234 &     0.5086826315 &    96.4358796119 &    135.220 &  58\\
II.B$^{i.c.}_{2}$ &     0.2517755120 &     0.2667371937 &    99.8166259190 &    302.976 &  122\\
\hline
\end{tabular*}
{\rule{\temptablewidth}{1pt}}
\end{center}
\end{table*}

\begin{table*}
\tabcolsep 0pt \caption{Initial conditions and periods  $T$ of the additional periodic three-body orbits for class I.B in case of the isosceles collinear configurations:  $\bm{r}_1(0)=(-1,0)=-\bm{r}_2(0)$,  $\dot{\bm{r}}_1(0)=(v_1,v_2)=\dot{\bm{r}}_2(0)$ and $\bm{r}_3(0)=(0,0)$, $\dot{\bm{r}}_3(0)=(-2v_1, -2v_2)$ when $G=1$ and $m_1=m_2=m_3=1$ by means of the search grid $2000\times 2000$ in the interval $T \in [0,200]$, where $T^*=T |E|^{3/2}$ is its scale-invariant period, $L_f$ is the length of the free group element.  Here, the superscript {\em i.c.} indicates the case of the initial conditions with {\em isosceles collinear}  configuration, due to the fact that there exist periodic orbits in many other cases.} \label{table-S6} \vspace*{-12pt}
\begin{center}
\def\temptablewidth{1\textwidth}
{\rule{\temptablewidth}{1pt}}
\begin{tabular*}{\temptablewidth}{@{\extracolsep{\fill}}lccccc}
\hline
Class and number  & $v_1$ & $v_2$  & $T$ & $T^*$ & $L_f$ \\
\hline
I.B$^{i.c.}_{56}$ &     0.4012891619 &     0.4985422646 &    68.5536700625 &     98.262 &  42\\
I.B$^{i.c.}_{57}$ &     0.2840814898 &     0.5556089363 &    83.9820579378 &    129.074 &  56\\
I.B$^{i.c.}_{58}$ &     0.4391638095 &     0.3451673262 &    72.9606345288 &    142.704 &  58\\
I.B$^{i.c.}_{59}$ &     0.3945382374 &     0.1424370298 &    55.7610531376 &    154.434 &  62\\
I.B$^{i.c.}_{60}$ &     0.4812679815 &     0.2803015236 &    93.0193545419 &    182.890 &  74\\
I.B$^{i.c.}_{61}$ &     0.4567265227 &     0.4110980415 &   117.6542984464 &    188.085 &  78\\
I.B$^{i.c.}_{62}$ &     0.1998578918 &     0.5751811878 &   112.6005586813 &    184.064 &  80\\
I.B$^{i.c.}_{63}$ &     0.4731193722 &     0.3817523595 &   121.3718607541 &    199.175 &  82\\
I.B$^{i.c.}_{64}$ &     0.4324559260 &     0.4468425955 &   125.9650168972 &    195.379 &  82\\
I.B$^{i.c.}_{65}$ &     0.4329467422 &     0.4287307557 &   126.6430122618 &    206.699 &  86\\
I.B$^{i.c.}_{66}$ &     0.1106525804 &     0.5843723566 &   117.2006330167 &    202.269 &  88\\
I.B$^{i.c.}_{67}$ &     0.2160100270 &     0.5721577857 &   125.2096376354 &    202.524 &  88\\
I.B$^{i.c.}_{68}$ &     0.4139613585 &     0.3600046586 &   109.4116313918 &    220.831 &  90\\
I.B$^{i.c.}_{69}$ &     0.1346825430 &     0.5825466616 &   129.3971645144 &    220.693 &  96\\
I.B$^{i.c.}_{70}$ &     0.4155669785 &     0.2988428103 &   107.8898275282 &    242.099 &  98\\
I.B$^{i.c.}_{71}$ &     0.4601989074 &     0.4074300958 &   148.0286393030 &    236.501 &  98\\
I.B$^{i.c.}_{72}$ &     0.4358120735 &     0.4591466107 &   157.2189925171 &    232.431 &  98\\
I.B$^{i.c.}_{73}$ &     0.4132841481 &     0.2851290367 &   109.6943890164 &    252.574 &  102\\
I.B$^{i.c.}_{74}$ &     0.1522366536 &     0.5809723053 &   141.6339896130 &    239.127 &  104\\
I.B$^{i.c.}_{75}$ &     0.0426815497 &     0.5900333430 &   136.8666399610 &    239.002 &  104\\
I.B$^{i.c.}_{76}$ &     0.4119987609 &     0.2643427092 &   110.6260549697 &    262.969 &  106\\
I.B$^{i.c.}_{77}$ &     0.5118118742 &     0.2366710013 &   136.7985866536 &    262.991 &  106\\
I.B$^{i.c.}_{78}$ &     0.4571979692 &     0.4150480887 &   161.5490315022 &    255.123 &  106\\
I.B$^{i.c.}_{79}$ &     0.4079238316 &     0.2420300626 &   110.8414293129 &    273.286 &  110\\
I.B$^{i.c.}_{80}$ &     0.4161585901 &     0.2988955475 &   121.2499818512 &    271.705 &  110\\
I.B$^{i.c.}_{81}$ &     0.1774072601 &     0.5821912520 &   157.3853633779 &    257.570 &  112\\
I.B$^{i.c.}_{82}$ &     0.5142298286 &     0.2568912491 &   152.2860975586 &    282.212 &  114\\
I.B$^{i.c.}_{83}$ &     0.4039097291 &     0.2177564200 &   111.0262814993 &    283.532 &  114\\
I.B$^{i.c.}_{84}$ &     0.4445769891 &     0.4487033431 &   188.8772873282 &    280.945 &  118\\
I.B$^{i.c.}_{85}$ &     0.3999556835 &     0.1909060542 &   111.2008319293 &    293.712 &  118\\
I.B$^{i.c.}_{86}$ &     0.4133120318 &     0.2701767161 &   124.4101367285 &    292.601 &  118\\
I.B$^{i.c.}_{87}$ &     0.1066778795 &     0.5948382311 &   165.7269258467 &    275.811 &  120\\
I.B$^{i.c.}_{88}$ &     0.4158199333 &     0.2990314521 &   134.3908162915 &    301.310 &  122\\
I.B$^{i.c.}_{89}$ &     0.3985640110 &     0.3698314086 &   145.9028456537 &    298.924 &  122\\
I.B$^{i.c.}_{90}$ &     0.4094842993 &     0.2509272653 &   124.6052699892 &    302.943 &  122\\
I.B$^{i.c.}_{91}$ &     0.5128091369 &     0.2180249116 &   154.2378088693 &    302.976 &  122\\
I.B$^{i.c.}_{92}$ &     0.3962177869 &     0.1601311749 &   111.3973050444 &    303.830 &  122\\
I.B$^{i.c.}_{93}$ &     0.4149780095 &     0.2904029941 &   136.9824352416 &    311.800 &  126\\
I.B$^{i.c.}_{94}$ &     0.4540683260 &     0.4147166692 &   190.2328210125 &    303.544 &  126\\
I.B$^{i.c.}_{95}$ &     0.4035798187 &     0.3517272858 &   147.4168577881 &    309.675 &  126\\
I.B$^{i.c.}_{96}$ &     0.3931011977 &     0.1222707697 &   111.6831379999 &    313.891 &  126\\
I.B$^{i.c.}_{97}$ &     0.1229330287 &     0.5846239911 &   172.1857297973 &    294.229 &  128\\
I.B$^{i.c.}_{98}$ &     0.2584123338 &     0.2319416805 &   103.4515732794 &    323.470 &  130\\
I.B$^{i.c.}_{99}$ &     0.4023586867 &     0.2076743922 &   124.9805591011 &    323.435 &  130\\
I.B$^{i.c.}_{100}$ &     0.3989023581 &     0.0587531871 &   113.4691210308 &    323.898 &  130\\
I.B$^{i.c.}_{101}$ &     0.4367428362 &     0.3606385790 &   172.3247110027 &    328.554 &  134\\
I.B$^{i.c.}_{102}$ &     0.4784306757 &     0.3771895698 &   199.5148149879 &    325.727 &  134\\
I.B$^{i.c.}_{103}$ &     0.4064178832 &     0.3648162353 &   161.5522956612 &    328.554 &  134\\
I.B$^{i.c.}_{104}$ &     0.1395147893 &     0.5832722105 &   184.5789530252 &    312.657 &  136\\
I.B$^{i.c.}_{105}$ &     0.0532000567 &     0.5910456775 &   180.2118985604 &    312.544 &  136\\
I.B$^{i.c.}_{106}$ &     0.4155474891 &     0.2921682541 &   150.5786020101 &    341.411 &  138\\
I.B$^{i.c.}_{107}$ &     0.4075222377 &     0.2396965085 &   138.5765198727 &    342.892 &  138\\
I.B$^{i.c.}_{108}$ &     0.3451744845 &     0.3808525684 &   155.6845054611 &    347.340 &  142\\
I.B$^{i.c.}_{109}$ &     0.4043102769 &     0.2202849054 &   138.7609319540 &    353.138 &  142\\
I.B$^{i.c.}_{110}$ &     0.4218640175 &     0.2770211780 &   153.8442992753 &    351.850 &  142\\
\hline
\end{tabular*}
{\rule{\temptablewidth}{1pt}}
\end{center}
\end{table*}

\begin{table*}
\tabcolsep 0pt \caption{Initial conditions and periods  $T$ of the additional periodic three-body orbits for class I.B, II.A and II.B  in case of the isosceles collinear configurations:  $\bm{r}_1(0)=(-1,0)=-\bm{r}_2(0)$,  $\dot{\bm{r}}_1(0)=(v_1,v_2)=\dot{\bm{r}}_2(0)$ and $\bm{r}_3(0)=(0,0)$, $\dot{\bm{r}}_3(0)=(-2v_1, -2v_2)$ when $G=1$ and $m_1=m_2=m_3=1$ by means of the search grid $2000\times 2000$ in the interval $T \in [0,200]$, where $T^*=T |E|^{3/2}$ is its scale-invariant period, $L_f$ is the length of the free group element.  Here, the superscript {\em i.c.} indicates the case of the initial conditions with {\em isosceles collinear}  configuration, due to the fact that there exist periodic orbits in many other cases. } \label{table-S6} \vspace*{-12pt}
\begin{center}
\def\temptablewidth{1\textwidth}
{\rule{\temptablewidth}{1pt}}
\begin{tabular*}{\temptablewidth}{@{\extracolsep{\fill}}lccccc}
\hline
Class and number  & $v_1$ & $v_2$  & $T$ & $T^*$ & $L_f$ \\
\hline
I.B$^{i.c.}_{111}$ &     0.3929297393 &     0.1198339408 &   125.6607884377 &    353.754 &  142\\
I.B$^{i.c.}_{112}$ &     0.1502500844 &     0.5819772796 &   196.4574240736 &    331.091 &  144\\
I.B$^{i.c.}_{113}$ &     0.2563482201 &     0.2261374828 &   115.3093440018 &    363.373 &  146\\
I.B$^{i.c.}_{114}$ &     0.4122588665 &     0.3610576585 &   177.1374986605 &    358.178 &  146\\
I.B$^{i.c.}_{115}$ &     0.4011284717 &     0.1992930016 &   138.9348056259 &    363.330 &  146\\
I.B$^{i.c.}_{116}$ &     0.3970105843 &     0.0653064641 &   127.2370563374 &    363.762 &  146\\
I.B$^{i.c.}_{117}$ &     0.4123635119 &     0.3484019438 &   177.9654310661 &    368.895 &  150\\
I.B$^{i.c.}_{118}$ &     0.0148838157 &     0.5682653432 &   184.4789881904 &    349.319 &  152\\
I.B$^{i.c.}_{119}$ &     0.2606923122 &     0.2911653502 &   130.7643465810 &    381.509 &  154\\
I.B$^{i.c.}_{120}$ &     0.4178518268 &     0.3580393477 &   193.1285092053 &    387.799 &  158\\
I.B$^{i.c.}_{121}$ &     0.4127521388 &     0.2678683319 &   165.8951639982 &    391.863 &  158\\
I.B$^{i.c.}_{122}$ &     0.4030019190 &     0.2119143431 &   152.7155517041 &    393.045 &  158\\
I.B$^{i.c.}_{123}$ &     0.3927437429 &     0.1178691817 &   139.6277439076 &    393.618 &  158\\
I.B$^{i.c.}_{124}$ &     0.4147362902 &     0.2949369415 &   177.1359109432 &    400.628 &  162\\
I.B$^{i.c.}_{125}$ &     0.3193090588 &     0.1018501668 &   126.8823101525 &    403.634 &  162\\
I.B$^{i.c.}_{126}$ &     0.4001323444 &     0.1921976871 &   152.8899791889 &    403.219 &  162\\
I.B$^{i.c.}_{127}$ &     0.4129154675 &     0.2844468071 &   178.2176770073 &    411.086 &  166\\
I.B$^{i.c.}_{128}$ &     0.3973680105 &     0.1704797689 &   153.0783295506 &    413.348 &  166\\
I.B$^{i.c.}_{129}$ &     0.4072545124 &     0.2381300512 &   166.3113175694 &    412.498 &  166\\
I.B$^{i.c.}_{130}$ &     0.4045775130 &     0.2219569575 &   166.4954609423 &    422.743 &  170\\
I.B$^{i.c.}_{131}$ &     0.3948281582 &     0.1458085489 &   153.3085182072 &    423.435 &  170\\
I.B$^{i.c.}_{132}$ &     0.4108695302 &     0.2585197447 &   179.8624087934 &    431.850 &  174\\
I.B$^{i.c.}_{133}$ &     0.3924554987 &     0.1162941173 &   153.5622330784 &    433.481 &  174\\
I.B$^{i.c.}_{134}$ &     0.4083876444 &     0.2446989126 &   180.0795834157 &    442.161 &  178\\
I.B$^{i.c.}_{135}$ &     0.3952920978 &     0.0730045437 &   155.0210535632 &    443.488 &  178\\
I.B$^{i.c.}_{136}$ &     0.4116924345 &     0.2628122619 &   193.6216330479 &    461.490 &  186\\
I.B$^{i.c.}_{137}$ &     0.3445561301 &     0.2603927098 &   174.5534838128 &    471.819 &  190\\
I.B$^{i.c.}_{138}$ &     0.4010075001 &     0.1984472605 &   180.6239886836 &    472.838 &  190\\
I.B$^{i.c.}_{139}$ &     0.3493341960 &     0.2098620100 &   166.9519921840 &    472.839 &  190\\
I.B$^{i.c.}_{140}$ &     0.3915508594 &     0.1152445699 &   167.3247662290 &    473.344 &  190\\
I.B$^{i.c.}_{141}$ &     0.3986254119 &     0.1807988654 &   180.8044298919 &    482.984 &  194\\
I.B$^{i.c.}_{142}$ &     0.4336920022 &     0.1476782466 &   192.7791207202 &    493.094 &  198\\
I.B$^{i.c.}_{143}$ &     0.3963541512 &     0.1614152259 &   181.0067311093 &    493.093 &  198\\
I.B$^{i.c.}_{144}$ &     0.4024848426 &     0.2085131141 &   194.4045903694 &    502.556 &  202\\
I.B$^{i.c.}_{145}$ &     0.3942996878 &     0.1395218939 &   181.2594734783 &    503.167 &  202\\
I.B$^{i.c.}_{146}$ &     0.3190182619 &     0.2561011852 &   181.2227405768 &    511.775 &  206\\
I.B$^{i.c.}_{147}$ &     0.3321031966 &     0.2075124897 &   175.9799936209 &    512.730 &  206\\
I.B$^{i.c.}_{148}$ &     0.4002334723 &     0.1929322974 &   194.5791562365 &    512.726 &  206\\
I.B$^{i.c.}_{149}$ &     0.3941006143 &     0.1369745839 &   195.2372295764 &    543.032 &  218\\
I.B$^{i.c.}_{150}$ &     0.3135653929 &     0.2823975233 &   199.6963199692 &    550.395 &  222\\
I.B$^{i.c.}_{151}$ &     0.3028596325 &     0.2513678491 &   190.0141857350 &    551.721 &  222\\
I.B$^{i.c.}_{152}$ &     0.3448466385 &     0.1310471303 &   182.8204654697 &    553.071 &  222\\
I.B$^{i.c.}_{153}$ &     0.3016985230 &     0.2653178570 &   199.5506880072 &    571.096 &  230\\
I.B$^{i.c.}_{154}$ &     0.2998118429 &     0.2294251194 &   191.7840057235 &    572.180 &  230\\
I.B$^{i.c.}_{155}$ &     0.3310536467 &     0.1351316515 &   192.5759642513 &    592.936 &  238\\
II.A$^{i.c.}_{1}$ &     0.4239199957 &     0.4890252375 &   121.7417821220 &    168.801 &  72\\
II.A$^{i.c.}_{2}$ &     0.4210931065 &     0.4930983422 &   135.8627697832 &    187.283 &  80\\
II.A$^{i.c.}_{3}$ &     0.3202131331 &     0.3028156498 &   145.3317595261 &    385.829 &  156\\
II.B$^{i.c.}_{3}$ &     0.5091186403 &     0.1142951579 &   120.8830891862 &    263.980 &  106\\
II.B$^{i.c.}_{4}$ &     0.2070614316 &     0.5772071599 &   183.2613461906 &    294.470 &  128\\
II.B$^{i.c.}_{5}$ &     0.2465911029 &     0.2579795310 &   111.2623782794 &    342.935 &  138\\
\hline
\end{tabular*}
{\rule{\temptablewidth}{1pt}}
\end{center}
\end{table*}

\begin{table*}
\tabcolsep 0pt \caption{Initial conditions and periods  $T$ of the additional periodic three-body orbits for class I.B, II.A and II.B  in case of the isosceles collinear configurations:  $\bm{r}_1(0)=(-1,0)=-\bm{r}_2(0)$,  $\dot{\bm{r}}_1(0)=(v_1,v_2)=\dot{\bm{r}}_2(0)$ and $\bm{r}_3(0)=(0,0)$, $\dot{\bm{r}}_3(0)=(-2v_1, -2v_2)$ when $G=1$ and $m_1=m_2=m_3=1$ by means of the search grid $4000\times 4000$ in the interval $T \in [0,200]$, where $T^*=T |E|^{3/2}$ is its scale-invariant period, $L_f$ is the length of the free group element.  Here, the superscript {\em i.c.} indicates the case of the initial conditions with {\em isosceles collinear}  configuration, due to the fact that there exist periodic orbits in many other cases.} \label{table-S6} \vspace*{-12pt}
\begin{center}
\def\temptablewidth{1\textwidth}
{\rule{\temptablewidth}{1pt}}
\begin{tabular*}{\temptablewidth}{@{\extracolsep{\fill}}lccccc}
\hline
Class and number  & $v_1$ & $v_2$  & $T$ & $T^*$ & $L_f$ \\
\hline
I.B$^{i.c.}_{156}$ &     0.3980393062 &     0.5038904789 &    82.2493251383 &    116.742 &  50\\
I.B$^{i.c.}_{157}$ &     0.2952861660 &     0.5534590851 &    97.3499809159 &    147.548 &  64\\
I.B$^{i.c.}_{158}$ &     0.4798380897 &     0.4113090190 &   106.5560380694 &    158.258 &  66\\
I.B$^{i.c.}_{159}$ &     0.4137787743 &     0.3437200842 &   111.0557458846 &    231.523 &  94\\
I.B$^{i.c.}_{160}$ &     0.3682877800 &     0.3998940275 &   116.9320547074 &    239.621 &  98\\
I.B$^{i.c.}_{161}$ &     0.4646477988 &     0.3883460870 &   149.4861493038 &    247.589 &  102\\
I.B$^{i.c.}_{162}$ &     0.4031162525 &     0.4617331826 &   155.9939695150 &    250.938 &  106\\
I.B$^{i.c.}_{163}$ &     0.4238959747 &     0.4719118617 &   170.7096365279 &    250.941 &  106\\
I.B$^{i.c.}_{164}$ &     0.4207783634 &     0.4487423577 &   176.2188318361 &    280.945 &  118\\
I.B$^{i.c.}_{165}$ &     0.4554244658 &     0.4087724748 &   176.4254157280 &    284.916 &  118\\
I.B$^{i.c.}_{166}$ &     0.5090827977 &     0.1065690769 &   138.5002592220 &    303.862 &  122\\
I.B$^{i.c.}_{167}$ &     0.3988858353 &     0.1828353597 &   125.1579036819 &    333.594 &  134\\
I.B$^{i.c.}_{168}$ &     0.4107436209 &     0.2578452663 &   138.3643307774 &    332.590 &  134\\
I.B$^{i.c.}_{169}$ &     0.5114846883 &     0.1521301286 &   172.1153250253 &    363.374 &  146\\
I.B$^{i.c.}_{170}$ &     0.4088301474 &     0.2472647559 &   152.3413800632 &    372.553 &  150\\
I.B$^{i.c.}_{171}$ &     0.3611442582 &     0.3727080944 &   171.2956907056 &    377.002 &  154\\
I.B$^{i.c.}_{172}$ &     0.2815716369 &     0.1833006566 &   120.7209378107 &    383.594 &  154\\
I.B$^{i.c.}_{173}$ &     0.3951862673 &     0.1497568031 &   139.3358032464 &    383.568 &  154\\
I.B$^{i.c.}_{174}$ &     0.4158315105 &     0.3472324558 &   193.3502089464 &    398.504 &  162\\
I.B$^{i.c.}_{175}$ &     0.4137016788 &     0.2715755177 &   179.7068265589 &    421.492 &  170\\
I.B$^{i.c.}_{176}$ &     0.4153946800 &     0.2957299894 &   190.7321992213 &    430.236 &  174\\
I.B$^{i.c.}_{177}$ &     0.3993116206 &     0.1861014901 &   166.8464696440 &    443.103 &  178\\
I.B$^{i.c.}_{178}$ &     0.4138089784 &     0.2864083402 &   191.9755354947 &    440.701 &  178\\
I.B$^{i.c.}_{179}$ &     0.3968143060 &     0.1656275217 &   167.0415888648 &    453.221 &  182\\
I.B$^{i.c.}_{180}$ &     0.4034485563 &     0.2148073995 &   180.4504761328 &    462.653 &  186\\
I.B$^{i.c.}_{181}$ &     0.4093563853 &     0.2502114533 &   193.8423157244 &    471.817 &  190\\
I.B$^{i.c.}_{182}$ &     0.4443735176 &     0.1849048718 &   194.9787589138 &    472.839 &  190\\
I.B$^{i.c.}_{183}$ &     0.3948446667 &     0.0755442635 &   168.9641243034 &    483.351 &  194\\
I.B$^{i.c.}_{184}$ &     0.3520864423 &     0.1736619942 &   169.5279904498 &    493.094 &  198\\
I.B$^{i.c.}_{185}$ &     0.3288872593 &     0.2366097357 &   173.0921124629 &    492.352 &  198\\
I.B$^{i.c.}_{186}$ &     0.3691268433 &     0.1230538137 &   175.3859687881 &    513.207 &  206\\
I.B$^{i.c.}_{187}$ &     0.3945254084 &     0.0775864607 &   182.9237531924 &    523.214 &  210\\
I.B$^{i.c.}_{188}$ &     0.3304190862 &     0.1753063381 &   177.6311331786 &    532.966 &  214\\
I.B$^{i.c.}_{189}$ &     0.3942881027 &     0.0792702101 &   196.8942769930 &    563.077 &  226\\
I.B$^{i.c.}_{190}$ &     0.3165116655 &     0.1757233721 &   187.3222444953 &    572.838 &  230\\
I.B$^{i.c.}_{191}$ &     0.3134547198 &     0.2019372816 &   197.1007090614 &    592.504 &  238\\
II.A$^{i.c.}_{4}$ &     0.0262032843 &     0.6907471490 &    46.5767068651 &     51.302 &  24\\
II.B$^{i.c.}_{6}$ &     0.0699202568 &     0.6190218805 &    95.2884025970 &    147.108 &  64\\
II.B$^{i.c.}_{7}$ &     0.3956548748 &     0.5097716060 &   109.8746683440 &    153.696 &  66\\
II.B$^{i.c.}_{8}$ &     0.3064804236 &     0.5513199378 &   111.1933380875 &    166.023 &  72\\
II.B$^{i.c.}_{9}$ &     0.5122100376 &     0.1633529632 &   155.0269037645 &    323.470 &  130\\
II.B$^{i.c.}_{10}$ &     0.5152772213 &     0.2342147738 &   189.7694426662 &    362.277 &  146\\
\hline
\end{tabular*}
{\rule{\temptablewidth}{1pt}}
\end{center}
\end{table*}

\begin{table*}
\tabcolsep 0pt \caption{Initial conditions and periods $T$ of the periodic three-body orbits for class II.C in case of the isosceles collinear configurations:  $\bm{r}_1(0)=(-1,0)=-\bm{r}_2(0)$,  $\dot{\bm{r}}_1(0)=(v_1,v_2)=\dot{\bm{r}}_2(0)$ and $\bm{r}_3(0)=(0,0)$, $\dot{\bm{r}}_3(0)=(-2v_1, -2v_2)$ when $G=1$ and $m_1=m_2=m_3=1$ by means of the search grid $1000\times 1000$ in the interval $T\in[0,100]$, where $T^*=T |E|^{3/2}$ is its scale-invariant period, $L_f$ is the length of the free group element.   } \label{table-S7} \vspace*{-12pt}
\begin{center}
\def\temptablewidth{1\textwidth}
{\rule{\temptablewidth}{1pt}}
\begin{tabular*}{\temptablewidth}{@{\extracolsep{\fill}}lccccc}
\hline
Class and number  & $v_1$ & $v_2$  & $T$ & $T^*$ & $L_f$ \\
\hline
II.C$^{i.c.}_{1}$ &     0.2827020949 &     0.3272089716 &    10.9633031497 &     29.602 &  12\\
II.C$^{i.c.}_{2}$ &     0.5647061130 &     0.5368389792 &    54.7501910522 &     30.615 &  14\\
II.C$^{i.c.}_{3}$ &     0.5225201635 &     0.2268083872 &    26.3193848826 &     49.643 &  20\\
II.C$^{i.c.}_{4}$ &     0.2016783093 &     0.4098955437 &    21.0205158972 &     53.923 &  22\\
II.C$^{i.c.}_{5}$ &     0.5512729728 &     0.5504821832 &    86.9417019465 &     48.666 &  22\\
II.C$^{i.c.}_{6}$ &     0.5233890828 &     0.3421147815 &    44.9302453896 &     68.687 &  28\\
II.C$^{i.c.}_{7}$ &     0.4737427040 &     0.4312766992 &    46.8946539631 &     67.014 &  28\\
II.C$^{i.c.}_{8}$ &     0.2433845587 &     0.2505166924 &    22.3333787897 &     69.623 &  28\\
II.C$^{i.c.}_{9}$ &     0.1739184309 &     0.1140843875 &    21.7535340992 &     79.380 &  32\\
II.C$^{i.c.}_{10}$ &     0.1448119296 &     0.5428981413 &    38.0630301501 &     73.656 &  32\\
II.C$^{i.c.}_{11}$ &     0.1359037483 &     0.5406786957 &    37.5276119206 &     73.655 &  32\\
II.C$^{i.c.}_{12}$ &     0.5547060649 &     0.2721678509 &    52.9480727303 &     83.484 &  34\\
II.C$^{i.c.}_{13}$ &     0.4458173723 &     0.4924872346 &    62.5459036895 &     79.776 &  34\\
II.C$^{i.c.}_{14}$ &     0.5368578744 &     0.3705761039 &    60.5607353967 &     81.946 &  34\\
II.C$^{i.c.}_{15}$ &     0.5196491506 &     0.3533207146 &    57.9902773954 &     87.485 &  36\\
II.C$^{i.c.}_{16}$ &     0.2449726867 &     0.5690681065 &    52.9395192839 &     82.896 &  36\\
II.C$^{i.c.}_{17}$ &     0.5164575299 &     0.1638309455 &    43.4746638318 &     89.583 &  36\\
II.C$^{i.c.}_{18}$ &     0.4052765310 &     0.4961906801 &    62.2999748903 &     89.021 &  38\\
II.C$^{i.c.}_{19}$ &     0.1868592137 &     0.2044548811 &    27.7253651277 &     94.814 &  38\\
II.C$^{i.c.}_{20}$ &     0.3026158825 &     0.3576846960 &    37.1769633336 &     92.900 &  38\\
II.C$^{i.c.}_{21}$ &     0.5044568709 &     0.3973785294 &    70.8348253915 &    100.524 &  42\\
II.C$^{i.c.}_{22}$ &     0.2661160110 &     0.5520357440 &    62.9864913052 &    101.369 &  44\\
II.C$^{i.c.}_{23}$ &     0.2865945055 &     0.5607550065 &    67.5864018030 &    101.366 &  44\\
II.C$^{i.c.}_{24}$ &     0.0882838085 &     0.3615155443 &    35.8170122339 &    107.796 &  44\\
II.C$^{i.c.}_{25}$ &     0.5159476361 &     0.1444933284 &    52.2126924444 &    109.534 &  44\\
II.C$^{i.c.}_{26}$ &     0.3970546540 &     0.4996659112 &    74.4041641681 &    107.502 &  46\\
II.C$^{i.c.}_{27}$ &     0.4168220336 &     0.3303332949 &    55.7893112814 &    118.396 &  48\\
II.C$^{i.c.}_{28}$ &     0.2663548692 &     0.3359619580 &    43.5341483568 &    118.413 &  48\\
II.C$^{i.c.}_{29}$ &     0.4293246983 &     0.3737393096 &    64.8733105441 &    122.532 &  50\\
II.C$^{i.c.}_{30}$ &     0.2542679242 &     0.2153811120 &    40.5931469775 &    129.480 &  52\\
II.C$^{i.c.}_{31}$ &     0.5059387393 &     0.3894005400 &    86.4180127242 &    124.733 &  52\\
II.C$^{i.c.}_{32}$ &     0.4110200317 &     0.4114629190 &    72.8712103518 &    131.908 &  54\\
II.C$^{i.c.}_{33}$ &     0.4158187604 &     0.3068035415 &    60.1507436615 &    133.196 &  54\\
II.C$^{i.c.}_{34}$ &     0.4480473854 &     0.4596449839 &    96.8391326694 &    137.607 &  58\\
II.C$^{i.c.}_{35}$ &     0.4177015960 &     0.3034552740 &    66.7511112061 &    147.998 &  60\\
II.C$^{i.c.}_{36}$ &     0.2710171846 &     0.5711013679 &    93.0418702843 &    138.097 &  60\\
II.C$^{i.c.}_{37}$ &     0.3499168589 &     0.2508093737 &    56.8351605485 &    154.045 &  62\\
II.C$^{i.c.}_{38}$ &     0.3783890881 &     0.5100736088 &    98.6023217127 &    144.458 &  62\\
II.C$^{i.c.}_{39}$ &     0.4270829576 &     0.3403174863 &    80.0382219001 &    162.799 &  66\\
II.C$^{i.c.}_{40}$ &     0.1703649368 &     0.5803609520 &    94.1583429936 &    156.387 &  68\\
II.C$^{i.c.}_{41}$ &     0.4762205782 &     0.2687092179 &    85.3959959932 &    173.320 &  70\\
II.C$^{i.c.}_{42}$ &     0.4318334213 &     0.1223893080 &    66.8092062353 &    174.367 &  70\\
II.C$^{i.c.}_{43}$ &     0.4146425708 &     0.3012162581 &    79.2835610170 &    177.602 &  72\\
II.C$^{i.c.}_{44}$ &     0.2710036741 &     0.3121152228 &    63.3936161404 &    177.616 &  72\\
II.C$^{i.c.}_{45}$ &     0.4234217026 &     0.2989326748 &    87.2610777514 &    192.404 &  78\\
II.C$^{i.c.}_{46}$ &     0.3282929621 &     0.1554041101 &    63.6556668110 &    194.300 &  78\\
II.C$^{i.c.}_{47}$ &     0.2950615495 &     0.2386718494 &    65.2298654293 &    193.976 &  78\\
II.C$^{i.c.}_{48}$ &     0.4069563704 &     0.3543774452 &    94.7746114120 &    196.579 &  80\\
II.C$^{i.c.}_{49}$ &     0.4182604701 &     0.2994822960 &    92.9809890454 &    207.206 &  84\\
II.C$^{i.c.}_{50}$ &     0.2881254754 &     0.2779750581 &    74.3601049133 &    213.349 &  86\\
II.C$^{i.c.}_{51}$ &     0.2833115180 &     0.2333212957 &    70.5062877151 &    213.934 &  86\\
II.C$^{i.c.}_{52}$ &     0.4473380629 &     0.1056541774 &    84.0347942344 &    214.234 &  86\\
II.C$^{i.c.}_{53}$ &     0.4121687384 &     0.3002886778 &    98.4332144983 &    222.009 &  90\\
II.C$^{i.c.}_{54}$ &     0.3351806004 &     0.3540343105 &    96.9055687195 &    231.480 &  94\\
II.C$^{i.c.}_{55}$ &     0.2761024483 &     0.2287323888 &    76.0753942955 &    233.888 &  94\\
II.C$^{i.c.}_{56}$ &     0.4502872480 &     0.0999875045 &    92.1831437265 &    234.168 &  94\\
II.C$^{i.c.}_{57}$ &     0.2793033126 &     0.2712019544 &    79.7728571001 &    233.344 &  94\\
II.C$^{i.c.}_{58}$ &     0.2723827379 &     0.2246817404 &    81.8883337850 &    253.840 &  102\\
II.C$^{i.c.}_{59}$ &     0.2726258099 &     0.2206843436 &    88.0168046313 &    273.789 &  110\\
II.C$^{i.c.}_{60}$ &     0.2769769990 &     0.2163211708 &    94.5261869045 &    293.737 &  118\\

\hline
\end{tabular*}
{\rule{\temptablewidth}{1pt}}
\end{center}
\end{table*}

\begin{table*}
\tabcolsep 0pt \caption{Initial conditions and periods $T$ of the additional periodic three-body orbits for class II.C in case of the isosceles collinear configurations:  $\bm{r}_1(0)=(-1,0)=-\bm{r}_2(0)$,  $\dot{\bm{r}}_1(0)=(v_1,v_2)=\dot{\bm{r}}_2(0)$ and $\bm{r}_3(0)=(0,0)$, $\dot{\bm{r}}_3(0)=(-2v_1, -2v_2)$ when $G=1$ and $m_1=m_2=m_3=1$ by means of the search grid $2000\times 2000$ in the interval $T \in [0,200]$, where $T^*=T |E|^{3/2}$ is its scale-invariant period, $L_f$ is the length of the free group element.  Here, the superscript {\em i.c.} indicates the case of the initial conditions with {\em isosceles collinear}  configuration, due to the fact that there exist periodic orbits in many other cases.} \label{table-S7} \vspace*{-12pt}
\begin{center}
\def\temptablewidth{1\textwidth}
{\rule{\temptablewidth}{1pt}}
\begin{tabular*}{\temptablewidth}{@{\extracolsep{\fill}}lccccc}
\hline
Class and number  & $v_1$ & $v_2$  & $T$ & $T^*$ & $L_f$ \\
\hline
II.C$^{i.c.}_{61}$ &     0.6696859765 &     0.4862798575 &   111.6735511043 &     33.168 &  16\\
II.C$^{i.c.}_{62}$ &     0.1467711144 &     0.1988238982 &    15.5852040984 &     54.959 &  22\\
II.C$^{i.c.}_{63}$ &     0.2980305901 &     0.6485348678 &    63.9306478929 &     61.240 &  26\\
II.C$^{i.c.}_{64}$ &     0.1066062988 &     0.2369740501 &    21.4593318053 &     74.728 &  30\\
II.C$^{i.c.}_{65}$ &     0.1189697937 &     0.5386581099 &    41.4367113206 &     82.848 &  36\\
II.C$^{i.c.}_{66}$ &     0.1349579489 &     0.1388158691 &    26.9173582655 &     99.303 &  40\\
II.C$^{i.c.}_{67}$ &     0.0190962965 &     0.7880982771 &   174.4738106853 &     88.413 &  40\\
II.C$^{i.c.}_{68}$ &     0.5313917017 &     0.3440533419 &    72.8917817632 &    107.762 &  44\\
II.C$^{i.c.}_{69}$ &     0.1983287250 &     0.1083646402 &    33.1578999528 &    119.204 &  48\\
II.C$^{i.c.}_{70}$ &     0.3961359419 &     0.5075990479 &    82.9095584181 &    116.741 &  50\\
II.C$^{i.c.}_{71}$ &     0.3890589703 &     0.4364126051 &    67.5933134073 &    121.028 &  50\\
II.C$^{i.c.}_{72}$ &     0.4662602934 &     0.4781121788 &    97.8521357163 &    122.573 &  52\\
II.C$^{i.c.}_{73}$ &     0.3912900930 &     0.5083913706 &    88.5152268416 &    125.981 &  54\\
II.C$^{i.c.}_{74}$ &     0.0475462969 &     0.5646588375 &    72.4005108986 &    137.919 &  60\\
II.C$^{i.c.}_{75}$ &     0.2744207740 &     0.5580117379 &    89.1716635083 &    138.312 &  60\\
II.C$^{i.c.}_{76}$ &     0.3631543956 &     0.4101572180 &    72.5369395460 &    146.759 &  60\\
II.C$^{i.c.}_{77}$ &     0.2855135908 &     0.5556857756 &    90.2667574352 &    138.311 &  60\\
II.C$^{i.c.}_{78}$ &     0.2575762954 &     0.2092185043 &    46.7559266660 &    149.424 &  60\\
II.C$^{i.c.}_{79}$ &     0.2825899895 &     0.5574593048 &    96.3976373998 &    147.549 &  64\\
II.C$^{i.c.}_{80}$ &     0.4617804462 &     0.3849338004 &   101.4930532952 &    170.969 &  70\\
II.C$^{i.c.}_{81}$ &     0.4456464329 &     0.3434928347 &    89.2721503494 &    172.311 &  70\\
II.C$^{i.c.}_{82}$ &     0.3379452918 &     0.5474555148 &   117.6340486503 &    166.030 &  72\\
II.C$^{i.c.}_{83}$ &     0.3129165204 &     0.5484741376 &   111.5227089594 &    166.023 &  72\\
II.C$^{i.c.}_{84}$ &     0.2968893204 &     0.5439638834 &   106.0964315543 &    166.027 &  72\\
II.C$^{i.c.}_{85}$ &     0.2512495838 &     0.5669129966 &   106.1149733422 &    165.791 &  72\\
II.C$^{i.c.}_{86}$ &     0.3992808324 &     0.5094959441 &   124.2440256690 &    172.173 &  74\\
II.C$^{i.c.}_{87}$ &     0.2254087369 &     0.3469436377 &    66.3522685919 &    185.770 &  76\\
II.C$^{i.c.}_{88}$ &     0.5234738074 &     0.3430125576 &   122.6793293886 &    187.099 &  76\\
II.C$^{i.c.}_{89}$ &     0.2645261109 &     0.5649476217 &   113.7789661432 &    175.026 &  76\\
II.C$^{i.c.}_{90}$ &     0.3753943858 &     0.5199969079 &   127.3458631987 &    181.409 &  78\\
II.C$^{i.c.}_{91}$ &     0.3734356168 &     0.5183824717 &   125.9373100767 &    181.411 &  78\\
II.C$^{i.c.}_{92}$ &     0.4243049239 &     0.4900563659 &   135.7270472138 &    187.284 &  80\\
II.C$^{i.c.}_{93}$ &     0.4778578506 &     0.4362351173 &   138.1973477797 &    191.760 &  80\\
II.C$^{i.c.}_{94}$ &     0.4478725342 &     0.4242755759 &   121.1471674140 &    191.761 &  80\\
II.C$^{i.c.}_{95}$ &     0.3680416437 &     0.5200194559 &   131.2837960007 &    190.649 &  82\\
II.C$^{i.c.}_{96}$ &     0.4439740574 &     0.4220558056 &   122.5262152360 &    197.395 &  82\\
II.C$^{i.c.}_{97}$ &     0.5049855347 &     0.4003409209 &   143.1429402354 &    201.045 &  84\\
II.C$^{i.c.}_{98}$ &     0.3619688366 &     0.5160511236 &   131.3618489148 &    196.510 &  84\\
II.C$^{i.c.}_{99}$ &     0.4963272196 &     0.2750082245 &   111.8364019177 &    212.500 &  86\\
II.C$^{i.c.}_{100}$ &     0.1600479685 &     0.5830974531 &   121.7551058696 &    202.368 &  88\\

\hline
\end{tabular*}
{\rule{\temptablewidth}{1pt}}
\end{center}
\end{table*}

\begin{table*}
\tabcolsep 0pt \caption{Initial conditions and periods $T_0$ of the additional periodic three-body orbits for class II.C in case of the isosceles collinear configurations:  $\bm{r}_1(0)=(-1,0)=-\bm{r}_2(0)$,  $\dot{\bm{r}}_1(0)=(v_1,v_2)=\dot{\bm{r}}_2(0)$ and $\bm{r}_3(0)=(0,0)$, $\dot{\bm{r}}_3(0)=(-2v_1, -2v_2)$ when $G=1$ and $m_1=m_2=m_3=1$ by means of the search grid $2000\times 2000$ in the interval $T \in [0,200]$, where $T^*=T |E|^{3/2}$ is its scale-invariant period, $L_f$ is the length of the free group element.  Here, the superscript {\em i.c.} indicates the case of the initial conditions with {\em isosceles collinear}  configuration, due to the fact that there exist periodic orbits in many other cases.} \label{table-S7} \vspace*{-12pt}
\begin{center}
\def\temptablewidth{1\textwidth}
{\rule{\temptablewidth}{1pt}}
\begin{tabular*}{\temptablewidth}{@{\extracolsep{\fill}}lccccc}
\hline
Class and number  & $v_1$ & $v_2$  & $T$ & $T^*$ & $L_f$ \\
\hline
II.C$^{i.c.}_{101}$ &     0.5102908231 &     0.3575901413 &   144.7874249037 &    223.383 &  92\\
II.C$^{i.c.}_{102}$ &     0.5032871392 &     0.2685375206 &   123.6355912506 &    232.553 &  94\\
II.C$^{i.c.}_{103}$ &     0.4831068889 &     0.3429708664 &   133.0041037254 &    231.494 &  94\\
II.C$^{i.c.}_{104}$ &     0.4600987572 &     0.4389868308 &   154.3254142371 &    225.270 &  94\\
II.C$^{i.c.}_{105}$ &     0.4135304700 &     0.2998696032 &   105.2365923011 &    236.811 &  96\\
II.C$^{i.c.}_{106}$ &     0.4300124259 &     0.3554749370 &   120.2233121235 &    235.641 &  96\\
II.C$^{i.c.}_{107}$ &     0.4128800187 &     0.4868642357 &   159.3592373143 &    230.096 &  98\\
II.C$^{i.c.}_{108}$ &     0.2337840221 &     0.4711247940 &   107.6906838559 &    232.442 &  98\\
II.C$^{i.c.}_{109}$ &     0.5021284225 &     0.4265371911 &   178.9211696576 &    234.551 &  98\\
II.C$^{i.c.}_{110}$ &     0.2164867425 &     0.5799872016 &   146.6171327324 &    230.040 &  100\\
II.C$^{i.c.}_{111}$ &     0.3579867923 &     0.3621287896 &   108.4386863974 &    245.065 &  100\\
II.C$^{i.c.}_{112}$ &     0.4031141443 &     0.5020964921 &   165.8327705497 &    233.483 &  100\\
II.C$^{i.c.}_{113}$ &     0.4481845220 &     0.3602811878 &   132.3386808547 &    245.065 &  100\\
II.C$^{i.c.}_{114}$ &     0.5500731263 &     0.3814634479 &   196.2411230097 &    243.818 &  102\\
II.C$^{i.c.}_{115}$ &     0.3562032573 &     0.5250003095 &   161.1788109414 &    236.834 &  102\\
II.C$^{i.c.}_{116}$ &     0.4118599465 &     0.3432155311 &   120.0533167933 &    251.614 &  102\\
II.C$^{i.c.}_{117}$ &     0.4801960806 &     0.2871406271 &   129.0160097094 &    251.594 &  102\\
II.C$^{i.c.}_{118}$ &     0.5371658731 &     0.3984943066 &   195.6689329179 &    243.818 &  102\\
II.C$^{i.c.}_{119}$ &     0.2725820526 &     0.2644346962 &    85.2269842957 &    253.331 &  102\\
II.C$^{i.c.}_{120}$ &     0.4500287021 &     0.4212626102 &   157.2987714029 &    249.489 &  104\\
II.C$^{i.c.}_{121}$ &     0.4438767435 &     0.4225775888 &   158.5428263415 &    255.123 &  106\\
II.C$^{i.c.}_{122}$ &     0.0931957583 &     0.5754064016 &   137.7843111275 &    248.248 &  108\\
II.C$^{i.c.}_{123}$ &     0.4175301746 &     0.3430957060 &   128.7472932221 &    266.417 &  108\\
II.C$^{i.c.}_{124}$ &     0.3600187937 &     0.5221797725 &   173.6222271714 &    255.315 &  110\\
II.C$^{i.c.}_{125}$ &     0.3806272915 &     0.3501131016 &   126.6161436477 &    280.062 &  114\\
II.C$^{i.c.}_{126}$ &     0.3837791731 &     0.3569518329 &   129.0860797033 &    280.060 &  114\\
II.C$^{i.c.}_{127}$ &     0.0715935321 &     0.5614606957 &   139.6739967734 &    266.646 &  116\\
II.C$^{i.c.}_{128}$ &     0.5084896856 &     0.3971173995 &   198.1578809274 &    277.336 &  116\\
II.C$^{i.c.}_{129}$ &     0.5129886961 &     0.1995185091 &   146.1306716825 &    293.285 &  118\\
II.C$^{i.c.}_{130}$ &     0.2239434477 &     0.5629809053 &   166.9610126801 &    276.188 &  120\\
II.C$^{i.c.}_{131}$ &     0.3646073207 &     0.2326179920 &   110.5223988488 &    298.378 &  120\\
II.C$^{i.c.}_{132}$ &     0.3758809553 &     0.3714191733 &   139.4763383031 &    298.922 &  122\\
II.C$^{i.c.}_{133}$ &     0.4079687070 &     0.3467946424 &   142.9574365649 &    300.211 &  122\\
II.C$^{i.c.}_{134}$ &     0.2582557078 &     0.2986633497 &   104.3188859584 &    302.238 &  122\\
II.C$^{i.c.}_{135}$ &     0.2613665653 &     0.3010646289 &   108.3619916341 &    311.829 &  126\\
II.C$^{i.c.}_{136}$ &     0.3955814740 &     0.3842771664 &   154.1379803565 &    308.316 &  126\\
II.C$^{i.c.}_{137}$ &     0.3455455677 &     0.3273880159 &   126.5531673809 &    310.791 &  126\\
II.C$^{i.c.}_{138}$ &     0.3385794074 &     0.2315252500 &   112.9484622797 &    318.336 &  128\\
II.C$^{i.c.}_{139}$ &     0.3578601157 &     0.1829928242 &   111.3983832975 &    318.716 &  128\\
II.C$^{i.c.}_{140}$ &     0.2662199199 &     0.2617288091 &   107.5138953175 &    322.958 &  130\\

\hline
\end{tabular*}
{\rule{\temptablewidth}{1pt}}
\end{center}
\end{table*}

\begin{table*}
\tabcolsep 0pt \caption{Initial conditions and periods $T$ of the additional periodic three-body orbits for class II.C in case of the isosceles collinear configurations:  $\bm{r}_1(0)=(-1,0)=-\bm{r}_2(0)$,  $\dot{\bm{r}}_1(0)=(v_1,v_2)=\dot{\bm{r}}_2(0)$ and $\bm{r}_3(0)=(0,0)$, $\dot{\bm{r}}_3(0)=(-2v_1, -2v_2)$ when $G=1$ and $m_1=m_2=m_3=1$ by means of the search grid $2000\times 2000$ in the interval $T \in [0,200]$, where $T^*=T |E|^{3/2}$ is its scale-invariant period, $L_f$ is the length of the free group element.  Here, the superscript {\em i.c.} indicates the case of the initial conditions with {\em isosceles collinear}  configuration, due to the fact that there exist periodic orbits in many other cases.} \label{table-S7} \vspace*{-12pt}
\begin{center}
\def\temptablewidth{1\textwidth}
{\rule{\temptablewidth}{1pt}}
\begin{tabular*}{\temptablewidth}{@{\extracolsep{\fill}}lccccc}
\hline
Class and number  & $v_1$ & $v_2$  & $T$ & $T^*$ & $L_f$ \\
\hline
II.C$^{i.c.}_{141}$ &     0.4458373407 &     0.4219085904 &   195.1709699750 &    312.851 &  130\\
II.C$^{i.c.}_{142}$ &     0.4161228347 &     0.3579128439 &   158.2418427194 &    319.125 &  130\\
II.C$^{i.c.}_{143}$ &     0.4104164725 &     0.3814951548 &   166.1586070391 &    323.146 &  132\\
II.C$^{i.c.}_{144}$ &     0.0693693379 &     0.6033646696 &   184.4392377108 &    303.372 &  132\\
II.C$^{i.c.}_{145}$ &     0.4735978920 &     0.2313177546 &   156.9836879900 &    337.752 &  136\\
II.C$^{i.c.}_{146}$ &     0.4799888062 &     0.2624795421 &   166.2000474547 &    337.042 &  136\\
II.C$^{i.c.}_{147}$ &     0.4637112431 &     0.2858923789 &   164.6023344184 &    336.170 &  136\\
II.C$^{i.c.}_{148}$ &     0.5206456110 &     0.2383685944 &   183.3096591850 &    342.274 &  138\\
II.C$^{i.c.}_{149}$ &     0.4175792025 &     0.3246498436 &   159.0527052142 &    340.388 &  138\\
II.C$^{i.c.}_{150}$ &     0.5093497308 &     0.1590840316 &   162.6587265245 &    343.423 &  138\\
II.C$^{i.c.}_{151}$ &     0.0683834774 &     0.5833653062 &   181.4427818832 &    321.742 &  140\\
II.C$^{i.c.}_{152}$ &     0.4639950316 &     0.3539989419 &   196.2961464240 &    352.779 &  144\\
II.C$^{i.c.}_{153}$ &     0.3518330832 &     0.3639937352 &   155.4205568976 &    354.011 &  144\\
II.C$^{i.c.}_{154}$ &     0.3153981646 &     0.2969797915 &   136.0121995709 &    366.662 &  148\\
II.C$^{i.c.}_{155}$ &     0.4099850242 &     0.3706053839 &   184.4454176663 &    367.599 &  150\\
II.C$^{i.c.}_{156}$ &     0.3231926176 &     0.3279135713 &   145.3798952880 &    369.993 &  150\\
II.C$^{i.c.}_{157}$ &     0.4771765515 &     0.1801229299 &   167.7213251594 &    378.199 &  152\\
II.C$^{i.c.}_{158}$ &     0.2973959601 &     0.2543194648 &   129.5705475970 &    377.705 &  152\\
II.C$^{i.c.}_{159}$ &     0.3134720455 &     0.1840666537 &   124.0700355615 &    378.530 &  152\\
II.C$^{i.c.}_{160}$ &     0.4160632302 &     0.3349978893 &   182.5440752184 &    384.788 &  156\\
II.C$^{i.c.}_{161}$ &     0.3358161544 &     0.3881164707 &   172.8143824669 &    386.358 &  158\\
II.C$^{i.c.}_{162}$ &     0.4870167590 &     0.2099302469 &   186.5727277418 &    397.676 &  160\\
II.C$^{i.c.}_{163}$ &     0.2969877047 &     0.2221255009 &   132.0216562587 &    398.149 &  160\\
II.C$^{i.c.}_{164}$ &     0.4156609919 &     0.3579596922 &   199.4409746332 &    402.608 &  164\\
II.C$^{i.c.}_{165}$ &     0.4871928568 &     0.1661495427 &   187.7801397568 &    418.099 &  168\\
II.C$^{i.c.}_{166}$ &     0.3638592241 &     0.3770170533 &   191.6726449737 &    416.030 &  170\\
II.C$^{i.c.}_{167}$ &     0.2985413068 &     0.3181960211 &   160.2142776167 &    429.197 &  174\\
II.C$^{i.c.}_{168}$ &     0.3102533155 &     0.3277636570 &   165.3209382776 &    429.196 &  174\\
II.C$^{i.c.}_{169}$ &     0.3098999619 &     0.3285015035 &   165.4253970148 &    429.196 &  174\\
II.C$^{i.c.}_{170}$ &     0.2767271690 &     0.2447992873 &   144.7823792155 &    437.609 &  176\\
II.C$^{i.c.}_{171}$ &     0.4388627404 &     0.2371915812 &   190.4374989777 &    442.162 &  178\\
II.C$^{i.c.}_{172}$ &     0.3625783597 &     0.3645069408 &   195.6153329373 &    436.274 &  178\\
II.C$^{i.c.}_{173}$ &     0.4277449853 &     0.2640042862 &   192.0213587611 &    441.494 &  178\\
II.C$^{i.c.}_{174}$ &     0.3729712940 &     0.2250121941 &   165.0090445138 &    442.699 &  178\\
II.C$^{i.c.}_{175}$ &     0.2888390504 &     0.2160604105 &   149.4648328243 &    457.995 &  184\\
II.C$^{i.c.}_{176}$ &     0.3666419820 &     0.1916460944 &   165.3779094251 &    463.044 &  186\\
II.C$^{i.c.}_{177}$ &     0.2884663698 &     0.1819781618 &   151.5859612492 &    478.215 &  192\\
II.C$^{i.c.}_{178}$ &     0.3749871963 &     0.1474830882 &   169.2093364950 &    483.234 &  194\\
II.C$^{i.c.}_{179}$ &     0.3017673304 &     0.3279588878 &   185.8778154983 &    488.400 &  198\\
II.C$^{i.c.}_{180}$ &     0.2920863313 &     0.2113564265 &   162.4404868020 &    497.887 &  200\\
II.C$^{i.c.}_{181}$ &     0.3363227387 &     0.1931223679 &   171.4994464900 &    502.925 &  202\\
II.C$^{i.c.}_{182}$ &     0.2965212394 &     0.3180127676 &   192.7713804348 &    518.004 &  210\\
II.C$^{i.c.}_{183}$ &     0.3263964865 &     0.1928214109 &   175.7091618817 &    522.865 &  210\\
II.C$^{i.c.}_{184}$ &     0.3066734671 &     0.2613669948 &   189.5879550733 &    541.439 &  218\\
II.C$^{i.c.}_{185}$ &     0.2992280886 &     0.1590598797 &   173.1881049209 &    548.070 &  220\\
II.C$^{i.c.}_{186}$ &     0.3117925543 &     0.1916312640 &   185.1588528533 &    562.745 &  226\\
II.C$^{i.c.}_{187}$ &     0.3033738636 &     0.2176185571 &   193.8770772063 &    582.359 &  234\\
II.C$^{i.c.}_{188}$ &     0.2604465076 &     0.2256869065 &   196.6479293263 &    617.213 &  248\\
\hline
\end{tabular*}
{\rule{\temptablewidth}{1pt}}
\end{center}
\end{table*}

\begin{table*}
\tabcolsep 0pt \caption{Initial conditions and periods $T$ of the additional periodic three-body orbits for class II.C in case of the isosceles collinear configurations:  $\bm{r}_1(0)=(-1,0)=-\bm{r}_2(0)$,  $\dot{\bm{r}}_1(0)=(v_1,v_2)=\dot{\bm{r}}_2(0)$ and $\bm{r}_3(0)=(0,0)$, $\dot{\bm{r}}_3(0)=(-2v_1, -2v_2)$ when $G=1$ and $m_1=m_2=m_3=1$ by means of the search grid $4000\times 4000$ in the interval $T \in [0,200]$, where $T^*=T |E|^{3/2}$ is its scale-invariant period, $L_f$ is the length of the free group element. Here, the superscript {\em i.c.} indicates the case of the initial conditions with {\em isosceles collinear}  configuration, due to the fact that there exist periodic orbits in many other cases. } \label{table-S7} \vspace*{-12pt}
\begin{center}
\def\temptablewidth{1\textwidth}
{\rule{\temptablewidth}{1pt}}
\begin{tabular*}{\temptablewidth}{@{\extracolsep{\fill}}lccccc}
\hline
Class and number  & $v_1$ & $v_2$  & $T$ & $T^*$ & $L_f$ \\
\hline
II.C$^{i.c.}_{189}$ &     0.0667327899 &     0.2912867569 &    13.3697110598 &     44.585 &  18\\
II.C$^{i.c.}_{190}$ &     0.0539725385 &     0.3917593746 &    15.2796210778 &     44.221 &  18\\
II.C$^{i.c.}_{191}$ &     0.4271293811 &     0.6282376589 &    58.8959853789 &     39.689 &  18\\
II.C$^{i.c.}_{192}$ &     0.1459197548 &     0.0434579988 &    16.8197113393 &     63.731 &  26\\
II.C$^{i.c.}_{193}$ &     0.0734343088 &     0.2956785083 &    22.4013173133 &     74.175 &  30\\
II.C$^{i.c.}_{194}$ &     0.5555876261 &     0.4214936039 &    75.1051252007 &     79.771 &  34\\
II.C$^{i.c.}_{195}$ &     0.1873991947 &     0.4829394618 &    40.3052946558 &     88.940 &  38\\
II.C$^{i.c.}_{196}$ &     0.3268704330 &     0.4439931746 &    46.4481084914 &     92.956 &  38\\
II.C$^{i.c.}_{197}$ &     0.0772553729 &     0.0962935023 &    25.8396411521 &     99.351 &  40\\
II.C$^{i.c.}_{198}$ &     0.4805076338 &     0.5832469030 &   131.5712571441 &     91.825 &  40\\
II.C$^{i.c.}_{199}$ &     0.0759936918 &     0.2997588242 &    31.5191656332 &    103.772 &  42\\
II.C$^{i.c.}_{200}$ &     0.2513336605 &     0.3012408423 &    37.4214190867 &    108.895 &  44\\
II.C$^{i.c.}_{201}$ &     0.2224798560 &     0.3218943929 &    38.7911677318 &    113.081 &  46\\
II.C$^{i.c.}_{202}$ &     0.1727411576 &     0.3365007421 &    39.3444245245 &    117.243 &  48\\
II.C$^{i.c.}_{203}$ &     0.3449378402 &     0.4918716974 &    72.6547174942 &    122.583 &  52\\
II.C$^{i.c.}_{204}$ &     0.3984837145 &     0.4527620517 &    76.7667464654 &    128.345 &  54\\
II.C$^{i.c.}_{205}$ &     0.3107903232 &     0.4359615931 &    67.9349980910 &    142.684 &  58\\
II.C$^{i.c.}_{206}$ &     0.0995077146 &     0.3957710385 &    51.8957289725 &    146.826 &  60\\
II.C$^{i.c.}_{207}$ &     0.2712629218 &     0.5766902452 &    95.1895194685 &    138.097 &  60\\
II.C$^{i.c.}_{208}$ &     0.3791031878 &     0.5143377515 &   100.3150104107 &    144.457 &  62\\
II.C$^{i.c.}_{209}$ &     0.2816496180 &     0.3514755712 &    60.0376470543 &    156.172 &  64\\
II.C$^{i.c.}_{210}$ &     0.3851685224 &     0.5086415312 &   106.2835399139 &    153.696 &  66\\
II.C$^{i.c.}_{211}$ &     0.4036022580 &     0.5028840213 &   109.6282058892 &    153.695 &  66\\
II.C$^{i.c.}_{212}$ &     0.3011541892 &     0.5487167377 &   102.8385494278 &    156.786 &  68\\
II.C$^{i.c.}_{213}$ &     0.3082337510 &     0.5516573762 &   105.5334666717 &    156.785 &  68\\
II.C$^{i.c.}_{214}$ &     0.5201916309 &     0.2123324421 &    87.2860842631 &    168.920 &  68\\
II.C$^{i.c.}_{215}$ &     0.2512381225 &     0.3272400470 &    61.3992062163 &    172.282 &  70\\
II.C$^{i.c.}_{216}$ &     0.3800232997 &     0.5162361284 &   114.2142913903 &    162.934 &  70\\
II.C$^{i.c.}_{217}$ &     0.3975219956 &     0.4882458860 &   112.4811763381 &    168.800 &  72\\
II.C$^{i.c.}_{218}$ &     0.1829585594 &     0.2103495753 &    55.5611081365 &    189.628 &  76\\
II.C$^{i.c.}_{219}$ &     0.3085347089 &     0.5520329973 &   118.2139479118 &    175.260 &  76\\
II.C$^{i.c.}_{220}$ &     0.4492249567 &     0.4587831231 &   127.0667084547 &    180.390 &  76\\
II.C$^{i.c.}_{221}$ &     0.3837232497 &     0.5140136806 &   127.4087704175 &    181.411 &  78\\
II.C$^{i.c.}_{222}$ &     0.3434370240 &     0.3432358104 &    80.1579217509 &    192.405 &  78\\
II.C$^{i.c.}_{223}$ &     0.3085741714 &     0.5449851912 &   121.1980464816 &    184.498 &  80\\
II.C$^{i.c.}_{224}$ &     0.5144801407 &     0.2857654410 &   117.3488690365 &    207.219 &  84\\
II.C$^{i.c.}_{225}$ &     0.2872301587 &     0.5602387584 &   128.9199840181 &    193.496 &  84\\
II.C$^{i.c.}_{226}$ &     0.3934771561 &     0.4902364154 &   130.3922423022 &    196.521 &  84\\
II.C$^{i.c.}_{227}$ &     0.5186642945 &     0.4541901634 &   178.6359517335 &    198.854 &  84\\
II.C$^{i.c.}_{228}$ &     0.3524931856 &     0.4505770398 &   110.5012506668 &    206.707 &  86\\
II.C$^{i.c.}_{229}$ &     0.3697593874 &     0.5206204356 &   138.5644185707 &    199.887 &  86\\
II.C$^{i.c.}_{230}$ &     0.1744019921 &     0.5732931066 &   119.3409051099 &    202.528 &  88\\
II.C$^{i.c.}_{231}$ &     0.2507324209 &     0.5609900202 &   132.5826540167 &    211.967 &  92\\
II.C$^{i.c.}_{232}$ &     0.3102568382 &     0.3782456948 &    98.4450118457 &    234.185 &  96\\
II.C$^{i.c.}_{233}$ &     0.3164477867 &     0.3698035086 &   100.1147034290 &    239.623 &  98\\
II.C$^{i.c.}_{234}$ &     0.5017058130 &     0.3990265330 &   164.4248192381 &    234.552 &  98\\
II.C$^{i.c.}_{235}$ &     0.4964835592 &     0.4057747351 &   164.5510136944 &    234.550 &  98\\
II.C$^{i.c.}_{236}$ &     0.4408929544 &     0.4698230681 &   167.8693941629 &    235.912 &  100\\
II.C$^{i.c.}_{237}$ &     0.2461487944 &     0.6145395833 &   178.2090341298 &    229.958 &  100\\
II.C$^{i.c.}_{238}$ &     0.4984243728 &     0.3329687902 &   148.3551137804 &    251.596 &  102\\
II.C$^{i.c.}_{239}$ &     0.3700247775 &     0.5202555395 &   164.0744043589 &    236.839 &  102\\
II.C$^{i.c.}_{240}$ &     0.0760602620 &     0.5596439595 &   124.7398513280 &    239.095 &  104\\
II.C$^{i.c.}_{241}$ &     0.2629848757 &     0.5682829440 &   157.2343331952 &    239.454 &  104\\
II.C$^{i.c.}_{242}$ &     0.4412121748 &     0.4675245937 &   173.2839445231 &    245.159 &  104\\
II.C$^{i.c.}_{243}$ &     0.2823167104 &     0.3047669596 &    93.9381222516 &    262.167 &  106\\
II.C$^{i.c.}_{244}$ &     0.4148849874 &     0.4294385631 &   152.4007166724 &    260.708 &  108\\
II.C$^{i.c.}_{245}$ &     0.2590526978 &     0.5675456785 &   161.7057059926 &    248.687 &  108\\

\hline
\end{tabular*}
{\rule{\temptablewidth}{1pt}}
\end{center}
\end{table*}

\begin{table*}
\tabcolsep 0pt \caption{Initial conditions and periods $T$ of the additional periodic three-body orbits for class II.C in case of the isosceles collinear configurations:  $\bm{r}_1(0)=(-1,0)=-\bm{r}_2(0)$,  $\dot{\bm{r}}_1(0)=(v_1,v_2)=\dot{\bm{r}}_2(0)$ and $\bm{r}_3(0)=(0,0)$, $\dot{\bm{r}}_3(0)=(-2v_1, -2v_2)$ when $G=1$ and $m_1=m_2=m_3=1$ by means of the search grid $4000\times 4000$ in the interval $T \in [0,200]$, where $T^*=T |E|^{3/2}$ is its scale-invariant period, $L_f$ is the length of the free group element. Here, the superscript {\em i.c.} indicates the case of the initial conditions with {\em isosceles collinear}  configuration, due to the fact that there exist periodic orbits in many other cases. } \label{table-S7} \vspace*{-12pt}
\begin{center}
\def\temptablewidth{1\textwidth}
{\rule{\temptablewidth}{1pt}}
\begin{tabular*}{\temptablewidth}{@{\extracolsep{\fill}}lccccc}
\hline
Class and number  & $v_1$ & $v_2$  & $T$ & $T^*$ & $L_f$ \\
\hline
II.C$^{i.c.}_{246}$ &     0.4127173212 &     0.4811313628 &   172.7271237236 &    254.407 &  108\\
II.C$^{i.c.}_{247}$ &     0.2984774772 &     0.3112529516 &    99.9639866944 &    270.552 &  110\\
II.C$^{i.c.}_{248}$ &     0.4548792495 &     0.0896454467 &   108.4524334532 &    274.035 &  110\\
II.C$^{i.c.}_{249}$ &     0.2665841319 &     0.3022483726 &   102.1824317103 &    291.781 &  118\\
II.C$^{i.c.}_{250}$ &     0.4844029847 &     0.3736255064 &   177.3515504868 &    286.659 &  118\\
II.C$^{i.c.}_{251}$ &     0.3411937664 &     0.3825680644 &   131.0523112095 &    293.481 &  120\\
II.C$^{i.c.}_{252}$ &     0.0586366068 &     0.5597110167 &   142.9605350754 &    275.837 &  120\\
\hline
\end{tabular*}
{\rule{\temptablewidth}{1pt}}
\end{center}
\end{table*}

\begin{sidewaystable}
\tabcolsep 0pt \caption{The free group elements for the periodic three-body orbits.} \label{table-S8} \vspace*{-12pt}
\begin{center}
\def\temptablewidth{1\textwidth}
{\rule{\temptablewidth}{1pt}}
\begin{tabular*}{\temptablewidth}{@{\extracolsep{\fill}}ll}
\hline
Class \& number  & \hspace{1.0cm}  free group element\\\hline
I.A$^{i.c.}_{1}$ & BabA\\
I.A$^{i.c.}_{2}$ & BAbabaBA\\
I.A$^{i.c.}_{3}$ & BaBabAbA\\
I.A$^{i.c.}_{4}$ & BabaBABabABAbabA\\
I.A$^{i.c.}_{5}$ & BAbabABAbabaBABabaBA\\
I.A$^{i.c.}_{6}$ & BabaBAbaBABabABAbaBAbabA\\
I.A$^{i.c.}_{7}$ & BAbaabABBAbabaBAABabbaBA\\
\hline
\end{tabular*}
{\rule{\temptablewidth}{1pt}}
\end{center}
\end{sidewaystable}

\begin{sidewaystable}
\tabcolsep 0pt \caption{The free group elements for the periodic three-body orbits.} \label{table-S8} \vspace*{-12pt}
\begin{center}
\def\temptablewidth{1\textwidth}
{\rule{\temptablewidth}{1pt}}
\begin{tabular*}{\temptablewidth}{@{\extracolsep{\fill}}ll}
\hline
Class \& number   & \hspace{1.0cm}  free group element\\\hline
I.A$^{i.c.}_{51}$ & BAbaBABabaBAbabABAbaBABabaBAbabaBAbabABAbaBABabaBAbabABAbaBA\\
I.A$^{i.c.}_{52}$ & BabABAbaBAbaBABabABabaBAbaBAbabABabABabaBAbaBAbabABabABAbaBAbaBABabA\\
I.A$^{i.c.}_{53}$ & BabaBABabaBABabaBABabaBABabaBABabaBABabABAbabABAbabABAbabABAbabABAbabABAbabA\\
I.A$^{i.c.}_{54}$ & BabABabABaaBAbaBAbaBAbaBAbaBBabABabABabABabABabAAbaBAbaBAbaBAbaBAbbABabABabA\\
I.A$^{i.c.}_{55}$ & BabABAbaBAbaBABabABabABabaBAbaBAbabABabABabaBAbaBAbabABabABabABAbaBAbaBABabA\\
I.A$^{i.c.}_{56}$ & BabaBAbabABabaBAbabABAbaBABabABAbaBABabABAbaBABabABAbaBABabaBAbabABabaBAbabA\\
I.A$^{i.c.}_{57}$ & BabABaaBAbaBAbaBBabABabABaaBAbaBAbaBBabABabABabAAbaBAbaBAbbABabABabAAbaBAbaBAbbABabA\\
\hline
\end{tabular*}
{\rule{\temptablewidth}{1pt}}
\end{center}
\end{sidewaystable}

\begin{sidewaystable}
\tabcolsep 0pt \caption{The free group elements for the periodic three-body orbits.} \label{table-S8} \vspace*{-12pt}
\begin{center}
\def\temptablewidth{1\textwidth}
{\rule{\temptablewidth}{1pt}}
\begin{tabular*}{\temptablewidth}{@{\extracolsep{\fill}}ll}
\hline
Class \& number   & \hspace{1.0cm}  free group element\\\hline
I.A$^{i.c.}_{91}$ & BAbabaBABAbabaBABababABABababABABababABABabaBABAbabaBABAbabaBABAbabaBABAbabABABababABABababABABababABAba\\&baBABAbabaBA\\
I.A$^{i.c.}_{92}$ & BAbabABAbabABAbabABABabaBABabaBABabaBABababABAbabABAbabABAbabaBABabaBABabaBABababABAbabABAbabABAbabABABa\\&baBABabaBABabaBA\\
I.A$^{i.c.}_{93}$ & BAbabABABabaBABAbabABABabaBABAbabaBABababABAbabaBABababABAbabaBABababABAbabaBABababABAbabaBABAbabABABaba\\&BABAbabABABabaBA\\
I.A$^{i.c.}_{94}$ & BabaBAbabABabaBAbabABabaBAbaBABabaBAbaBABabABAbaBABabABAbaBABabABAbaBABabABAbaBABabABAbaBAbabABAbaBAbabA\\
\hline
\end{tabular*}
{\rule{\temptablewidth}{1pt}}
\end{center}
\end{sidewaystable}

\begin{sidewaystable}
\tabcolsep 0pt \caption{The free group elements for the periodic three-body orbits.} \label{table-S8} \vspace*{-12pt}
\begin{center}
\def\temptablewidth{1\textwidth}
{\rule{\temptablewidth}{1pt}}
\begin{tabular*}{\temptablewidth}{@{\extracolsep{\fill}}ll}
\hline
Class \& number   & \hspace{1.0cm}  free group element\\\hline
I.A$^{i.c.}_{116}$ & BAbabABABabaBABAbabaBABabaBABAbabABABababABAbabaBABabaBABAbabaBABababABAbabaBABababABAbabaBABAbabABAbab\\&aBABababABABabaBABAbabABAbabaBABAbabABABabaBA\\
I.A$^{i.c.}_{117}$ & BabAAbaBAbaBAbbABabABabAAbaBAbaBBabABabABaaBAbaBAbbABabABabAAbaBAbaBAbbABabABaaBAbaBAbaBBabABabABaaBAba\\&BAbbABabABabAAbaBAbaBBabABabABaaBAbaBAbaBBabA\\
I.A$^{i.c.}_{118}$ & BabaBABabaBAbabABAbaBABabaBABabaBAbabABAbaBABabaBABabaBAbabABAbaBABabaBABabABAbabABAbaBABabaBAbabABAbab\\&ABAbaBABabaBAbabABAbabABAbaBABabaBAbabABAbabA\\
I.A$^{i.c.}_{119}$ & BabaBABabaBAbabABAbabABabaBABabaBAbabABAbaBABabaBABabABAbabABAbaBABabaBABabABAbabABAbaBABabaBABabABAbab\\
\hline
\end{tabular*}
{\rule{\temptablewidth}{1pt}}
\end{center}
\end{sidewaystable}

\begin{sidewaystable}
\tabcolsep 0pt \caption{The free group elements for the periodic three-body orbits.} \label{table-S8} \vspace*{-12pt}
\begin{center}
\def\temptablewidth{1\textwidth}
{\rule{\temptablewidth}{1pt}}
\begin{tabular*}{\temptablewidth}{@{\extracolsep{\fill}}ll}
\hline
Class \& number   & \hspace{1.0cm}  free group element\\\hline
I.A$^{i.c.}_{131}$ & BAbabaBABAbabaBABAbabABABababABABababABABababABABababABABababABAbabaBABAbabaBABAbabaBABAbabaBABAbabaBAB\\&ababABABababABABababABABababABABababABABabaBABAbabaBABAbabaBA\\
I.A$^{i.c.}_{132}$ & BAbabABABabaBABababABAbabaBABababABAbabABABababABAbabABABabaBABAbabABABabaBABababABAbabaBABababABAbabAB\\&ABabaBABAbabABABabaBABababABABabaBABababABAbabaBABababABAbabABABabaBA\\
I.A$^{i.c.}_{133}$ & BAbabaBABAbabaBABAbababABABababABABababABABababABABababABABababABABAbabaBABAbabaBABAbabaBABAbabaBABAbab\\&aBABABababABABababABABababABABababABABababABABababaBABAbabaBABAbabaBA\\
I.A$^{i.c.}_{134}$ & BAbabABABabaBABAbabABABabaBABAbabABABabaBABAbabaBABababABAbabaBABababABAbabaBABababABAbabaBABababABAbab\\
\hline
\end{tabular*}
{\rule{\temptablewidth}{1pt}}
\end{center}
\end{sidewaystable}

\begin{sidewaystable}
\tabcolsep 0pt \caption{The free group elements for the periodic three-body orbits.} \label{table-S8} \vspace*{-12pt}
\begin{center}
\def\temptablewidth{1\textwidth}
{\rule{\temptablewidth}{1pt}}
\begin{tabular*}{\temptablewidth}{@{\extracolsep{\fill}}ll}
\hline
Class \& number   &   \hspace{1.0cm}  free group element\\\hline
I.A$^{i.c.}_{157}$ & BabABabAAbaBAbaBAbaBAbaBBabABabABabABaaBAbaBAbaBAbaBAbbABabABabABabABaaBAbaBAbaBAbaBAbbABabABabABabAAbaBAb\\&aBAbaBAbaBBabABabA\\
I.A$^{i.c.}_{158}$ & BabaBAbabABAbaBABabABAbaBABabABAbabABabaBAbabABabaBAbabABAbaBABabABAbaBABabaBAbabABabaBAbabABabaBABabABAba\\&BABabABAbaBABabaBAbabA\\
I.A$^{i.c.}_{159}$ & BabABabAAbaBAbaBAbaBAbaBBabABabABabABabABaaBAbaBAbaBAbaBAbbABabABabABabABaaBAbaBAbaBAbaBAbbABabABabABabABa\\&bAAbaBAbaBAbaBAbaBBabABabA\\
I.A$^{i.c.}_{160}$ & BAbabABABabaBABAbabABABabaBABAbabABAbabaBABababABAbabaBABababABAbabaBABababABAbabaBABababABAbabaBABabaBABA\\
\hline
\end{tabular*}
{\rule{\temptablewidth}{1pt}}
\end{center}
\end{sidewaystable}

\begin{sidewaystable}
\tabcolsep 0pt \caption{The free group elements for the periodic three-body orbits.} \label{table-S8} \vspace*{-12pt}
\begin{center}
\def\temptablewidth{1\textwidth}
{\rule{\temptablewidth}{1pt}}
\begin{tabular*}{\temptablewidth}{@{\extracolsep{\fill}}ll}
\hline
Class \& number & \hspace{1.0cm} free group element \\ \hline

I.A$^{i.c.}_{181}$ & BAbabABABabaBABAbabABABabaBABAbabABABabaBABAbabABAbabaBABababABAbabaBABababABAbabaBABababABAbabaBABababABA\\&babaBABababABAbabaBABababABAbabaBABabaBABAbabABABabaBABAbabABABabaBABAbabABABabaBA\\
I.A$^{i.c.}_{182}$ & BAbabaBABAbabaBABAbabaBABababABABababABABababABABababABABababABABababABABabaBABAbabaBABAbabaBABAbabaBABAba\\&baBABAbabaBABAbabABABababABABababABABababABABababABABababABABababABAbabaBABAbabaBABAbabaBA\\
I.A$^{i.c.}_{183}$ & BAbabaBABAbabaBABababABABababABABababABABababABABababABABababABABababABABababABABabaBABAbabaBABAbabaBABAba\\&baBABAbabABABababABABababABABababABABababABABababABABababABABababABABababABAbabaBABAbabaBA\\
I.A$^{i.c.}_{184}$ & BAbabaBABababABABabaBABAbabaBABababABABababABAbabaBABAbabABABababABABabaBABAbabaBABababABABabaBABAbabaBABA\\
\hline
\end{tabular*}
{\rule{\temptablewidth}{1pt}}
\end{center}
\end{sidewaystable}

\begin{sidewaystable}
\tabcolsep 0pt \caption{The free group elements for the periodic three-body orbits.} \label{table-S8} \vspace*{-12pt}
\begin{center}
\def\temptablewidth{1\textwidth}
{\rule{\temptablewidth}{1pt}}
\begin{tabular*}{\temptablewidth}{@{\extracolsep{\fill}}ll}
\hline
Class \& number   & \hspace{1.0cm}  free group element\\\hline
I.B$^{i.c.}_{16}$ 	&	 BabABabABaaBAbaBAbaBAbaBAbaBAbbABabABabA\\
I.B$^{i.c.}_{17}$ 	&	 BAbabaBABAbabABABababABABababABABabaBABAbabaBA\\
I.B$^{i.c.}_{18}$ 	&	 BabaBABabaBABabaBABabaBAbabABAbabABAbabABAbabA\\
I.B$^{i.c.}_{19}$ 	&	 BabABabABabAAbaBAbaBAbaBAbaBAbaBAbaBBabABabABabA\\
I.B$^{i.c.}_{20}$ 	&	 BAbabABAbabABAbabABAbabABABabaBABabaBABabaBABabaBA\\
I.B$^{i.c.}_{21}$ 	&	 BabaBAbabABAbabABAbabABabaBAbabABabaBABabaBABabaBAbabA\\
I.B$^{i.c.}_{22}$ 	&	 BAbabABAbabaBABababABAbabABABabaBABababABAbabaBABabaBA\\
\hline
\end{tabular*}
{\rule{\temptablewidth}{1pt}}
\end{center}
\end{sidewaystable}

\newpage

\begin{sidewaystable}
\tabcolsep 0pt \caption{The free group elements for the periodic three-body orbits.} \label{table-S8} \vspace*{-12pt}
\begin{center}
\def\temptablewidth{1\textwidth}
{\rule{\temptablewidth}{1pt}}
\begin{tabular*}{\temptablewidth}{@{\extracolsep{\fill}}ll}
\hline
Class \& number   & \hspace{1.0cm}  free group element\\\hline
I.B$^{i.c.}_{56}$ & BabABabABabaBAbaBAbaBAbaBAbaBAbabABabABabA\\
I.B$^{i.c.}_{57}$ & BabABabABabABaaBAbaBAbaBAbaBAbaBAbaBAbaBAbbABabABabABabA\\
I.B$^{i.c.}_{58}$ & BabaBABabaBABabaBABabaBABabaBAbabABAbabABAbabABAbabABAbabA\\
I.B$^{i.c.}_{59}$ & BAbabaBABAbabaBABababABABababABABababABABababABAbabaBABAbabaBA\\
I.B$^{i.c.}_{60}$ & BababABAbabABAbabABAbabABAbabABAbabaBAbabaBABabaBABabaBABabaBABabaBABababA\\
I.B$^{i.c.}_{61}$ & BabaBAbabABabABAbaBABabABAbaBAbabABabaBAbabABabaBAbaBABabABAbaBABabABabaBAbabA\\
I.B$^{i.c.}_{62}$ & BabABaaBAbaBAbaBAbaBBabABabABabABaaBAbaBAbaBAbbABabABabABabAAbaBAbaBAbaBAbbABabA\\
\hline
\end{tabular*}
{\rule{\temptablewidth}{1pt}}
\end{center}
\end{sidewaystable}

\begin{sidewaystable}
\tabcolsep 0pt \caption{The free group elements for the periodic three-body orbits.} \label{table-S8} \vspace*{-12pt}
\begin{center}
\def\temptablewidth{1\textwidth}
{\rule{\temptablewidth}{1pt}}
\begin{tabular*}{\temptablewidth}{@{\extracolsep{\fill}}ll}
\hline
Class \& number   &  \hspace{1.0cm}  free group element \\ \hline
I.B$^{i.c.}_{91}$ & BababABAbabaBABababABAbabABABabaBABAbabABAbabaBABababABAbabaBAbabaBABababABAbabaBABabaBABAbabABABabaBABaba\\&bABAbabaBABababA\\
I.B$^{i.c.}_{92}$ & BAbabaBABAbabABABababABABababABAbabaBABAbabaBABAbabABABababABABababABABabaBABAbabaBABAbabaBABababABABababA\\&BABabaBABAbabaBA\\
I.B$^{i.c.}_{93}$ & BAbabABAbabABAbabABAbabaBABabaBABabaBABababABAbabABAbabABAbabABABabaBABabaBABabaBABababABAbabABAbabABAbaba\\&BABabaBABabaBABabaBA\\
I.B$^{i.c.}_{94}$ & BabaBAbaBABabABAbaBABabABabaBAbabABabaBAbaBABabABAbaBABabABabaBAbabABabABAbaBABabABAbaBAbabABabaBAbabABabA\\
\hline
\end{tabular*}
{\rule{\temptablewidth}{1pt}}
\end{center}
\end{sidewaystable}

\begin{sidewaystable}
\tabcolsep 0pt \caption{The free group elements for the periodic three-body orbits.} \label{table-S8} \vspace*{-12pt}
\begin{center}
\def\temptablewidth{1\textwidth}
{\rule{\temptablewidth}{1pt}}
\begin{tabular*}{\temptablewidth}{@{\extracolsep{\fill}}ll}
\hline
Class \& number   & \hspace{1.0cm}  free group element\\\hline
I.B$^{i.c.}_{113}$ & BAbabaBABababABABabaBABababABABababABAbabaBABAbabaBABabaBABAbabaBABababABABababABAbabaBABAbabABAbabaBABAb\\&abaBABababABABababABAbabABABababABAbabaBA\\
I.B$^{i.c.}_{114}$ & BabaBABabaBABabABAbabABAbabABabaBABabaBABabaBAbabABAbabABAbaBABabaBABabaBAbabABAbabABAbaBABabaBABabaBAbab\\&ABAbabABAbabABabaBABabaBABabABAbabABAbabA\\
I.B$^{i.c.}_{115}$ & BAbabABABababABAbabaBABAbabABABababABAbabaBABAbabABABababABAbabaBABAbabABABabaBABAbabaBABababABABabaBABAb\\&abaBABababABABabaBABAbabaBABababABABabaBA\\
I.B$^{i.c.}_{116}$ & BAbabaBABAbabaBABAbabaBABAbabaBABAbababABABababABABababABABababABABababABABababABABababABABababABABababAB\\
\hline
\end{tabular*}
{\rule{\temptablewidth}{1pt}}
\end{center}
\end{sidewaystable}

\begin{sidewaystable}
\tabcolsep 0pt \caption{The free group elements for the periodic three-body orbits.} \label{table-S8} \vspace*{-12pt}
\begin{center}
\def\temptablewidth{1\textwidth}
{\rule{\temptablewidth}{1pt}}
\begin{tabular*}{\temptablewidth}{@{\extracolsep{\fill}}ll}
\hline
Class \& number   & \hspace{1.0cm}  free group element\\\hline
I.B$^{i.c.}_{136}$ & BAbabABAbabaBABabaBABAbabABAbabaBABabaBABAbabABAbabABABabaBABababABAbabABABabaBABababABAbabABABabaBABabab\\&ABAbabABABabaBABababABAbabABABabaBABabaBABAbabABAbabaBABabaBABAbabABAbabaBABabaBA\\
I.B$^{i.c.}_{137}$ & BAbabABABabaBABababABAbabaBABabaBABAbabABABabaBABabaBABAbabABABabaBABababABAbabaBABabaBABAbabABABabaBABAb\\&abABAbabaBABababABAbabABABabaBABAbabABAbabABABabaBABAbabABAbabaBABababABAbabABABabaBA\\
I.B$^{i.c.}_{138}$ & BAbabABABababABAbabaBABAbabABABababABAbabaBABAbabABABababABAbabaBABAbabABABababABAbabaBABAbabABABabaBABAb\\&abaBABababABABabaBABAbabaBABababABABabaBABAbabaBABababABABabaBABAbabaBABababABABabaBA\\
I.B$^{i.c.}_{139}$ & BAbabaBABababABAbabaBABAbabABABababABAbabaBABAbabABABababABAbabaBABAbabABABababABAbabaBABababABABababABAb\\
\hline
\end{tabular*}
{\rule{\temptablewidth}{1pt}}
\end{center}
\end{sidewaystable}

\begin{sidewaystable}
\tabcolsep 0pt \caption{The free group elements for the periodic three-body orbits.} \label{table-S8} \vspace*{-12pt}
\begin{center}
\def\temptablewidth{1\textwidth}
{\rule{\temptablewidth}{1pt}}
\begin{tabular*}{\temptablewidth}{@{\extracolsep{\fill}}ll}
\hline
Class \& number   & \hspace{1.0cm}  free group element\\\hline
I.B$^{i.c.}_{153}$ & BAbabABABabaBABababABAbabaBABabaBABababABAbabaBABabaBABAbabABABabaBABababABAbabaBABabaBABababABAbabaBABab\\&aBABAbabABABabaBABAbabABAbabaBABababABAbabABAbabaBABababABAbabABABabaBABAbabABAbabaBABababABAbabABAba\\&baBABababABAbabABABabaBA\\
I.B$^{i.c.}_{154}$ & BAbabaBABababABAbabaBABababABABabaBABAbabABABabaBABAbabaBABababABABabaBABAbabABABabaBABAbabaBABababABAbab\\&aBABababABABababABAbabaBABababABAbabaBABAbabABABabaBABAbabABABababABAbabaBABAbabABABabaBABAbabABABaba\\&bABAbabaBABababABAbabaBA\\
I.B$^{i.c.}_{155}$ & BAbabaBABAbabaBABAbabaBABAbabaBABAbabaBABAbabaBABAbabaBABAbabABABababABABababABABababABABababABABababABAB\\
\hline
\end{tabular*}
{\rule{\temptablewidth}{1pt}}
\end{center}
\end{sidewaystable}

\begin{sidewaystable}
\tabcolsep 0pt \caption{The free group elements for the periodic three-body orbits.} \label{table-S8} \vspace*{-12pt}
\begin{center}
\def\temptablewidth{1\textwidth}
{\rule{\temptablewidth}{1pt}}
\begin{tabular*}{\temptablewidth}{@{\extracolsep{\fill}}ll}
\hline
Class \& number   & \hspace{1.0cm}  free group element\\\hline
I.B$^{i.c.}_{176}$ & BAbabABAbabABAbabABAbabABAbabaBABabaBABabaBABabaBABabaBABababABAbabABAbabABAbabABAbabABABabaBABabaBABabaBA\\&BabaBABababABAbabABAbabABAbabABAbabABAbabaBABabaBABabaBABabaBABabaBA\\
I.B$^{i.c.}_{177}$ & BAbabaBABababABABabaBABAbabaBABAbabABABababABAbabaBABAbabABABababABABabaBABAbabaBABababABABababABAbabaBABA\\&babABABababABABabaBABAbabaBABababABABabaBABAbabaBABAbabABABababABAbabaBA\\
I.B$^{i.c.}_{178}$ & BAbabABAbabABAbabABABabaBABabaBABababABAbabABAbabABAbabaBABabaBABabaBABAbabABAbabABAbabABABabaBABabaBABaba\\&BABAbabABAbabABAbabaBABabaBABabaBABababABAbabABAbabABABabaBABabaBABabaBA\\
I.B$^{i.c.}_{179}$ & BAbabaBABababABABababABABabaBABAbabaBABAbabaBABababABABababABABabaBABAbabaBABAbabaBABababABABababABAbabaBA\\
\hline
\end{tabular*}
{\rule{\temptablewidth}{1pt}}
\end{center}
\end{sidewaystable}

\begin{sidewaystable}
\tabcolsep 0pt \caption{The free group elements for the periodic three-body orbits.} \label{table-S8} \vspace*{-12pt}
\begin{center}
\def\temptablewidth{1\textwidth}
{\rule{\temptablewidth}{1pt}}
\begin{tabular*}{\temptablewidth}{@{\extracolsep{\fill}}ll}
\hline
Class \& number   & \hspace{1.0cm}  free group element\\\hline

II.B$^{i.c.}_{1}$ & BabABabABabABabaBAbaBAbaBAbaBAbaBAbaBAbaBAbabABabABabABabA\\
II.B$^{i.c.}_{2}$ & BAbabABABabaBABababABAbabaBABabaBABababABAbabaBABabaBABAbabABABabaBABAbabABAbabaBABababABAbabABAbabaBABaba\\&bABAbabABABabaBA\\
II.B$^{i.c.}_{3}$ & BAbabABABababABABababABABabaBABAbabaBABAbabaBABAbabABABabaBABAbabaBABAbabaBABAbabABABababABABababABABabaBA\\
II.B$^{i.c.}_{4}$ & BabABaaBAbaBAbaBAbaBBabABabABabAAbaBAbaBAbaBBabABabABabABaaBAbaBAbaBAbbABabABabABabAAbaBAbaBAbaBBabABabABa\\&bAAbaBAbaBAbaBAbbABabA\\
II.B$^{i.c.}_{5}$ & BAbabABABabaBABababABAbabaBABAbabABAbabABABababABAbabaBABabaBABAbabABABabaBABAbabABAbabaBABababABABabaBABa\\
\hline
\end{tabular*}
{\rule{\temptablewidth}{1pt}}
\end{center}
\end{sidewaystable}

\begin{sidewaystable}
\tabcolsep 0pt \caption{The free group elements for the periodic three-body orbits.} \label{table-S8} \vspace*{-12pt}
\begin{center}
\def\temptablewidth{1\textwidth}
{\rule{\temptablewidth}{1pt}}
\begin{tabular*}{\temptablewidth}{@{\extracolsep{\fill}}ll}
\hline
Class \& number   & \hspace{1.0cm}  free group element\\\hline
II.C$^{i.c.}_{36}$ 	&	 BabABabABaaBAbaBAbaBAbaBBabABabABabAAbaBAbaBAbaBAbbABabABabA\\
II.C$^{i.c.}_{37}$ 	&	 BAbabABABabaBABababABAbabaBABababABAbabaBABababABAbabABABabaBA\\
II.C$^{i.c.}_{38}$ 	&	 BabABabABabABabaBAbaBAbaBAbaBAbaBAbaBAbaBAbaBAbabABabABabABabA\\
II.C$^{i.c.}_{39}$ 	&	 BabaBABabaBABabaBABabaBABabaBABababABAbabABAbabABAbabABAbabABAbabA\\
II.C$^{i.c.}_{40}$ 	&	 BabABaaBAbaBAbaBBabABabABabAAbaBAbaBAbaBBabABabABabAAbaBAbaBAbbABabA\\
II.C$^{i.c.}_{41}$ 	&	 BababABAbabABAbabABABabaBABabaBABababABAbabABAbabABABabaBABabaBABababA\\
II.C$^{i.c.}_{42}$ 	&	 BAbabaBABAbabABABababABABababABABababABABababABABababABABabaBABAbabaBA\\
\hline
\end{tabular*}
{\rule{\temptablewidth}{1pt}}
\end{center}
\end{sidewaystable}

\begin{sidewaystable}
\tabcolsep 0pt \caption{The free group elements for the periodic three-body orbits.} \label{table-S8} \vspace*{-12pt}
\begin{center}
\def\temptablewidth{1\textwidth}
{\rule{\temptablewidth}{1pt}}
\begin{tabular*}{\temptablewidth}{@{\extracolsep{\fill}}ll}
\hline
Class \& number   & \hspace{1.0cm}  free group element\\\hline
II.C$^{i.c.}_{76}$ & BabaBAbabABABabABAbabABabaBABabABAbabABabaBABabABABabaBAbabA\\
II.C$^{i.c.}_{77}$ & BabABabABabABaaBAbaBAbaBAbaBAbaBAbaBAbaBAbaBAbbABabABabABabA\\
II.C$^{i.c.}_{78}$ & BAbabaBABababABABababABABabaBABAbabABABababABABababABAbabaBA\\
II.C$^{i.c.}_{79}$ & BabABabABabABaaBAbaBAbaBAbaBAbaBAbaBAbaBAbaBAbaBAbbABabABabABabA\\
II.C$^{i.c.}_{80}$ & BabaBAbabABAbaBABabaBAbabaBABabABAbaBABabABAbabaBAbabABAbaBABabaBAbabA\\
II.C$^{i.c.}_{81}$ & BabaBABabaBABabaBABabaBABabaBABabABABabABAbabABAbabABAbabABAbabABAbabA\\
II.C$^{i.c.}_{82}$ & BabABabABabABabABabABabABabAAbaBAbaBAbaBAbaBBabABabABabABabABabABabABabA\\
\hline
\end{tabular*}
{\rule{\temptablewidth}{1pt}}
\end{center}
\end{sidewaystable}

\begin{sidewaystable}
\tabcolsep 0pt \caption{The free group elements for the periodic three-body orbits.} \label{table-S8} \vspace*{-12pt}
\begin{center}
\def\temptablewidth{1\textwidth}
{\rule{\temptablewidth}{1pt}}
\begin{tabular*}{\temptablewidth}{@{\extracolsep{\fill}}ll}
\hline
Class \& number   & \hspace{1.0cm}  free group element\\\hline
II.C$^{i.c.}_{121}$ & BabaBAbaBABabABAbaBAbabABabaBAbaBABabABAbaBABabABabaBAbabABabABAbaBABabABAbaBAbabABabaBAbaBABabABAbaBAba\\&bA\\
II.C$^{i.c.}_{122}$ & BabAAbaBAbaBAbbABabABaaBAbaBAbbABabABabAAbaBAbaBAbbABabABaaBAbaBAbaBBabABabABaaBAbaBAbbABabABaaBAbaBAbaB\\&BabA\\
II.C$^{i.c.}_{123}$ & BabaBABabaBABabaBABabaBABabaBABabaBABabaBABabaBABabaBABAbabABAbabABAbabABAbabABAbabABAbabABAbabABAbabABA\\&babA\\
II.C$^{i.c.}_{124}$ & BabABabABabABabABabABabABabABabaBAbaBAbaBAbaBAbaBAbaBAbaBAbaBAbaBAbaBAbaBAbaBAbabABabABabABabABabABabABa\\
\hline
\end{tabular*}
{\rule{\temptablewidth}{1pt}}
\end{center}
\end{sidewaystable}

\begin{sidewaystable}
\tabcolsep 0pt \caption{The free group elements for the periodic three-body orbits.} \label{table-S8} \vspace*{-12pt}
\begin{center}
\def\temptablewidth{1\textwidth}
{\rule{\temptablewidth}{1pt}}
\begin{tabular*}{\temptablewidth}{@{\extracolsep{\fill}}ll}
\hline
Class \& number   & \hspace{1.0cm}  free group element\\\hline
II.C$^{i.c.}_{146}$ & BababABAbabABABabaBABabaBABAbabABAbabABAbabaBABabaBABababABAbabABAbabaBABabaBABababABAbabABAbabaBABabaBA\\&BabaBABAbabABAbabABABabaBABababA\\
II.C$^{i.c.}_{147}$ & BababABAbabABAbabABAbabABAbabABAbabaBABabaBABabaBABabaBABabaBABabaBABAbabABAbabABAbabABAbabABAbabABAbaba\\&BABabaBABabaBABabaBABabaBABababA\\
II.C$^{i.c.}_{148}$ & BababABAbabaBABabaBABababABAbabABABabaBABabaBABAbabABAbabABABabaBABAbaBABAbabABABabaBABabaBABAbabABAbabA\\&BABabaBABababABAbabABAbabaBABababA\\
II.C$^{i.c.}_{149}$ & BabaBABabaBABabaBABabaBABAbabABAbabABAbabABAbaBABabaBABabaBABabaBABababABAbabABAbabABAbabABAbaBABabaBABa\\
\hline
\end{tabular*}
{\rule{\temptablewidth}{1pt}}
\end{center}
\end{sidewaystable}

\begin{sidewaystable}
\tabcolsep 0pt \caption{The free group elements for the periodic three-body orbits.} \label{table-S8} \vspace*{-12pt}
\begin{center}
\def\temptablewidth{1\textwidth}
{\rule{\temptablewidth}{1pt}}
\begin{tabular*}{\temptablewidth}{@{\extracolsep{\fill}}ll}
\hline
Class \& number   & \hspace{1.0cm}  free group element\\\hline
II.C$^{i.c.}_{168}$ & BAbabABabaBABAbabABabaBABAbabABabaBABAbabABababABAbaBABababABAbaBABababABAbaBABababABAbaBABababABAbaBABa\\&babABAbaBABababABAbaBABababABabaBABAbabABabaBABAbabABabaBABAbabABabaBA\\
II.C$^{i.c.}_{169}$ & BAbabABabaBABAbabABabaBABAbabABabaBABAbabABabaBABAbaBABababABAbaBABababABAbaBABababABAbaBABababABAbaBABa\\&babABAbaBABababABAbaBABAbabABabaBABAbabABabaBABAbabABabaBABAbabABabaBA\\
II.C$^{i.c.}_{170}$ & BAbabABABabaBABAbabABABabaBABAbabaBABabaBABAbabABABababABAbabaBABabaBABAbabaBABababABAbabaBABababABAbaba\\&BABAbabABAbabaBABababABABabaBABAbabABAbabaBABAbabABABabaBABAbabABABabaBA\\
II.C$^{i.c.}_{171}$ & BAbabABABabaBABababABAbabaBABababABAbabABABabaBABAbabABABabaBABababABAbabaBABabaBABAbabABABabaBABAbabABA\\
\hline
\end{tabular*}
{\rule{\temptablewidth}{1pt}}
\end{center}
\end{sidewaystable}

\begin{sidewaystable}
\tabcolsep 0pt \caption{The free group elements for the periodic three-body orbits.} \label{table-S8} \vspace*{-12pt}
\begin{center}
\def\temptablewidth{1\textwidth}
{\rule{\temptablewidth}{1pt}}
\begin{tabular*}{\temptablewidth}{@{\extracolsep{\fill}}ll}
\hline
Class \& number   & \hspace{1.0cm}  free group element\\\hline
II.C$^{i.c.}_{189}$ & BAbabABAbaBABabaBA\\
II.C$^{i.c.}_{190}$ & BabaBABBabAABAbabA\\
II.C$^{i.c.}_{191}$ & BaBAbaBAbaBAbaBAbA\\
II.C$^{i.c.}_{192}$ & BAbaabABBBAbabaBAAABabbaBA\\
II.C$^{i.c.}_{193}$ & BAbabABAbaBABababABAbaBABabaBA\\
II.C$^{i.c.}_{194}$ & BabABAbaBAbaBAbaBAbaBAbaBAbaBABabA\\
II.C$^{i.c.}_{195}$ & BabABabaBAbaBAbaBAbaBAbaBAbaBAbabABabA\\
\hline
\end{tabular*}
{\rule{\temptablewidth}{1pt}}
\end{center}
\end{sidewaystable}

\begin{sidewaystable}
\tabcolsep 0pt \caption{The free group elements for the periodic three-body orbits.} \label{table-S8} \vspace*{-12pt}
\begin{center}
\def\temptablewidth{1\textwidth}
{\rule{\temptablewidth}{1pt}}
\begin{tabular*}{\temptablewidth}{@{\extracolsep{\fill}}ll}
\hline
Class \& number   & \hspace{1.0cm}  free group element\\\hline
II.C$^{i.c.}_{229}$ & BabABabABabABabABabABabABabaBAbaBAbaBAbaBAbaBAbaBAbaBAbaBAbabABabABabABabABabABabABabA\\
II.C$^{i.c.}_{230}$ & BabABaaBAbaBAbaBAbaBBabABabABabABaaBAbaBAbaBAbaBAbaBAbbABabABabABabAAbaBAbaBAbaBAbbABabA\\
II.C$^{i.c.}_{231}$ & BabABabABaaBAbaBAbaBAbaBAbaBAbaBAbbABabABabABabABabABabABaaBAbaBAbaBAbaBAbaBAbaBAbbABabABabA\\
II.C$^{i.c.}_{232}$ & BabaBABabaBAbabABabaBABabaBABabaBAbabABabaBAbabABabaBAbabABabaBAbabABAbabABAbabABabaBAbabABAbabA\\
II.C$^{i.c.}_{233}$ & BAbaBABabaBAbabABabaBABabaBABabABAbabABabaBABabaBAbabABAbabABabaBABabABAbabABAbabABabaBAbabABAbaBA\\
II.C$^{i.c.}_{234}$ & BabaBAbaBABabABabaBAbaBAbabABabABAbaBABabABabaBAbaBAbabABabABAbaBABabABabaBAbaBAbabABabABAbaBAbabA\\
II.C$^{i.c.}_{235}$ & BabaBAbaBABabABabABAbaBAbabABabABAbaBABabABabaBAbaBAbabABabABAbaBABabABabaBAbaBABabABabABAbaBAbabA\\
\hline
\end{tabular*}
{\rule{\temptablewidth}{1pt}}
\end{center}
\end{sidewaystable}

\clearpage

\begin{sidewaystable}
\tabcolsep 0pt \caption{The free group elements for the periodic three-body orbits.} \label{table-S8} \vspace*{-12pt}
\begin{center}
\def\temptablewidth{1\textwidth}
{\rule{\temptablewidth}{1pt}}
\begin{tabular*}{\temptablewidth}{@{\extracolsep{\fill}}ll}
\hline
Class \& number   & \hspace{1.0cm}  free group element\\\hline
II.C$^{i.c.}_{259}$ & BAbabABabaBABabaBABabaBAbabABAbaBABabABAbabABAbabABabaBABabABAbabABAbaBABabaBABabABAbabABabaBABabaBABabA\\&BAbaBABabaBAbabABAbabABAbabABabaBA\\
II.C$^{i.c.}_{260}$ & BAbabaBABababABAbabaBABababABABabaBABAbabaBABabaBABAbabaBABababABABabaBABAbabABABababABAbabaBABAbabABAba\\&baBABAbabABABababABAbabaBABababABAbabaBA\\
II.C$^{i.c.}_{261}$ & BAbabABABabaBABababABAbabaBABAbabABABabaBABababABAbabaBABAbabABAbabaBABababABAbabaBABabaBABAbabaBABababA\\&BAbabABABabaBABAbabaBABababABAbabABABabaBA\\
II.C$^{i.c.}_{262}$ & BabAAbaBAbaBBabABabAAbaBAbaBAbbABabABaaBAbaBBabABabABaaBAbaBAbbABabABaaBAbaBAbbABabABaaBAbaBAbbABabABabA\\
\hline
\end{tabular*}
{\rule{\temptablewidth}{1pt}}
\end{center}
\end{sidewaystable}

\begin{sidewaystable}
\tabcolsep 0pt \caption{The free group elements for the periodic three-body orbits.} \label{table-S8} \vspace*{-12pt}
\begin{center}
\def\temptablewidth{1\textwidth}
{\rule{\temptablewidth}{1pt}}
\begin{tabular*}{\temptablewidth}{@{\extracolsep{\fill}}ll}
\hline
Class \& number   & \hspace{1.0cm}  free group element\\\hline
II.C$^{i.c.}_{284}$ & BAbabABABabABABabaBABAbabABAbabaBABababABAbabABAbabaBABababABAbabABAbabaBABababABAbabABABabaBABAbabABABa\\&bABABabaBABAbabABABabaBABababABAbabaBABabaBABababABAbabaBABabaBABababABAbabaBABabaBABAbabABABabABABabaBA\\
II.C$^{i.c.}_{285}$ & BAbabaBABAbabaBABAbabaBABAbabaBABAbabaBABAbabaBABABababABABababABABababABABababABABababABABababABABababA\\&BABababABABababABABababABABababABABababABABababABABababABABAbabaBABAbabaBABAbabaBABAbabaBABAbabaBABAbabaBA\\
II.C$^{i.c.}_{286}$ & BAbabABABabaBABababABAbabaBABabaBABababABABabaBABabaBABAbabABABabaBABababABAbabaBABabaBABAbabABABabaBABa\\&babABAbabABABabaBABAbabABAbabaBABababABAbabABABabaBABAbabABAbabABABababABAbabABAbabaBABababABAbabABABabaBA\\
II.C$^{i.c.}_{287}$ & BAbabABAbabABAbabABAbabABABabaBABabaBABabaBABabaBABababABAbabABAbabABAbabABAbabaBABabaBABabaBABabaBABaba\\
\hline
\end{tabular*}
{\rule{\temptablewidth}{1pt}}
\end{center}
\end{sidewaystable}

\end{document}